\documentclass[11pt,letterpaper,usenames,dvipsnames]{article}
\usepackage{jheppub}

\allowdisplaybreaks[2]  
\usepackage{pifont}
\newcommand{\cmark}{\text{\ding{51}}}
\newcommand{\xmark}{\text{\ding{55}}}
\usepackage{bm} 
\usepackage{dsfont} 

\usepackage{tikz} 
\usetikzlibrary{arrows,snakes,shapes.arrows,decorations.markings}
     \tikzset{>=triangle 90}
     \tikzstyle{gr}=[draw,circle,green!50!black,fill=green!50!black,scale=.6]
     \tikzstyle{Bl}=[draw,circle,blue,scale=.6]
     \tikzstyle{R}=[draw,circle,fill=red,scale=.6]
     \tikzstyle{bl}=[draw,circle,fill=black,scale=.35]
     \tikzstyle{bbc}=[draw,circle,fill=black,scale=.75]
     \tikzstyle{bbcs}=[draw,circle,fill=black,scale=.5]
     \tikzstyle{rc}=[circle,fill=red,scale=.6]
     \tikzstyle{wc}=[draw,circle,scale=.75]

\usepackage{array} 
\usepackage{multirow} 
\usepackage{colortbl} 
\usepackage{stfloats}  

\usepackage{xcolor}   

\usepackage{url} 

\newcommand{\beq}{\begin{equation}}
\newcommand{\eeq}{\end{equation}}
\def\del{{\partial}}

\def\til{\widetilde}

\def\vev#1{{\langle{#1}\rangle}} 

\def\^{\wedge}
\def\I{\mathds{1}}
\def\Tr{\mathop{\rm Tr}}

\def\U{\mathop{\rm u}}
\def\SU{\mathop{\rm su}}

\def\SO{\mathop{\rm so}}
\def\SL{\mathop{\rm SL}}
\def\GL{\mathop{\rm GL}} 
\def\Sp{\mathop{\rm sp}}

\def\C{\mathbb{C}} 

\def\N{\mathbb{N}} 
 
\def\R{\mathbb{R}} 
\def\Z{\mathbb{Z}} 
\def\ff{{\mathfrak f}}

\def\be{{\bf e}}

\def\tM{{\til M}}
\def\tm{{\til m}}
\def\bm{{\bf m}}
\def\tN{{\til N}}
\def\bn{{\bf n}}

\def\bR{{\bf R}}
\def\tr{{\til r}}

\def\bz{{\boldsymbol{z}}}

\def\cN{{\mathcal N}}

\def\cS{{\mathcal S}}

\def\a{{\alpha}}

\def\ba{{\boldsymbol\a}}
\def\b{{\beta}}
\def\bb{{\boldsymbol\b}}

\def\g{{\gamma}}

\def\G{{\Gamma}}

\def\D{{\Delta}}

\def\th{{\theta}}
\def\k{{\kappa}}
\def\l{{\lambda}}
\def\L{{\Lambda}}
\def\m{{\mu}}

\def\s{{\sigma}}
\def\S{{\Sigma}}
\def\t{{\tau}}

\def\f{{\phi}}

\def\w{{\omega}}
\def\bw{{\boldsymbol\w}}
\def\ccw{\cellcolor{white}}

\def\rcy{\rowcolor{black!25!yellow!10}}

\def\ccr{\cellcolor{red!15}}

\title{Geometric constraints on the space of N=2 SCFTs II: Construction of special K\"ahler geometries and RG flows}
\author{Philip C. Argyres,}
\author{Matteo Lotito,}
\author{Yongchao L\"u}
\author{and Mario Martone}
\affiliation{University of Cincinnati,
Physics Department, PO Box 210011, Cincinnati OH 45221}
\emailAdd{philip.argyres@gmail.com}
\emailAdd{lotitomo@mail.uc.edu}
\emailAdd{lychaoaa@gmail.com}
\emailAdd{martonmo@ucmail.uc.edu}

\abstract{This is the second in a series of three papers on systematic analysis of rank 1 Coulomb branch geometries of four dimensional $\cN{=}2$ SCFTs.  In \cite{Argyres:2015ffa} we developed a strategy for classifying physical rank-1 CB geometries of $\cN{=}2$ SCFTs.  Here we show how to carry out this strategy computationally to construct the Seiberg-Witten curves and one-forms for all the rank-1 SCFTs.  Explicit expressions are given for all 28 cases, with the exception of the $N_f{=}4$ $\SU(2)$ gauge theory and the $E_n$ SCFTs which were constructed in \cite{Seiberg:1994rs,Seiberg:1994aj} and \cite{Minahan:1996fg,Minahan:1996cj}.

}
 
\begin{document}
\maketitle

\section{Introduction}

Since the work of Seiberg and Witten \cite{Seiberg:1994rs,Seiberg:1994aj}, a rich variety of techniques have been brought to bear in the study of the strong coupling dynamics of four dimensional supersymmetric $\cN=2$ theories.  These techniques extend well beyond the usual lagrangian methods. While a complete list of lagrangian field theories is known \cite{Bhardwaj:2013qia}, there is by now a large list of non-lagrangian theories but still not a complete classification. Recently, progress has been made in trying to study and classify conformal fixed points both using the revived conformal bootstrap approach \cite{Beem:2014zpa,Liendo:2015ofa}, $\cS$-duality techniques \cite{Chacaltana:2010ks,Chacaltana:2011ze,Chacaltana:2014jba,Chacaltana:2014nya,Chacaltana:2016shw}, and geometric engineering constructions \cite{Xie:2015rpa,Xie:2015xva}.  

In this work we present results which make a classification of $\cN=2$ SCFTs through a geometric approach possible. This is the second paper of a series of three.  In the first \cite{Argyres:2015ffa} we outlined a program for a classification based on the study of the Coulomb branch geometries of $\cN=2$ SCFTs. The Coulomb branch (CB) of $\cN=2$ supersymmetric theories plays a special role as it cannot be lifted by any $\cN=2$ supersymmetry preserving deformation of the theory.   The low energy $\cN=2$ supersymmetry on the CB constrains its geometry to be rigid special K\"ahler.  Our goal is to classify the possible rigid special K\"ahler geometries.  Such a classification would restrict the possible $\cN=2$ theories, but would not prove the existence of a theory for each geometry.

The difficulty of this classification stems in part from that fact that physically interesting special K\"ahler geometries have singularities. In \cite{Argyres:2015ffa} we showed that there are physical consistency requirements on the kinds of singularities that can occur beyond mere special K\"ahlerness.  There we organized the classification problem by studying the scale-invariant geometries (associated to supercconformal field theories, SCFTs) and their deformations (associated to $\cN=2$ preserving deformations of the SCFT by relevant operators) for rank-1 theories (ones with a one-dimensional CB).  In \cite{Argyres:2015ffa} we extensively describe our definition of {\it physical consistency} and carry out the complete classification of rank 1 CB geometries (reported in table 1 of that paper) which includes all the known rank 1 $\cN=2$ SCFTs plus sixteen new ones. 

In \cite{Argyres:2015ffa} we claimed the existence of the above classification of deformations, but showed no proof of it. This task will be carried out here. We will not only show the existence of such deformations but we will explicitly write down the Seiberg-Witten (SW) curve and SW one-form for each case. The curves and the one-forms will depend explicitly on the mass parameters which initiate the RG-flow. We will present a general technique which allows one to write down SW curves for any {\it sub-maximal} deformation (more below) of scale-invariant Kodaira singularities. The explicit expression for the curve then allows us to reconstruct the maximal flavor symmetry algebra, $F$, of the theory.  The results of this construction are reported here in table \ref{table:theories} (which just reproduces table 1 from \cite{Argyres:2015ffa}).

\begin{table}
\centering
$\begin{array}{|c|c|c|c|}
\hline
\multicolumn{4}{|c|}{\qquad \text{\bf Deformations of planar, rank 1, scale-invariant CBs satisfying}\qquad \,} \\
\multicolumn{4}{|c|}{\text{\bf low energy supersymmetry and Dirac quantization constraints}} \\
\hline\hline
\ \ \text{\#}\ \ \,  &
\quad\qquad \text{SCFT}\quad\qquad\,  &
\quad\qquad \text{deformation}\quad\qquad\, &
\text{max flavor} 
\\
& \text{singularity} & \text{pattern} & \text{symmetry $F$}
\\
\hline
1. & &\{{I_1}^{10}\} & E_8  
\\
2. & &\{{I_1}^6,I_4\} & \Sp(10) 
\\
\rcy \ccw
3. & \ccw & \{{I_1}^2,{I_4}^2\} & \ccr \Sp(4) 
\\
\rcy \ccw
4. & \ccw &\{{I_1}^4,I^*_0\} &F_4 
\\
\rcy \ccw
5. & \ccw &\{{I_1}^3,I^*_1\} &\ccr \Sp(6)\ \text{or}\ \SO(7) 
\\
\rcy \ccw
6. & \ccw &\{I_3,I^*_{1\ Q{=}\sqrt3}\} &\SU(2)
\\
\rcy \ccw
7. & \ccw &\{{I_1}^2,I^*_2\} &\Sp(4) 
\\
\rcy \ccw
8. & \ccw &\{I_1,I^*_3\} &\SU(2) 
\\
\rcy \ccw
9. & \ccw &\{I_2,IV^*_{Q{=}\sqrt2}\} &\SU(2)
\\
\rcy \ccw
10. & \ccw &\{{I_1}^{2},IV^*\} &G_2 
\\ 
\rcy \ccw 
11. &\ccw \multirow{-11}{*}{$II^*$}   
&\{I_1,III^*\} &\SU(2) 
\\
\hline 
12. & \ccw &\{{I_1}^9\} &E_7   
\\ 
13. & \ccw &\{{I_1}^5,I_4\} &\ \Sp(6)\oplus\Sp(2)\ \,
\\
\rcy \ccw
14. & \ccw &\{{I_1}^3,I^*_0\} &\SO(7) 
\\
\rcy \ccw
15. & \ccw &\{{I_1}^2,I^*_1\} &\ccr \SU(2)\oplus\SU(2) 
\\
\rcy \ccw
16. & \ccw &\{I_2,I^*_{1\ Q{=}\sqrt2}\} &\SU(2) 
\\
\rcy \ccw
17. & \ccw &\{I_1,I^*_2\} &\SU(2) 
\\
\rcy \ccw 
18. & \ccw \multirow{-7}{*}{$III^*$}  
&\{I_1,IV^*\} &\SU(2) 
\\
\hline
19. & &\{{I_1}^8\} &E_6  
\\
20. & &\{{I_1}^4,I_4\} &\Sp(4)\oplus\U(1) 
\\
\rcy \ccw
21. & \ccw &\{{I_1}^2,I^*_0\} &\SU(3) 
\\
\rcy \ccw 
22. & \ccw \multirow{-5}{*}{$IV^*$} 
&\{I_1,I^*_1\} &\U(1) 
\\
\hline
23. & &\{{I_1}^6\} &\SO(8)
\\
24. & &\{{I_2}^3\} &\Sp(2) 
\\
\rcy \ccw
25. & \ccw \multirow{-3}{*}{$I^*_0$} 
&\{{I_1}^2,I_4\} &\Sp(2) 
\\
\hline
26. & IV 
&\{{I_1}^4\} &\SU(3)
\\
\hline
27. & III
&\{{I_1}^3\} &\SU(2)
\\
\hline
28. & II 
&\{{I_1}^2\} &-
\\
\hline
\end{array}$
\caption{\label{table:theories} 
The 28 families of deformed planar rank-1 CB geometries consistent with the low energy Dirac quantization condition.  Column 2 lists the Kodaira type of the scale invariant CB geometry, column 3 the resulting singularity types under a generic relevant deformation, and column 4 the maximal flavor symmetry of the SCFT. The values for the central charges for the entries in the table are known and can be found in table 1 of \cite{Argyres:2015ccharges}.}
\end{table}

This table lists the geometries in terms of the Kodaira type ($II$, $II^*$, etc., and reviewed in section \ref{sec:math} below) of the SCFT singularity, and the set of Kodaira singularities into which it is deformed.  The shaded entries are the new geometries found by our construction.  In each case we find a single geometry associated to each deformation pattern.  
We also list the maximal flavor symmetry associated with each geometry, which is uniquely determined except for the case of entry number 5, where an ambiguity persists.
Note that by $\Sp(2r)$ we mean the algebra of rank $r$.  The $Q=\sqrt q$ subscripts appearing on some of the singularities in the deformation column record the unit of charge quantization \cite{Argyres:2015ffa} in these theories; the $I^*_n$, $III^*$, or $IV^*$ singularities appearing without such subscripts all have $Q=1$, while for $I_n$ singularities they are $Q=\sqrt n$.  

In the last column, three maximal flavor symmetry entries (numbers 3, 5, and 15) are shaded red.  These indicate theories which do not satisfy the an RG flow self-consistency test (described below) for this maximal flavor assignment.  There are possible sub-maximal flavor assignments, described in \cite{Argyres:2016xua}, for which the RG flow test is also carried out.  There it is found that no flavor symmetry assignment for entry 3 (the $II^*\to\{{I_1}^2,{I_4}^2\}$ geometry) is consistent, but there are consistent assignments for the other two.

The paper is organized as follows.  Section \ref{sec:math} presents a brief review of the basic mathematical ingredients needed to carry out our construction and introduce a more systematic definition of deformation of the initial scale-invariant singularity.  (The first part of this section is not intended to be self-contained but serves more as a reminder and to set up notation --- a much more detailed discussion of these topics can be found in \cite{Argyres:2015ffa}.)  

Section \ref{sec:defs}, \ref{section:one-form} and \ref{RGflow} are the heart of the paper.  In section \ref{sec:defs} we present the construction of the SW curves.  In particular we carefully define and discuss {\it maximal} and {\it sub-maximal} deformations of Kodaira singularities. The string web description of neutral BPS state is reviewed \cite{DeWolfe:1998zf,DeWolfe:1998eu} and the operation of coalescence of singularities is introduced. These are the necessary ingredients to go from the {\it deformation pattern} associated to each $\cN=2$ SCFTs to the explicit form of the SW curve. We carry out our construction explicitly in two cases --- $II^*\to\{{I_1}^6,I_4\}$ and $II^*\to\{{I_1}^2,{I_4}^2\}$ --- which will be used as the two working examples throughout the paper.   Section \ref{section:one-form} describes how to construct the SW one-form.  Our method is a slight generalization of one introduced by Minahan and Nemeschansky \cite{Minahan:1996fg,Minahan:1996cj}.  The existence of the SW one-form generally involves finding solutions of an over-constrained system of equations. We find a solution for each one of the deformation patterns.  We report the full list of resulting SW curves and one-forms in appendix \ref{app:curves}.  

Section \ref{RGflow} describes a further physical condition any CB geometry must satisfy to be identified as an $\cN=2$ SCFT.   This {\it RG flow constraint} requires that for particular values of the mass parameters determined by the detailed form of the SW curve, the singularity should split appropriately in a way dictated by the unbroken flavor group.  Since evaluating this constraint requires the explicit construction of the CB geometry, it was only  briefly discussed in \cite{Argyres:2015ffa}, and is fully discussed here.  Part of this discussion leads to the possible interpretation of some rank 1 CB geometries as the result of weakly gauging $\U(1)$ or $\SU(2)$ subgroups of the flavor group of rank-0 SCFTs.  

We conclude in section \ref{sec:end} with a list of open questions.  For the purposes of keeping the length of the paper limited, we omitted the derivation of the results reported in the appendices. We will be happy to provide any interested reader with a Mathematica notebook where all the results are derived.  Finally, we will, where convenient, switch between the standard and Dynkin names for the classical simple Lie algebras: 
\begin{align}\label{}
A_n\equiv\SU(n+1),\quad B_n\equiv \SO(2n+1),\quad C_n\equiv\Sp(2n),\quad D_n\equiv\SO(2n).
\end{align}

\section{Review of rank 1 special K\"ahler geometry}\label{sec:math}

In this section we give a quick review of rank-1 Coulomb branch (CB) geometries, summarizing what we described in section 2 and appendix A of \cite{Argyres:2015ffa}.  We focus on the features of the objects we are aiming to construct in this paper --- the Seiberg-Witten (SW) curves and one-forms --- which provide all the information needed to specify the low-energy physics on the Coulomb branch.  We also review the normalization of electric and magnetic charges, and the constraints they satisfy coming from Dirac quantization and the {\it safely irrelevant conjecture} \cite{Argyres:2015ffa}. We then define some simple analytic and topological invariants of rank-1 CB geometries.  Together with the constraints coming from the safely irrelevant conjecture and from Dirac quantization, these invariants allow us to restrict the set of physical deformations of scale invariant rank-1 CB geometries to the list of 28 possibilities, given in table \ref{table:theories} above.

\subsection{Basics of rank-1 SK geometries}

We assume our rank-1 (i.e., 1 complex-dimensional) Coulomb branch is isomorphic as a complex space to the complex plane.  Such a ``planar" CB is parametrized by some complex coordinate $u \in \C$ which is a holomorphic function of the vev $\langle \phi \rangle$ of the scalar component of the low energy $\U(1)$ vector multiplet.

As recalled in \cite{Argyres:2015ffa}, CBs of $\cN=2$ SCFTs  correspond to scale invariant geometries with one singular point which can always be chosen to be located at the origin. $u=0$ is also the only vacuum in the theory which preserves scale invariance.  In vacua with $u\neq 0$ the distance to the origin represents a scale in the theory, and in these vacua scale invariance is broken spontaneously.  Electric-magnetic duality of the low energy $\U(1)$ gauge theory introduces further constraints which restrict the allowed scale invariant geometries to be a finite set of special K\"ahler geometries: complex cones characterized by a set of values of their deficit angles. These spaces follow Kodaira's classification of elliptic surfaces \cite{KodairaI, KodairaII}, and are listed in table \ref{Table:Kodaira} below.

In \cite{Argyres:2015ffa} we explained in detail that a given scale invariant CB does not correspond to a unique SCFT: one must specify in addition (at least) the set of allowed mass deformations. Mass deformation parameters $m_i$ have scaling dimension $\Delta(m_i)=1$.  When the deformation parameters vanish, $m_i=0$, the CB geometry is scale invariant and has a global internal symmetry $\U(2)_R\oplus F$, where $\U(2)_R$ is the R-symmetry and $F$ is the flavor symmetry.  Turning on masses explicitly breaks the flavor symmetry.  Since the masses appear as vevs of vector multiplets upon weakly gauging $F$, they can be thought of as linear coordinates on $\ff_\C$, the complexified Cartan subalgebra of $F$.  Thus we write $\bm = m_i \be^i\in \ff_\C$ with $\{\be^i\}$ a basis of $\ff$, and we will call $\bm$ ``linear masses".   At generic masses the flavor symmetry algebra is broken to a rank$(F)$ abelian algebra, $F\to\oplus_i^{\text{rank}(F)} \U(1)_i$ whose generators, $\{\be^i\}$, form a basis of $\ff$.  States can be classified by their flavor charges $\w^i$ under each $\U(1)_i$ factor.   
When the mass dimension of the CB vev $u$, $\Delta(u)$, is between 1 and 2, $1<\Delta(u)<2$, there is the possibility of a deformation parameter $\m$ with dimension $\Delta(\m)=2-\Delta(u)$, described in \cite{Argyres:1995xn}. When $\Delta(u)=2$ there can also be a marginal deformation parameter $\t$, with $\Delta(\t)=0$. For a complete analysis of $\cN=2$ SUSY preserving deformations see \cite{Argyres:2015ffa}. We will often suppress mention of $\m$ and $\t$ for simplicity of notation in what follows.

For a rank 1 theory, the leading low energy physics on the CB is that of a free $\U(1)$ $\cN=2$ gauge theory coupled to massive charged sources.  This coupling depends holomorphically on the CB coordinate and is ambiguous up to fractional-linear $SL(2,\Z)$ EM-duality transformations.  It thus describes in an EM-duality invariant way a complex 2-torus fibered over the CB.  We will indicate this torus fiber by $\S(u,\bm)$, and the total space of the fiber bundle by $\S$.  The fiber can be written as an elliptic curve in Weierstrass form as
\begin{align}\label{rank1curve}
\S(u,\bm): \qquad y^2 &= x^3 + f(u,\bm)\,x + g(u,\bm),
\end{align}
where $f,g$ are polynomial functions of the complex coordinate $u$ and of the mass deformation parameters $\bm$.  We will also refer to $\S$ as the SW curve.

The scalar and fermion kinetic terms of the $\U(1)$ gauge multiplet on the CB are furthermore determined by a meromorphic one-form $\l(u)$, the SW one-form.  $\l$ satisfies the following two special K\"ahler (SK) conditions:
\begin{align}\label{SWform}
\text{(I)}\quad\del_u \l &= \k\frac{{\rm d}x}{y} + {\rm d}\f, &
\text{(II)}\quad\text{Res}(\l) &\in \{\bw(\bm)\ |\ \bw\in\L_F\}.
\end{align}
Here $\k$ is an arbitrary non-zero numerical constant, $\f$ is an arbitrary meromorphic function on the fiber, Res$(\l)$ means the residue of $\l$ at any of its poles, and $\L_F$ is the root lattice of $F$.  Elements of the root lattice and masses are dual, $\bw\in\ff^*$ and $\bm\in\ff_\C$, and $\bw(\bm) := \w^i m_i$ is the dual pairing.  For a given choice of $\k$, the first SK condition in \eqref{SWform} determines the normalization of $\l$.  The second SK condition does not determine a normalization since the normalization of $\L_F$ depends on that of the Killing form on $F$ which has not been specified.  This normalization is actually arbitrary since the only scales in the problem are given by $u$ and $\bm$, whose overall normalizations have no independent definition in a CFT.  (In the case of a lagrangian theory, however, this normalization can be compared to a conventional one at weak coupling.)  In the following we will also refer to the pair $(\Sigma,\lambda)$ as a SW geometry.

For a given value of $\bm$, the elliptic curve is singular at values of $u$ corresponding to the zeros of the discriminant of the right side of \eqref{rank1curve}
\begin{align}\label{disc}
D_x \equiv 4\, f^3 + 27\, g^2 = 0.
\end{align}
These singularities physically correspond to points on the CB where $\U(1)$-charged states become massless.  

\begin{table}
\centering
$\begin{array}{|c|l|c|c|c|c|c|}
\hline
\multicolumn{7}{|c|}{\text{\bf Possible scaling behaviors near singularities of a rank 1 CB}}\\
\hline\hline
\text{Name} & \multicolumn{1}{c|}{\text{planar SW curve}} & \ \text{ord}_0(D_{x})\ \ &\ \D(u)\ \ & M_{0} & \text{deficit angle} 
& \t_0 \\
\hline
II^* &\parbox[b][0.45cm]{4cm}{$\ y^2=x^3+u^5$} 
&10 &6 &ST &5\pi/3 &\ e^{2\pi i/3}\ \\
III^* &\ y^2=x^3+u^3x &9 &4 &S &3\pi/2 & i\\
IV^* &\ y^2=x^3+u^4 &8 &3 &(ST)^2 &4\pi/3 &e^{2\pi i/3}\\
I_0^* &\ y^2=\prod_{i=1}^3\left(x-e_i(\t)\, u\right)
&6 &2 &-I &\pi &\t\\
IV &\ y^2=x^3+u^2 &4 &3/2 &(ST)^{-2} &2\pi/3 &e^{2\pi i/3}\\
III &\ y^2=x^3+u x &3 &4/3 &S^{-1} &\pi/2 & i\\
II &\ y^2=x^3+u &2 &6/5 &(ST)^{-1} &\pi/3 &e^{2\pi i/3}\\
\hline
\hline
I^*_n\ \ (n{>}0) &
\parbox[b][0.45cm]{5cm}{
$\ y^2=x^3+ux^2+\L^{-2n}u^{n+3}\ \ $}
& n+6 & 2 & -T^n & 2\pi\ \text{(cusp)} 
& i\infty\\
I_n\ \ (n{>}0)    &\ y^2=(x-1)(x^2+\L^{-n}u^n)  
& n     & 1 & T^n & 2\pi\ \text{(cusp)} 
& i\infty\\[0.5mm]
\hline
\end{array}$
\caption{\label{Table:Kodaira} Scaling forms of rank 1 planar special K\"ahler singularities, labeled by their Kodaira type (column 1), a representative family of elliptic curves with singularity at $u=0$ (column 2), order of vanishing of the discriminant of the curve at $u=0$ (column 3), mass dimension of $u$ (column 4), a representative of the $SL(2,\Z)$ conjugacy class of the monodromy around $u=0$ (column 5), the deficit angle of the associated conical geometry (column 6), and the value of the low energy $\U(1)$ coupling at the singularity (column 7).  The first seven rows are scale invariant.  The last two rows give infinite series of singularities which have a further dimensionful parameter $\L$ so are not scale invariant; they are IR free since $\t_0=i\infty$.}
\end{table}

As already mentioned, for $\bm=0$ scale-invariance restricts the geometry to have only one singular point, but this need not be the case for $\bm\neq0$.  In fact for generic values of the linear mass parameters $\bm$, the initial singularity splits into ones of lesser order \cite{Argyres:2015ffa}.  For our study we only need to understand the vicinity of a singularity. Upon scaling-in closely enough to a given singularity, the geometry becomes locally approximately scale-invariant and thus ``locally Kodaira''.  It follows that turning on $\bm$ splits the initial singularity into lesser Kodaira singularities.  Kodaira singularities play a central role in this work, and explicit expressions for their curves are reported in table \ref{Table:Kodaira}.  The freedom to shift and rescale the $x$, $y$, and $u$ variables appropriately has been used to put the curves in Weierstrass form, to put the singularity at $u=0$, and to fix the normalizations of the terms.  The first seven entries are scale-invariant singularities and the last two are infinite series of singularities depending on an extra dimension-one parameter, $\L$.  The $I_0^*$ singularity is actually a 1-complex-dimensional family of singularities depending on a dimensionless parameter, $\t$.

For the scale-invariant singularities, the equation \eqref{SWform} for the SW one-form is solved by $\l \sim u\, {\rm d}x/y$, which is proportional to the holomorphic one-form of the fiber.  The discriminants of these singularities are homogeneous in $u$, $D_x \sim  u^{n}$ for some $n:=\text{ord}_0(D_x)$ listed in table \ref{Table:Kodaira}.  In particular, the only singular fiber on the Coulomb branch is at the origin.

The non-scale-invariant $I_n$ and $I^*_n$ singularities have singular fibers not only at $u=0$ but also at points $u\sim\L^{\Delta(u)}$.  We are only interested in the vicinity of the origin, $|u|\ll \L^{\Delta(u)}$, since these singularities have the interpretation as IR free field theories near $u=0$, with $\L$ playing the role of the strong coupling scale (Landau pole).  The SW one-form is again of the form $\l\sim u\, {\rm d}x/y$ near $u=0$, and ord${}_0 (D_x)$ is given in the table.

\subsection{Charge normalization, Dirac quantization, and the safely irrelevant conjecture}\label{sec2.2}

Here we will reproduce the main formulae which set the relative normalization of the electric and magnetic charges of the massless states associated with CB singularities.  A detailed discussion can again be found in Appendix A of \cite{Argyres:2015ffa}.

Denote by $\bz$ the row vector of the physical magnetic and electric charges of a particle, $\bz := (p,q)$.  The general solution of the Dirac-Schwinger-Zwanziger quantization condition \cite{Dirac:1931kp, Schwinger:1969ib, Zwanziger:1968rs} (in an appropriate normalization of the charges) is that $\bz$ is of the form
\begin{align}\label{zvect}
\bz := (p,q) = \sqrt P \,(c,d),
\qquad\text{with}\qquad c,d\in \Z,
\end{align}
with $P\in\N$.  In particular, for $P$ not a perfect square, the electric and magnetic charges are not integral.

The $\SL(2,\Z)$ group of electric-magnetic (EM) duality transformations acts linearly on charge vectors.  It is useful to specify a particular set of generators of $\SL(2,\Z)$, 
\begin{align}\label{}
 S&:=\begin{pmatrix}0 & -1 \\ 1 & 0 \end{pmatrix},& 
 T&:= \begin{pmatrix}1 & 1 \\ 0 & 1 \end{pmatrix} ,
\end{align}
which satisfy the relations $S^2=(ST)^3=-I$.  There are two  basic EM-duality invariants built from the charges.  One is the EM duality \emph{invariant charge}, $Q$, defined by
\begin{align}\label{Qdef}
Q^2 := \gcd(p^2,q^2)
\qquad \text{and} \qquad 
Q>0.
\end{align}
Thus $Q=\sqrt{P}\gcd(c,d)$.  The second is the (EM duality invariant) \emph{charge inner product} given by
\begin{align}\label{dszform}
\vev{\bz_1,\bz_2}  
:=  - \bz_1 S \bz_2^T 
= \det
\begin{pmatrix} p_1 & q_1 \\ p_2 & q_2 \end{pmatrix} 
= P\,\det
\begin{pmatrix} c_1 & d_1 \\ c_2 & d_2 \end{pmatrix}.
\end{align}

Charges and the charge inner product are encoded in the SW curve as follows.  The Dirac quantization condition specifies a choice of polarization on the  torus fiber \cite{Donagi:1995cf}.   A polarization is equivalent to a non-degenerate integral antisymmetric pairing, $\vev{\cdot,\cdot}$, of 1-cycles on the torus.  Thus for a 2-torus a polarization is a positive integral multiple of the intersection form for 1-cycles, and this multiple is $P$.  A \emph{canonical basis} $\{\a,\b\}$ of 1-cycles is one where the polarization is given by
\begin{align}\label{polarization}
\vev{\a,\a} = \vev{\b,\b} = 0, \qquad 
\vev{\a,\b} = -\vev{\b,\a} = P. 
\end{align}
The subgroup of $\GL(2,\Z)$ transformations of $H_1(\S,\Z)$ which preserve the polarization is the EM duality group, ${\rm Sp}(2,\Z)\simeq \SL(2,\Z)$, independent of the value of $P$.  A charge vector $\bz$ determines a homology class of 1-cycles on the torus fiber by $[\g]=c\,[\a]+d\,[\b]$.  The polarization \eqref{polarization} on the fiber thus induces the charge inner product \eqref{dszform} by $\vev{\bz_1,\bz_2} := \vev{\g_1,\g_2}$.

The \emph{safely irrelevant conjecture} \cite{Argyres:2015ffa} states that 4d $\cN=2$ field theories do not have dangerously irrelevant operators.  Evidence for this conjecture and its implications are described in detail in \cite{Argyres:2015ffa}.  One of the main implications is that the singularity of a scale invariant CB geometry at $u=0$ splits under a generic relevant deformation, $\bm\neq0$, into a set of other singularities at $u=u_a$ corresponding to scale-invariant or IR free field theories which themselves are ``frozen", i.e., have no further relevant deformations;\footnote{It is possible that in the vicinity of an $\cN=2$ supersymmetry-preserving flow between fixed points, new $\cN=2$ relevant directions at the IR fixed point can only be turned on by nearby non-($\cN=2$)-supersymmetric flows.  This situation could result in an $\cN=2$ flow satisfying the safely irrelevant conjecture but whose generic IR singularities are not frozen.  Though this is a logical possibility, we do not know of any cases in which this happens.  We thank Thomas Dumitrescu for pointing this out.} see figure \ref{f1} below.  If $\U(1)$-charged states with charges $\bz_a$ become massless at the $u_a$ singularities, then the Dirac quantization condition implies that their invariant charges, $Q_a$, must all be commensurate \cite{Argyres:2015ffa}.  Most deformations involve splitting into  $I_{n_a}$ singularities, and the safely irrelevant conjecture then implies that these singularities must be due to massless hypermultiplets of invariant charge $Q_a = \sqrt{n_a}$.  Dirac quantization then implies that the $\sqrt{n_a}$'s be commensurable.  A similar, but slightly more involved argument applies to $I^*_n$ singularities as well; see \cite{Argyres:2015ffa}.

\subsection{Deformations}

Turning on non-zero mass parameters splits the initial scale-invariant singularity at $u=0$ into lesser ones at a set of points $u=u_a$ \cite{Argyres:2015ffa}.  We will now define three useful analytic or topological invariants of such deformations of increasing specificity: the orders of the vanishing of the discriminant of the curve at the singularities; the set of $\SL(2,\Z)$ conjugacy classes of the monodromies around the singularities; and the equivalence class of the set of all $\SL(2,\Z)$ monodromies around the singularities under the action of the braid group and global $\SL(2,\Z)$ conjugation.  The second of these, which we call the \emph{deformation pattern} will turn out to uniquely specify the deformed SW geometry.  We have no a priori argument for this fact; only the detailed constructions of the rest of this paper justify this statement.

\paragraph{Orders of vanishing of the discriminant at the singularities.}  

The splitting of the initial singularity means that the order ord${}_0(D_x)$ zero (at $u=0$) of $D_x$ of the scale-invariant theory (with $\bm=0$) splits into some number, $Z$, of zeros at $u=u_a$, $a=1,\ldots,Z$, for generic values of the deformation parameters $\bm\neq0$.  We will indicate the order of vanishing of the discriminant at these non-zero singular locations, $u=u_a$, as ord${}_a(D_x)$.  For a given deformation the discriminant thus has the form:
\beq\label{discfac1}
D_x=\prod_{a=1,...,Z}(u-u_a)^{{\rm ord}_a(D_x)}
\eeq
Though the locations, $u_a$, of the zeros depend on the deformation parameters, $\bm$, their integer multiplicities, ord${}_a(D_x)$, will be unchanged under small changes of the deformation parameters.   Thus the set of integers, $\{$ord${}_a(D_x)\}$, is an invariant of the deformed curve.

\paragraph{The deformation pattern.}  

The singularities at $u=u_a$ will be one of the Kodaira types in table \ref{Table:Kodaira}.  For a given initial scale invariant singularity we define the deformation pattern of a deformation of its CB geometry to be the list of the Kodaira types of the singularities resulting from the splitting obtained by turning on that particular deformation.  For example,
\begin{align}\label{pair}
II^*\quad\to\quad\{{I_1}^2,IV^*\}
\end{align}
is the deformation which splits an initial type $II^*$ Kodaira singularity into three singularities, two of type $I_1$ and one of type $IV^*$.

Each scale invariant singularity has an associated monodromy $K_0\in\SL(2,\Z)$ which corresponds to the transformation which 1-cycles undergo while traversing a simple closed path in the CB encircling $u=0$.  Under a change in choice of canonical 1-cycle basis by an element $g\in\SL(2,\Z)$, $K_0$ changes by $K_0 \to g K_0 g^{-1}$, so only the conjugacy class of $K_0$ in $\SL(2,\Z)$ is an invariant property of the singularity.  Representative monodromies of the Kodaira singularities are listed in table \ref{Table:Kodaira}.   Note that the monodromy of each type of Kodaira singularity is in a distinct $\SL(2,\Z)$ conjugacy class.   Thus the deformation pattern is equivalent to specifying the set of $\SL(2,\Z)$ conjugacy classes of the singularities resulting from the deformation.

\paragraph{EM duality monodromies.}

\begin{figure}[tbp]
\centering
\begin{tikzpicture}[decoration={markings,
mark=at position .5 with {\arrow{>}}}]
\clip (0,-.75) rectangle (15,4);
\fill[color=black!05] (0,0) rectangle (6,4);
\node at (3,-0.5) {$m_i=0$};
\node[Bl] (or) at (3,3) {};
\node[R] (br) at (3,2) {};
\node[red] at (3,2.35) {$u{=}0$};
\draw[decorate,decoration=snake,red!50] (br) -- (3,0);
\draw[thick,blue,postaction={decorate}] (or) .. controls (0,0) and (6,0) .. (or);
\node[blue] at (4.25,1) {$K_0$};
\node[single arrow, draw, black] at (7.5,2) {{\small deformation}};
\fill[color=black!05] (9,0) rectangle (15,4);
\node at (12,-0.5) {$m_i\neq 0$};
\node[Bl] (or1) at (12,3) {};
\node[R] (br1) at (10.5,2) {};
\node[red] at (10.5,2.35) {$u_1$};
\node[R] (br2) at (12,1.5) {};
\node[red] at (12,1.85) {$u_2$};
\node[R] (br3) at (13.5,2) {};
\node[red] at (13.5,2.35) {$u_3$};
\draw[decorate,decoration=snake,red!50] (br1) -- (10,0);
\draw[decorate,decoration=snake,red!50] (br2) -- (12,0);
\draw[decorate,decoration=snake,red!50] (br3) -- (14,0);
\draw[thick,blue,postaction={decorate}] (or1) .. controls (8,3) and (10,-1) .. (or1);
\node[blue] at (10,3) {$K_1$};
\draw[thick,blue,postaction={decorate}] (or1) .. controls (10.5,0) and (13.5,0) .. (or1);
\node[blue] at (11.3,1) {$K_2$};
\draw[thick,blue,postaction={decorate}] (or1) .. controls (14,-1) and (16,3) .. (or1);
\node[blue] at (14,3) {$K_3$};
\end{tikzpicture}
\caption{Singularities and their monodromies on the $u$-plane without ($m_i=0$) and with ($m_i\neq 0$) generic mass deformation.  The solid points with coordinates $u_a$ are the singularities shown with a choice of ``branch cuts" emanating from them.    The $K_a\in SL(2,\Z)$ are EM duality monodromies associated to the closed paths looping around these singularities starting from a conventional base point given by the open circle.\label{f1}}
\end{figure}
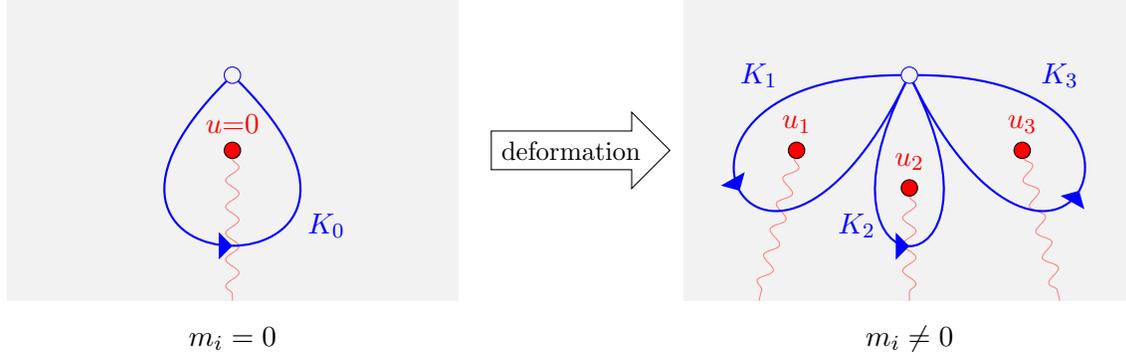

A more precise invariant of a deformed CB geometry is the set of EM duality monodromies (and not just their $\SL(2,\Z)$ conjugacy classes) around each of the singularities at generic values of the deformation parameters.  This set of monodromies can be specified up to an overall $\SL(2,\Z)$ conjugation by picking a base point, an ordering of the singularities, and a set of simple closed paths encircling each singularity in the same sense.  A convenient way of specifying the ordering and paths is by choosing a set of non-intersecting ``branch cuts" emanating from each singularity and going to $u=\infty$ parallel to the negative imaginary $u$-axis.  Then the singularities are ordered according to increasing Re$(u)$-values of the cuts at Im$(u)\to-\infty$, and a basis of (homotopy classes of) closed paths on the punctured $u$-plane, $\{\g_a\}$, are defined by demanding that $\g_a$ crosses only the $a$th branch cut just once counterclockwise; see figure \ref{f1}.  

Denote the $\SL(2,\Z)$ monodromy around $\g_a$ by $K_a$ for $a=1,\ldots,Z$.  These monodromies are specified up to a common $\SL(2,\Z)$ conjugation, reflecting the freedom to choose an arbitrary EM duality basis at the base point.  Furthermore, the ordering of the singularities is arbitrary, and can be changed by moving cuts across neighboring singularities (therefore passing them through neighboring cuts).  Upon dragging the monodromy paths, one finds that neighboring monodromies can get interchanged or conjugated by each other, giving an action of the braid group on $Z$ strands on $\SL(2,\Z)$ matrices \cite{DeWolfe:1998zf,DeWolfe:1998eu}.  We will describe this action in more detail below.  Thus it is really only the set of $\{K_a\}$ monodromies up to an overall $\SL(2,\Z)$ conjugation and braid equivalences which characterizes the deformation of the curve.

\subsection{Constraints on possible deformations}

The simplest constraint on deformations of a given singularity is that the total number of zeros of the discriminant counted with multiplicity will stay the same under deformation,
\beq\label{OrdDis}
\sum_{a=1}^Z \text{ord}_a(D_x) = \text{ord}_0(D_x).
\eeq  
This is because, as discussed in \cite{Argyres:2015ffa}, relevant deformations do not deform the CB geometry at large $u$, and so, in particular, under mass deformations no ``extra" zeros of the discriminant can come in from $u=\infty$.  Since, from table \ref{Table:Kodaira}, scale invariant Kodaira singularities have $\text{ord}_0(D_x) \le 10$, it follows from \eqref{OrdDis} that only a finite (though fairly large) number of deformation patterns are possible. 

The safely irrelevant conjecture, discussed at length in \cite{Argyres:2015ffa}, limits the generic singularities that appear upon deformation to be of types $I_{n>0}$, $I^*_{n\ge0}$, $II^*$, $III^*$, or $IV^*$.  The safely irrelevant conjecture also constrains the IR free $I_{n>0}$ and $I^*_{n>0}$ singularities that appear in the deformation to have massless charged hypermultiplets of only specific $\U(1)$ charges, and so the Dirac quantization condition outlined above can be applied.  As described in detail in \cite{Argyres:2015ffa}, this eliminates many possible deformation patterns.  

Another constraint is that the EM duality monodromies, $\{K_a\}$, of a deformation satisfy
\begin{align}\label{K0}
K_0 = K_Z K_{Z-1} \cdots K_2 K_1 ,
\end{align}
where $K_0$ is the monodromy around the undeformed singularity.  Some of the remaining possible deformation patterns may fail to exist because there is no corresponding solution to the monodromy constraint \eqref{K0}.  Establishing directly whether or not \eqref{K0} can be satisfied is algebraically challenging.  But a simple necessary condition is that the trace of \eqref{K0} be satisfied.  

In the case where all the singularities that appear in the deformation pattern are of $I_{n_a}$ type, we know from the safely irrelevant conjecture that they correspond to IR free $\U(1)$ theories each with a single massless hypermultiplet with charge vector $\bz_a$ with invariant charge $Q_a=\sqrt{n_a}$.\footnote{As we will discuss further in section \ref{RGflow}, an alternative possible non-lagrangian interpretation of a ``frozen" or non-deformable $I_n$ singularity is as a rank-0 interacting SCFT with flavor group $F=\U(1)$ or $\SU(2)$, a $\U(1)$ subgroup of which is weakly gauged.}  The EM duality monodromy around such a singularity is
\begin{equation}\label{zmonod}
K_a=\I-S\bz_a^T \bz_a
\end{equation}
in a matrix notation where $\bz_a$ is a row vector and the $T$ superscript denotes transpose.  This can be deduced from the one-loop beta function of the $\U(1)$ theory with a hypermultiplet of electric charge $Q_a$ together with an appropriate EM duality transformation \cite{Seiberg:1994rs,Seiberg:1994aj}.  It is also easy to check that \eqref{zmonod} parameterizes all $\SL(2,\Z)$ elements in the $T^{n_a}$ conjugacy class.

Taking the trace of \eqref{K0} using \eqref{zmonod} and the definition \eqref{dszform} of the charge inner product, it follows that the trace of the total monodromy is \cite{DeWolfe:1998eu}
\begin{align}\label{traceK}
\mbox{Tr}(K_0) = 2+
\sum_{k=2}^Z \, \sum_{a_1>a_2\cdots>a_k}
\langle \bz_{a_1},\bz_{a_2}\rangle\ 
\langle \bz_{a_2},\bz _{a_3}\rangle
\cdots
\langle \bz_{a_k},\bz_{a_1}\rangle .
\end{align}
Note that in each term on the right side of \eqref{traceK} any $\bz_a$ appears twice or not at all.  Since each $\bz_a$ has a common factor of $Q_a$, if $\bz_a$ appears in a term, then the term will be divisible by $Q^2_a$.  This observation can be used to rule out some deformation patterns.  For example, the $II^*\to\{I_1,I_9\}$ deformation pattern respects both the Dirac quantization condition and \eqref{OrdDis}.  But Tr$(K_0)=1$ for a $II^*$ singularity by table \ref{Table:Kodaira}, while \eqref{traceK} implies Tr$(K_0)=2 \mod 9$, showing that there do not exist $\SL(2,\Z)$ monodromies realizing this deformation pattern. 

This can be easily extended to cases where there is one type $I_{n_1}^*$ singularity appearing in the deformation pattern along with type $I_{n_a}$ singularities.   Simply go to an EM duality basis in which the monodromy of the $I_{n_1}^*$ is $K_1=-T^{n_1} = -(\I-S\bz_1^T \bz_1)$ with $\bz_1=(0,\sqrt{n_1})$.  One then gets a similar formula for the trace as \eqref{traceK}, but with an overall minus sign on the right side.  This also serves to eliminate a number of potential deformation patterns.  Adding this constraint we rule out all but 28 deformation patterns of the scale-invariant Kodaira singularities, which are listed in table \ref{table:theories} in the introduction.  

In the next two sections we construct explicit SW geometries realizing each of these deformation patterns.  (This therefore shows that there in fact  exist monodromies satisfying \eqref{K0} --- and not just its trace --- for each of these deformation patterns.)   In fact, we find a unique SW geometry for each deformation pattern.  SW curves and one-forms for those geometries that have not appeared before in the literature are given in the appendices.  

From the explicit construction of the SW curves and one-forms, the maximal flavor symmetries of the corresponding theories can be deduced, as will be discussed in detail below.  These flavor symmetries for all these deformations are listed in table \ref{table:theories} in the introduction.

It is important to note that not all of these geometries are physical.   We have so far only imposed the constraint from the safely irrelevant conjecture for \emph{generic} values of the deformation parameters.  In section 5 we check whether it is satisfied for \emph{all} values of the deformation parameters, and find that three of the geometries fail this test, again as indicated in table \ref{table:theories}.

\section{Construction of Seiberg-Witten curves}\label{sec:defs}

In this section and the next we construct explicit SW geometries corresponding to the 28 commensurate deformation patterns in table \ref{table:theories}.  We do this in two steps:  in this section we construct the curves and in the next we construct the one-forms.

Consider a deformation pattern $\{{T_1}^{n_1}, \ldots, {T_p}^{n_p}\}$ where $T_a$ denote distinct Kodaira singularity types, and $n_a$ count the number of times each type occurs in the deformation pattern.  Then a necessary condition for a SW curve to realize this deformation pattern is that its discriminant must factorize as
\begin{align}\label{discfac}
D_x(u,\bm) = \prod_{a=1}^p  \left[P_{n_a}(u,\bm)\right]^{\text{ord}_a}
\end{align}
where the $P_n$ are polynomials in $u$ and the $\bm$ of   degree $n$ in $u$ and ord$_a$ is the order of the vanishing of the discriminant of the $T_a$ singularity.   This is because, as one performs monodromies in the space of the $\bm$, only singularities of the same Kodaira type can be permuted among themselves, so zeros in the discriminant corresponding to different Kodaira types must belong to different polynomial factors.  Note that   \emph{maximal deformation} patterns --- those of the form $\{{I_1}^n\}$ --- require no special factorization of the discriminant, and so curves for them are easy to write down (and are given in table \ref{t4} below).

It is in principle possible to systematically search for families of curves \eqref{rank1curve} in Weierstrass form with polynomial coefficients $f(u,\bm)$ and $g(u,\bm)$ such that the discriminant of the right side has a given factorization pattern as in \eqref{discfac}.  In fact, we have done this for many (non-maximal deformation) entries in table \ref{table:theories} where we were able to find a solution for the curve.  Only in a few of those cases were we also able to show that there were no other inequivalent solutions.   As the dimension $\D(u)$ of the Coulomb branch vev increases, such searches rapidly become calculationally intractable \cite{Argyres:2010py}, and in many cases we were not able to uncover any solutions by this direct method.  

However, a less direct approach using known properties of the maximal deformations of the Kodaira singularities yields an easy existence proof and straightforward construction of curves realizing each of the allowed deformation patterns in table \ref{table:theories}.  In each case it yields only a single solution for the curve, and that solution coincides with that found by the direct factorization search in the cases where a solution was found in that way.  But this indirect approach has the disadvantage that it cannot be used to rule out the existence of additional solutions.

To describe this approach we first need to describe how the flavor symmetry of the SCFT is encoded in the SW curve.

\paragraph{Mass deformations and the flavor Weyl group.}

When the (linear mass) deformation parameters $\bm=\sum_{i=1}^R m_i \be^i$ are turned on they necessarily appear in the curve \eqref{rank1curve} in homogeneous polynomial combinations.  Let $M_{d_i}=M_{d_i}(\bm)$, $i=1,\ldots,R$ be an algebraically independent basis of these polynomials of degrees $\{d_i\}$.  (Though a basis of the $M_d$ is not unique, the set of their degrees is.)  Since the curve only depends on the linear masses through the $M_d$, then the curve will be the same for different values of the $\bm$ which give the same values of the $\{M_{d_i}\}$.  These identifications on the space of $\bm$ form a discrete group which, by the Chevalley-Shephard-Todd theorem \cite{Shephard1954,Chevalley1955} is a complex reflection group acting linearly on the $\bm$ and completely determined by the set $\{d_1,\ldots,d_R\}$ of degrees of the $M_{d_i}$.

From the field theory point of view, the flavor symmetry algebra $F$ is a reductive Lie algebra by the Coleman-Mandula theorem \cite{Coleman:1967ad}.  The linear masses transform in the complexified adjoint representation of $F$, so (generically) explicitly break $F\to \text{Weyl}(F)\ltimes \U(1)^{\text{rank}(F)}$.  Here Weyl$(F)$ is the Weyl group of $F$, which acts on the $\bm$ as the complexification of a real crystallographic reflection group \cite{humphreys1990coxeter}.  
The SW curve deformation parameters $M_{d_i}$, being homogeneous polynomials of degree $d_i$ in the linear masses, will have scaling (mass) dimension $d_i$.  Thus we can read off from the SW curve the rank of the flavor algebra from the number of algebraically independent deformation parameters, rank$(F)=R:=|\{M_{d_i}\}|$, and deduce the Weyl group of the flavor symmetry from the set of dimensions $\{d_i\}$ of the deformation parameters.  The flavor symmetry itself can be largely, but not completely, reconstructed from its Weyl group data.  For each $d_i=1$ there is a $\U(1)$ factor in $F$ upon which the Weyl group acts trivially.  The Weyl group is the direct product of the Weyl groups of each simple factor of $F$.  But the Weyl group cannot distinguish between $\SO(2n+1)$ and $\Sp(2n)$ factors.

Another ambiguity in the identification of the flavor symmetry from the discrete Weyl group, $\G$, determined from the curve comes from the possibility that $\G$ might actually be larger than Weyl$(F)$.  That is, it could happen that $\G = \G' \ltimes \text{Weyl}(F)$, so that the actual flavor Lie algebra, $F$, is of smaller dimension than that deduced from $\G$.  This possibility is analyzed in detail in \cite{Argyres:2016xua}.  In this paper we will focus on determining the maximal allowed flavor algebra from the CB geometry found by assuming that $\G=\text{Weyl}(F)$.  This focus has no effect on the actual determination of the CB curve and one-form.

It is interesting to note that not all complex reflection groups are Weyl groups of reductive Lie algebras.  So it is possible to have  curves describing deformations whose complex reflection group symmetries are not Weyl$(F)$ for any field theory symmetry $F$.  A simple example is the curve $y^2=x^3 + (u^4 + u^2 M_6 + u M_9 + M_{12})$ which is a 3-parameter deformation of the $IV^*$ singularity.  Its discriminant is $D_x \sim (u^4+u^2 M_6 + uM_9 + M_{12})^2$, so (generically) has 4 singularities on the CB each with ord$_a(D_x)=2$.  It is not hard to check that the monodromy, $K_a$, around each of the 4 generic singularities is in the $\SL(2,\Z)$ conjugacy class $[ST]$, and so each corresponds to a type $II$ Kodaira singularity; see table \ref{Table:Kodaira}.  Thus this curve describes a deformation with pattern $IV^*\to\{II^{\,4}\}$.  Its set of deformation parameter dimensions, $\{6,9,12\}$, is not the set of degrees of Weyl-invariant polynomials of any reductive Lie algebra.\footnote{It is, in fact, the set of degrees of polynomial invariants of the complex reflection group $W(L_3)$ of order 648, which is number 25 in the Shephard-Todd classification of complex reflection groups \cite{Shephard1954,Cohen1976}.}  Many similar examples can be constructed with non-Weyl reflection groups.  They all have the property that their deformation pattern contains at least one singularity of Kodaira type $II$, $III$, or $IV$.  But precisely these singularities were ruled out of physical deformation patterns by the safely irrelevant conjecture, as argued in \cite{Argyres:2015ffa}.  As a result, all deformation patterns compatible with the safely irrelevant conjecture turn out to have complex reflection group symmetries which are Weyl groups of reductive Lie algebras.  This can be interpreted as additional evidence for the correctness of the safely irrelevant conjecture.

\subsection{Maximal deformations and the string web picture}
\label{StrWeb}

A \emph{maximal deformation} is one where each zero of the curve discriminant has multiplicity one at generic values of the $M_d$, and so has generic number of zeros $Z=\text{ord}_0(D_x)$.  Since the $I_1$ singularity is the only one with discriminant vanishing to order one (see table \ref{Table:Kodaira}), the deformation pattern of a maximal deformation of a scale invariant singularity with ord$_0(D_x)=n$, will be of the form
\beq\label{Maxi}
*\to\{{I_1}^n\}.
\eeq
This implies that for a maximal deformation the discriminant satisfies no particular factorization condition.  Thus these curves are easy to write down: they are simply the most general complex deformations of the Kodaira curves which do not increase the order of the discriminant.  After using the freedom to redefine $y$ and $x$ and shift $u$, these deformations are shown in table \ref{t4}.  It is straightforward to obtain the flavor group for each one of these deformation patterns, as discussed above. The result is listed in the last column of table \ref{t4}.  (Since these algebras are all simply-laced, they are uniquely specified by their Weyl groups. For more details on Weyl groups of simple Lie algebras see, e.g., the appendices of \cite{Argyres:2012ka}.)

\begin{table}[tbp]
\centering\small
\begin{tabular}{c|lr}
\hline
singularity & generically deformed curve & $F$ \\ 
\hline\\[-4mm]
$II^*$ &
$y^2 = x^3 + x\,(M_2 u^3{+}
M_8 u^2{+}M_{14} u{+}M_{20})
+ (u^5{+}M_{12} u^3{+}M_{18} u^2
{+} M_{24} u {+}M_{30})$ & 
$E_8$ \\
$III^*$ &
$y^2 = x^3 + x\,( u^3{+}M_8 u{+}M_{12}) 
+ (M_2 u^4{+}M_6 u^3{+}M_{10} u^2
{+}M_{14} u{+}M_{18})$ & 
$E_7$ \\
$IV^*$ &
$y^2 = x^3 + x\,(M_2 u^2{+}M_5 u{+}M_8) 
+ ( u^4{+}M_6 u^2{+}M_9 u{+}M_{12})$ & 
$E_6$ \\
$I_0^*$ &
$y^2 = x^3 + x\,(\tau  u^2{+}M_2 u{+}M_4) 
+ (u^3{+}\til M_4 u{+}M_6)$ & 
$\SO(8)$ \\
$IV$ &
$y^2 = x^3 + x\,(M_{1/2} u{+}M_2) 
+ (u^2{+}M_3)$ & 
$\SU(3)$ \\
$III$ &
$y^2 = x^3 + x\,u + (M_{2/3} u{+}M_2)$ & 
$\SU(2)$ \\
$II$ &
$y^2 = x^3 + x\,M_{4/5} + u$ & 
--- \\[1mm]
\hline\\[-4mm]
$I_{n\ge1}$  &
$y^2 = (x-1) (x^2 + \L^{-n} [ u^{n}{+}M_2 u^{n-2}{+}\cdots{+}M_{n}])$ 
& 
$\U(n)$ \\[.5mm]
$I_{n\ge1}^*$ &
$y^2 = x^3 + u x^2 + \L^{-n} \til M_{n+4} x
+ \L^{-2n} (u^{n+3}{+}M_2 u^{n+2}{+}\cdots
{+}M_{2n+6})$ & 
\hspace{-10mm} $\SO(2n{+}8)$ \\[1mm]
\hline
\end{tabular}
\caption{Maximal deformations of the Kodaira singularities along with the associated flavor algebra.  The subscript on the deformation parameters, $M_d$, is their mass scaling dimension.\label{t4}}
\end{table}

Note that the $II$, $III$ and $IV$ singularities have deformation parameters $M_d$ with fractional dimension $d=2-\D(u)$ which do not transform under any flavor symmetry.  These correspond \cite{Argyres:1995xn,Argyres:2015ffa} to a deformation by the relevant operator $M_d\, \int d^4\th \, U$ (written in an $\cN{=}2$ superspace notation).  Here $U$ is the operator in the CFT whose vev is the Coulomb branch parameter, $\vev{U}=u$.  

For a given choice of basis of cycles (specified as in figure \ref{f1}) and choice of EM duality basis, one can easily compute \cite{Dasgupta:1996ij,Sen:1996vd} the set of $\SL(2,\Z)$ monodromies $\{K_a\}$ for generic deformations $\{M_d\neq0\}$ given the explicit curves in table \ref{t4}.  In all these cases each $K_a$ is found to be conjugate to $T$.  From table \ref{Table:Kodaira} the $I_1$ singularity also has monodromy conjugate to $T$.  Thus this result is consistent with the deformation pattern in \eqref{Maxi}.  Since they are all conjugate to $T$, they can be parameterized as in \eqref{zmonod} as $K_a = \I-S\bz_a^T \bz_a$, in terms of a set of EM charge vectors, $\{\bz_a\}$, each with invariant charge $Q_a =1$.  This just means that each charge vector $\bz_a$ is given by a pair of coprime integers.

The resulting set of $z_a$'s for the maximal deformations in \ref{t4} are (for a particular choice of EM duality basis and choice of basis cycles)
\begin{align}\label{Ftheory}
\{\bz_a\} 
&= \{(0,1)^n,\, (1,-2),\, (1,1)\}
& &\text{for\ the}\ IV^*,\ III^*,\ II^*\ {\rm cases},\ n\in\{6,7,8\}, 
\nonumber\\
&= \{(0,1)^n,\, (1,1)\}
& &\text{for\ the}\ II,\ III,\ IV\ {\rm cases},\ n\in\{1,2,3\},
\nonumber\\
&= \{(0,1)^{n+4},\, (1,-1),\, (1,1)\}
& &\text{for the $I^*_n$ cases,}\ n\in\{0,1,\ldots\},
\nonumber\\
&= \{(0,1)^n\}
& &\text{for the $I_n$ cases,}\ n\in\{1,2,\ldots\}.
\end{align}
where $(p,q)^n$ denote a sequence of $n$ consecutive equal EM charge vectors $\bz=(p,q)$.  Note, for later use, that in this basis the total monodromies \eqref{K0} of these singularities are
\begin{align}\label{FtheoryK0}
II^*: \qquad K_0 &= \left(\begin{smallmatrix} -3 & -13\\ 1 & 4
\end{smallmatrix}\right) = T^{-3} (ST) T^3 ,
\nonumber\\
III^*: \qquad K_0 &= \left(\begin{smallmatrix} -3 &-10\\ 1 & 3
\end{smallmatrix}\right) = T^{-3} (S) T^3 ,
\nonumber\\
IV^*: \qquad K_0 &= \left(\begin{smallmatrix} -3 & -7 \\ 1 & 2
\end{smallmatrix}\right) = T^{-2} (ST)^2 T^2 ,
\nonumber\\
IV: \qquad  K_0 &= \left(\begin{smallmatrix} 2 & 7 \\ -1 & -3
\end{smallmatrix}\right) = T^{-2} (ST)^{-2} T^2 ,
\\
III: \qquad  K_0 &= \left(\begin{smallmatrix} 2 & 5 \\ -1 & -2
\end{smallmatrix}\right) = T^{-2} (S)^{-1} T^2 ,
\nonumber\\
II: \qquad K_0 &= \left(\begin{smallmatrix} 2 & 3 \\ -1 & -1
\end{smallmatrix}\right) = T^{-1} (ST)^{-1} T^1,
\nonumber\\
I_n^*: \qquad K_0 &= \left(\begin{smallmatrix} -1 &-n\\ 0& -1
\end{smallmatrix}\right) = -T^n ,
\nonumber\\
I_n: \qquad K_0 &= \left(\begin{smallmatrix} 1 & n \\ 0 & 1
\end{smallmatrix}\right) = T^n .
\nonumber
\end{align}
The convenient bases shown in \eqref{Ftheory} are ones discovered and studied in \cite{DeWolfe:1998eu}.  

\paragraph{Linear masses, coalescing singularities, and the string web picture of neutral BPS states.} 

Knowing the SW curve in terms of the Weyl-invariant mass parameters $\{M_a\}$ as in table \ref{t4} does not tell us what it is in terms of the linear masses $\bm$.  This is because there are (infinitely) many Weyl-invariant polynomials $M_a(\bm)$ that do not differ simply by linear redefintions of the $\bm$.  The particular dependence of the SW curve (and one-form) on the linear masses is an important part of the low energy effective action.  For instance, the linear masses enter in the central charge, and so are ``observed" through the BPS spectrum.  Also, the way the linear masses enter the SW curve is strongly constrained by the fact that the SW curve should degenerate in special ways for special values of the $\bm$ related to the roots of the flavor symmetry algebra.  We will explain this connection below in some detail.  It will be important not so much as a method for determining the $M_a(\bm)$ polynomials (demanding linearity of the residues of the SW form turns out to be more efficient \cite{Seiberg:1994rs,Minahan:1996cj}), but because it gives a physical and geometrical picture encoding the linear mass dependence.  The algebraic form of the $M_a(\bm)$ polynomials (many examples of which are given in the appendices) are not particularly enlightening, but we will see that the geometrical picture gives a powerful tool for constructing deformed special K\"ahler geometries.

Recall that at generic masses, the flavor symmetry algebra $F$ is broken to a Cartan subalgebra $\ff$, and states are classified by their flavor charge vectors $\bw \in \L_F \subset \ff^*$ in the root lattice of $F$.  An EM-neutral BPS one-particle state with quark number $\bw$ has central charge and thus BPS mass equal to $\bw(\bm)$.  But at special values of the linear masses, namely those on a hyperplane $H_\ba = \{\bm\in\ff_\C | \ba(\bm)=0 \}$ for any root $\ba$ of $F$, the unbroken flavor symmetry will be enhanced by a nonabelian factor that includes the $\SU(2)$ factor generated by the generator associated to $\ba$ (in a Cartan basis).  This is reflected in an additional ``accidental" massless EM-neutral BPS states with quark number $\bn \propto \ba$, and an associated collision of some simple singularities.

The reflection $\s_\ba \in \text{Weyl}(F)$ which fixes the hyperplane $H_\ba$ is given by the action on $\ff_\C$
\begin{align}\label{Weylrefl}
\s_\ba :\ \  \bm \ \mapsto\ \bm - 2 \ba(\bm) \, \ba^* .
\end{align}
Following a path in mass parameter space (i.e., in $\ff_\C$) joining a general point close to $H_\ba$ to its image under $\s_\ba$ and which does not go through $H_\bb$ for any root $\bb$ (which is possible since the $H_\bb$ are complex codimension one in $\ff_\C$) induces a motion of the $I_1$ singularities on the Coulomb branch in which they do not collide and return to their original configuration.   They return to their original configuration because $\S(\bm)=\S(M_d)$ is Weyl-invariant.  It is clear (e.g., by taking the path arbitrarily close to $H_\ba$) that the motion only rearranges the subset of the $I_1$ singularities which collide at $H_\ba$.\footnote{The singularities do not all have to collide at the same point on the Coulomb branch, but can occur as the simultaneous collision of subsets at different points.}  This subset of $I_1$ singularities and their pattern of rearrangement can be encoded in a branched path (i.e., topologically a tree) in the Coulomb branch connecting these singularities \cite{DeWolfe:1998eu, DeWolfe:1998zf}.  We call these branched paths ``string webs" since that is how they appear in F-theory constructions.

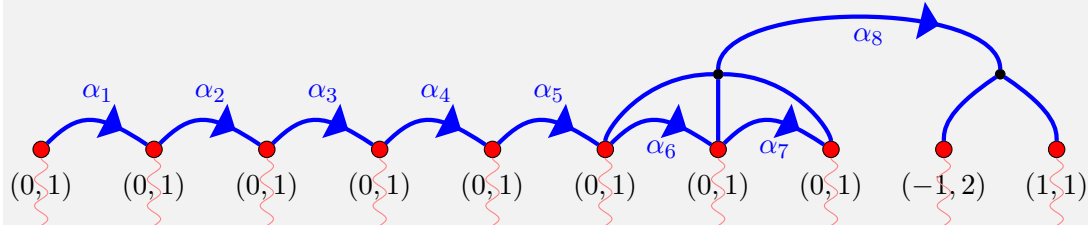
\begin{figure}[htbp]
\centering
\begin{tikzpicture}[decoration={markings,
mark=at position .75 with {\arrow{>}}}]
\clip (0.5,0.5) rectangle (15,3.5);
\fill[color=black!05] (0.5,0) rectangle (15,4);
\node[R] (br1) at (1,1.5) {};
\node at (1,1) {$(0,1)$};
\draw[decorate,decoration=snake,red!50] (br1) -- (1,0);
\node[R] (br2) at (2.5,1.5) {};
\node at (2.5,1) {$(0,1)$};
\draw[decorate,decoration=snake,red!50] (br2) -- (2.5,0);
\node[R] (br3) at (4,1.5) {};
\node at (4,1) {$(0,1)$};
\draw[decorate,decoration=snake,red!50] (br3) -- (4,0);
\node[R] (br4) at (5.5,1.5) {};
\node at (5.5,1) {$(0,1)$};
\draw[decorate,decoration=snake,red!50] (br4) -- (5.5,0);
\node[R] (br5) at (7,1.5) {};
\node at (7,1) {$(0,1)$};
\draw[decorate,decoration=snake,red!50] (br5) -- (7,0);
\node[R] (br6) at (8.5,1.5) {};
\node at (8.5,1) {$(0,1)$};
\draw[decorate,decoration=snake,red!50] (br6) -- (8.5,0);
\node[R] (br7) at (10,1.5) {};
\node at (10,1) {$(0,1)$};
\draw[decorate,decoration=snake,red!50] (br7) -- (10,0);
\node[R] (br8) at (11.5,1.5) {};
\node at (11.5,1) {$(0,1)$};
\draw[decorate,decoration=snake,red!50] (br8) -- (11.5,0);
\node[R] (br9) at (13,1.5) {};
\node at (13,1) {$(-1,2)$};
\draw[decorate,decoration=snake,red!50] (br9) -- (13,0);
\node[R] (br10) at (14.5,1.5) {};
\node at (14.5,1) {$(1,1)$};
\draw[decorate,decoration=snake,red!50] (br10) -- (14.5,0);
\draw[ultra thick,blue,postaction={decorate}] (br1) .. controls (1.5,2) and (1.75,2) ..(br2);
\node[blue] at (1.75,2.25) {$\a_1$};
\draw[ultra thick,blue,postaction={decorate}] (br2) .. controls (3,2) and (3.25,2) .. (br3);
\node[blue] at (3.25,2.25) {$\a_2$};
\draw[ultra thick,blue,postaction={decorate}] (br3) .. controls (4.5,2) and (4.75,2) ..(br4);
\node[blue] at (4.75,2.25) {$\a_3$};
\draw[ultra thick,blue,postaction={decorate}] (br4) .. controls (6,2) and (6.25,2) ..(br5);
\node[blue] at (6.25,2.25) {$\a_4$};
\draw[ultra thick,blue,postaction={decorate}] (br5) .. controls (7.5,2) and (7.75,2) ..(br6);
\node[blue] at (7.75,2.25) {$\a_5$};
\draw[ultra thick,blue,postaction={decorate}] (br6) .. controls (9,2) and (9.25,2) ..(br7);
\node[blue] at (9.25,1.5) {$\a_6$};
\draw[ultra thick,blue,postaction={decorate}] (br7) .. controls (10.5,2) and (10.75,2) ..(br8);
\node[blue] at (10.75,1.5) {$\a_7$};
\node[bl] (jnc1) at (10,2.5) {};
\node[bl] (jnc2) at (13.75,2.5) {};
\draw[ultra thick,blue] (br6) .. controls (8.5,1.75) and (9,2.5) ..(jnc1);
\draw[ultra thick,blue] (br7) -- (jnc1);
\draw[ultra thick,blue] (br8) .. controls (11.5,1.75) and (11,2.5) ..(jnc1);
\draw[ultra thick,blue,postaction={decorate}] (jnc1) .. controls (10,3.5) and (13.75,3.5) ..(jnc2);
\draw[ultra thick,blue] (jnc2) .. controls (13,2) and (13,1.75) .. (br9);
\draw[ultra thick,blue] (jnc2) .. controls (14.5,2) and (14.5,1.75) .. (br10);
\node[blue] at (12,3) {$\a_8$};
\end{tikzpicture}
\caption{A presentation of the singularities on the Coulomb branch of the maximally deformed $II^*$ singularity, with the singularities and their cuts in red, labelled by their EM charge vectors.  In blue is a basis found in \cite{DeWolfe:1998eu} of eight neutral oriented ``string webs" connecting the singularities corresponding to simple roots of $E_8$.\label{f2}}
\end{figure}

In the F-theory realization of SW geometries \cite{Sen:1996vd,Banks:1996nj}, the Coulomb branch is the transverse space to a collection of $(p,q)$-7branes and an $I_1$ singularity associated with a massless hypermultiplet of EM charge $\bz_a$ is a $(p_a,q_a)=\bz_a$ 7brane.  Then the above branched path is a ``neutral" string web, whose ends can be (integer multiples of) $(p_a,q_a)$-strings ending on the $(p_a,q_a)$-7branes.  See figure \ref{f2}.

(This string realization, of course, has a translation purely in terms of the low energy field theory without reference to strings, but it is a little complicated to describe in terms of the formulation of SW geometry that we have been using.  The natural setting is not the Coulomb branch, but the total space, $\S$, of the elliptic fibration over the Coulomb branch.  Neutral string webs on the CB lift to non-trivial cycles in $H_2(\S)$.  The periods of the holomorphic closed $(2,0)$ form $\w = du\, dx/y$ on $\S$, which is related to the exterior derivative (on $\S$) of the SW one form \cite{Seiberg:1994aj,Donagi:1995cf} compute the $\bw(\bm)$ linear mass term contribution to the central charge.\footnote{It is also worth pointing out that these string or field theory constructions are realizations of the classical association of simply-laced Lie algebras to singularities \cite{Brieskorn:1970}. The sub-maximal deformations of the Kodaira singularities discussed in this paper point to a generalization of this classical association to non-simply laced Lie algebras, and will be explored in detail elsewhere \cite{Argyres:2016new}.})

In any case, the end result is that using the string web technology developed in \cite{DeWolfe:1998eu, DeWolfe:1998zf} we can associate to each root $\ba$ of $F$ a set of singularities which collide when $\bm$ satisfies $\ba(\bm)=0$.  For example, figure \ref{f2} shows a set of ten $I_1$ singularities in the $u$-plane with a set of neutral string webs (in blue) associated to eight simple roots $\ba_i$.  Thus, for linear masses satisfying $\ba_3(\bm)=0$, the third and fourth $I_1$ singularities (counting from the left) collide, while for $\bm$ such that $\ba_8(\bm)=0$, the last five singularities collide.

We know that SW curves depending on linear masses in this way exist for the maximal deformations of the Kodaira singularities, thanks to the explicit constructions of \cite{Seiberg:1994rs,Minahan:1996cj}.  Furthermore, the authors of \cite{DeWolfe:1998eu} have computed a basis of string webs corresponding to simple roots of their associated flavor algebras for the presentations of the singularities given above in \eqref{Ftheory}.  For example, this basis is shown for the $E_8$ maximal deformation in figure \ref{f2} \cite{DeWolfe:1998eu} where the simple roots correspond to the Dynkin diagram nodes as
\begin{align}\label{}
\begin{tikzpicture}[scale=0.6]
\node[bbcs] (r1) at (1,.5) {};
\node at (1,0) {$\a_1$};
\node[bbcs] (r2) at (2,.5) {};
\node at (2,0) {$\a_2$};
\node[bbcs] (r3) at (3,.5) {};
\node at (3,0) {$\a_3$};
\node[bbcs] (r4) at (4,.5) {};
\node at (4,0) {$\a_4$};
\node[bbcs] (r5) at (5,.5) {};
\node at (5,0) {$\a_5$};
\node[bbcs] (r6) at (6,.5) {};
\node at (6,0) {$\a_6$};
\node[bbcs] (r7) at (7,.5) {};
\node at (7,0) {$\a_7$};
\node[bbcs] (r8) at (5,1.15) {};
\node at (5.6,1.1) {$\a_8$};
\draw[thick] (r1) -- (r2);
\draw[thick] (r2) -- (r3);
\draw[thick] (r3) -- (r4);
\draw[thick] (r4) -- (r5);
\draw[thick] (r5) -- (r6);
\draw[thick] (r6) -- (r7);
\draw[thick] (r5) -- (r8);
\end{tikzpicture}
\end{align}
In the basis $\bn = n_i\be^i$ of the $E_8$ Cartan subalgebra used in \cite{Minahan:1996fg} (we are now denoting the linear mass parameters of the $E_8$ deformation by $n_i$, $i=1,\ldots,8$), the simple roots are given by 
\begin{align}\label{E8roots}
\ba_i &= \be_{i+1}-\be_{i+2}, \quad i=1,\ldots,6, &
\ba_7 &= \frac12 \biggl( \be_1 - \sum_{i=2}^7\be_i  + \be_8 \biggr) , &
\ba_8 &= \be_7 + \be_8
\end{align}
where the $\{\be_i\}$ are the dual basis to $\{\be^i\}$.  

\subsection{Construction of sub-maximal SW curves\label{s3.2}} 

We will now show how to use the information contained in the linear mass dependence of the maximal deformations of the scale invariant Kodaira singularities to construct deformations which realize all the (non-maximal) deformation patterns shown in table \ref{table:theories}.  We will call these new SW curves the \emph{sub-maximal} deformations of the Kodaira singularities.

The basic idea is very simple:  we coalesce sets of $I_1$ singularities of the maximal deformation curves to get the sub-maximal curves by appropriately tuning the linear masses of the maximal deformation curves.  With the explicit dependence of the maximal curves on the linear masses together with an identification, as in figure \ref{f2}, of flavor roots with string webs connecting singularities, we can engineer the coalesence of set of singularities by restricting the linear masses to a subspace which is annihilated by some set of roots.  

For instance, by restricting the $E_8$ linear masses, $\bn$, to a 5-dimensional subspace described by $\ba_1(\bn)=\ba_2(\bn)=\ba_3(\bn)=0$, we engineer a deformation of the $II^*$ singularity with 5 mass parameters for which the generic singularities are an $I_4$ and six $I_1$'s.  This is apparent from figure \ref{f2} since setting the first three roots to zero coalesces the four left-most singularities.  (Recall that the singularities are \emph{ordered} sets, the ordering reflecting the adjacency of the singularities as determined by the choice of ``cuts" in figures \ref{f1} or \ref{f2}.)  The 4 adjacent $I_1$ singularities coalesce to form an $I_4$ singularity because their EM charge vectors are parallel, or ``mutually local".  In this case they are all $\bz=(0,1)$, so their monodromies are each $T$, so the monodromy around all four of them is $T^4$.  Thus we find that four adjacent $\bz=(0,1)$ singularities can be coalesced to form a single $\bz=(0,2)$ singularity (since a singularity with invariant charge $Q=2$ has monodromy conjugate to $T^4$).  The fact that they were parallel to the particular choice $(0,1)$ was inconsequential since any $(p,q)$ with $\gcd(p,q)=1$ can be made so by an appropriate choice of EM duality basis.   

\paragraph{Proof of existence of SW curves for sub-maximal deformations.}

The presentations of the maximal deformations in \eqref{Ftheory} and the above discussion make it obvious that all the sub-maximal deformation patterns of the form $\{Q_a\} = \{{I_1}^n,{I_4}^m\}$ for the $I_0^*$ and $II^*$, $III^*$ and $IV^*$ singularities shown in table \ref{table:theories} can be realized by coalescing adjacent $(0,1)$ charges in groups of 4 into $(0,2)$ charges in \eqref{Ftheory}, thus realizing the following deformation patterns:
\begin{align}\label{subgrouping1}
\{\bz_a\} 
&= \{(0,2)^2,\, (1,-2),\, (1,1)\}
& &\text{for}\ II^*\to\{{I_1}^2,{I_4}^2\} ,
\nonumber\\
&= \{(0,2),\, (0,1)^{n-4},\, (1,-2),\, (1,1)\}
& &\text{for}\ (IV^*,III^*,II^*)\to\{{I_1}^{n-2},I_4\},\ n\in\{6,7,8\},
\nonumber\\
&= \{(0,2),\, (1,-1),\, (1,1)\}
& &\text{for}\ I_0^*\to\{{I_1}^2,I_4\}.
\end{align}

Similarly, coalescing all but 1 or 2 $\bz=(0,1)$ singularities in ``reverse'' of a $III^*$ and $IV^*$ maximal deformation, we  can obtain the following deformation patterns:
\begin{align}\label{subgrouping2}
\{\bz_a\} 
&= \{(0,1),K_0(III^*)\}
& &\text{for}\ II^*\to\{I_1,III^*\} ,
\nonumber\\
&= \{(0,1)^2,K_0(IV^*)\}
& &\text{for}\ II^*\to\{{I_1}^2,IV^*\} ,
\nonumber\\
&= \{(0,1),K_0(IV^*)\}
& &\text{for}\ III^*\to\{I_1,IV^*\} .
\end{align}
Here $K_0(*)$ refers to the specific total monodromy of the singularity in the basis computed in \eqref{FtheoryK0}.
By further coalescing the two $(0,1)$ singularities in the second line of \eqref{subgrouping2}, we obtain 
\begin{align}\label{subgrouping3}
\{\bz_a\} 
&= \{(0,\sqrt2),K_0(IV^*)\}
& &\text{for}\ II^*\to\{I_2,IV^*\} .
\end{align}
So, by referring to figure \ref{f2} we see that we obtain the submaximal deformations of the $II^*$ singularity in \eqref{subgrouping2} and \eqref{subgrouping3} by choosing $\bn$ such that $\ba_i(\bn)=0$ for simple roots labelled by various sets of $i$.  Explicitly, the $\{I_1,III^*\}$ deformation has $i\in\{2,\ldots,8\}$; the $\{I_1^2,IV^*\}$ deformation has $i\in\{3,\ldots,8\}$; and the $\{I_2,IV^*\}$ deformation has $i\in\{1,3,\ldots,8\}$.

So far we have shown the existence of curves realizing all but twelve of the deformation patterns listed in table \ref{table:theories}.  For these last twelve deformation patterns, it is more convenient to choose a different presentation of the charge vectors for the maximal deformations than the one shown in \eqref{Ftheory}.  Different presentations are related by a change of basis of the monodromy cycles.  This can be achieved by choosing a different set of ``cuts" emanating from the singularities as in figure \ref{f1}.  

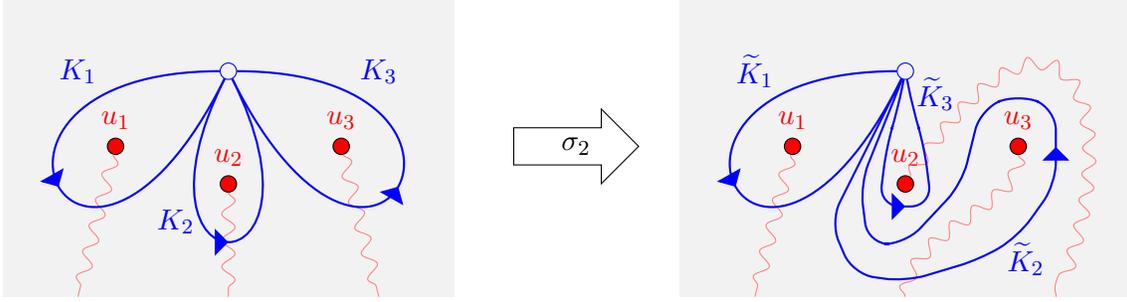
\begin{figure}[tbp]
\centering
\begin{tikzpicture}[decoration={markings,
mark=at position .5 with {\arrow{>}}}]
\clip (0,0) rectangle (15,4);
\fill[color=black!05] (0,0) rectangle (6,4);
\node[Bl] (or1) at (3,3) {};
\node[R] (br1) at (1.5,2) {};
\node[red] at (1.5,2.35) {$u_1$};
\node[R] (br2) at (3,1.5) {};
\node[red] at (3,1.85) {$u_2$};
\node[R] (br3) at (4.5,2) {};
\node[red] at (4.5,2.35) {$u_3$};
\draw[decorate,decoration=snake,red!50] (br1) -- (1,0);
\draw[decorate,decoration=snake,red!50] (br2) -- (3,0);
\draw[decorate,decoration=snake,red!50] (br3) -- (5,0);
\draw[thick,blue,postaction={decorate}] (or1) .. controls (-1,3) and (1,-1) .. (or1);
\node[blue] at (1,3) {$K_1$};
\draw[thick,blue,postaction={decorate}] (or1) .. controls (1.5,0) and (4.5,0) .. (or1);
\node[blue] at (2.3,1) {$K_2$};
\draw[thick,blue,postaction={decorate}] (or1) .. controls (5,-1) and (7,3) .. (or1);
\node[blue] at (5,3) {$K_3$};
\node[single arrow, draw, black] at (7.5,2) {\hspace{.5cm}$\s_2$  \ \hspace{.7cm}};
\fill[color=black!05] (9,0) rectangle (15,4);
\node[Bl] (or1) at (12,3) {};
\node[R] (br1) at (10.5,2) {};
\node[red] at (10.5,2.35) {$u_1$};
\node[R] (br2) at (12,1.5) {};
\node[red] at (12,1.85) {$u_2$};
\node[R] (br3) at (13.5,2) {};
\node[red] at (13.5,2.35) {$u_3$};
\draw[decorate,decoration=snake,red!50] (br1) -- (10,0);
\draw[decorate,decoration=snake,red!50] (br2) .. controls (14,4.5) and (15,2.5) .. (14,0);
\draw[decorate,decoration=snake,red!50] (br3) .. controls (13.5,1.25) and (12,1) .. (12,0);
\draw[thick,blue,postaction={decorate}] (or1) .. controls (8,3) and (10,-1) .. (or1);
\node[blue] at (10,3) {$\til K_1$};
\draw[thick,blue,postaction={decorate},rounded corners=8pt] 
(or1) -- (11.9,2.5) -- (11.75,2) --
(11.65,1.2) -- (12.35,1.2) --
(12.25,2) -- (12.1,2.5) -- (or1);
\node[blue] at (12.4,2.7) {$\til K_3$};
\draw[thick,blue,postaction={decorate},rounded corners=8pt] (or1) -- (11,1) -- (11.3,0.3) -- (12,0.2) -- (13,0.5) -- 
(13.5,0.8) -- (14,1.5) -- (14,2.5) -- (13.5,2.7) -- 
(13,2.5) -- (12.75,1.5) -- (12,0.7) -- 
(11.5,0.7) -- (11.4,1.5) -- (or1);
\node[blue] at (13.6,.5) {$\til K_2$};
\end{tikzpicture}
\caption{Change of monodromy basis associated with $\s_2$ which moves the 2nd cut through the 3rd singularity.  The new monodromies are related to the old by $\til K_1 = K_1$, $\til K_2 = K_2^{-1} K_3 K_2$, and $\til K_3 = K_2$.\label{f3}}
\end{figure}

All possible cut positions can be generated by successively moving a single cut through the singularity of one of its neighboring cuts.  Denote by $\s_j$ the operation of moving the $j$th cut through the $(j+1)$th singularity, and by $\s_j^{-1}$ moving the $(j+1)$th cut through the $j$th singularity; see figure \ref{f3}.  Following how the basis of monodromy contours changes shows that the new monodromies $\{\til K_a\}$ after the $\s_j$ move are related to the old ones by $\til K_j = K_j^{-1} K_{j+1} K_j$, $\til K_{j+1} = K_j$, and the rest remain unchanged.  This translates to the following action on the set of EM charges describing the monodromies in case they are all of type $I_n$ \cite{DeWolfe:1998eu, DeWolfe:1998zf}
\begin{align}\label{braidgens}
\s_j  &: 
\begin{cases}
    \bz_j &\!\!\!\to \ 
    \bz_{j+1}+\vev{\bz_{j+1},\bz_j}\,\bz_j ,\\
    \bz_{j+1} &\!\!\!\to \ 
    \bz_j ,\\
    \bz_k &\!\!\!\to \ 
    \bz_k \ \ \text{for $k\neq j, j{+}1$},
\end{cases} &
\s_j^{-1} &: 
\begin{cases}
    \bz_j &\!\!\!\to\ 
    \bz_{j+1},\\
    \bz_{j+1} &\!\!\!\to\ 
    \bz_j -\vev{\bz_j,\bz_{j+1}}\,\bz_{j+1},\\
    \bz_k &\!\!\!\to \ 
    \bz_k \ \ \text{for $k\neq j, j{+}1$}.
\end{cases}
\end{align}
The $\s_j$ satisfy braid relations, and thus give an action of the braid group on the set of EM charge vectors $\{\bz_a\}$. Applying $\sigma^{-1}_{n+1}\circ\sigma_n$ to the $IV^*$, $III^*$ and $II^*$ cases in \eqref{Ftheory} with $n=6,7,8$, respectively, we obtain
\begin{align}\label{Ftheory1}
\{\bz_a\} 
&= \{(0,1)^n,\, (1,-1),\, (1,1),\, (1,2)\}
& &\text{for the $IV^*$, $III^*$, $II^*$ cases, $n\in\{5,6,7\}$}.
\end{align}
This presentation makes it clear that the nine deformation patterns involving only $I_m^*$ and $I_1$ singularities in table \ref{table:theories} can also be realized since, by \eqref{Ftheory}, these singularities are realized by coalescing the $m+6$ adjacent singularities in \eqref{Ftheory1} previous to the rightmost $(2,1)$ singularity.  In this way we obtain the deformation patterns
\begin{align}\label{subgrouping4}
\{\bz_a\} 
&= \{(0,1)^{n-m-1},\, K_0(I_m^*),\, (1,2)\}
\quad\text{for}\ X\to\{{I_1}^{n-m},I_m^*\},\ 0\le m<n, 
\nonumber\\
& \quad \text{with}\ X=IV^*,III^*,II^*\ \text{for}\ n=2,3,4,
\ \text{respectively.}
\end{align}
The braidings used to obtain the presentation in \eqref{Ftheory1} effectively leave the basis of string webs corresponding to simple roots of the maximal deformation flavor algebra unchanged.  For instance, in the case of the $E_8$ basis shown in figure \ref{f2}, the braiding can be thought of as a rearrangement of the red ``cuts" which, though it changes the way we compute the adjacency of the singularities, does not affect the location of the singularities or string webs in the figure.  Note that some of the new cuts will intersect some of the string webs, and, to the extent that it is more convenient to work with string webs which do not intersect cuts, it may be advantageous to deform the webs by passing them through some of the singularities.  Upon passing a strand of a string web through a singularity, generally a new strand connected to that singularity is created;  the rules for this are described in detail in \cite{DeWolfe:1998eu, DeWolfe:1998zf}.  

The two remaining deformation patterns involving $I^*_n$ singularities --- numbers 6 and 16 in \ref{table:theories} --- could be found in a similar way by a suitable braiding.  But in these cases it is easier to construct the curve by direct factorization of the discriminant.  The result is given in the appendix.

The final deformation pattern --- number 24 in table \ref{table:theories} --- can be realized by braiding the maximal deformation of the $I_0^*$ in \eqref{Ftheory} by $\s_4\circ\s_3^{-1}\circ\s_5\circ\s_4^{-1}$ to obtain the presentation 
\begin{align}\label{Ftheory2}
\{\bz_a\} 
&= \{(1,0)^2,\, (1,-1)^2,\, (0,1)^2\}
& &\text{for the $I_0^*$ case.}
\end{align}
Now we can coalesce each pair of adjacent parallel charge vectors to get the submaximal deformation pattern
\begin{align}\label{subgrouping5}
\{\bz_a\} 
&= \{(\sqrt2,0),\, (\sqrt2,-\sqrt2),\, (0,\sqrt2)\}
& &\text{for}\ I_0^*\to\{{I_2}^3\} .
\end{align}

We have thus shown that all the deformation patterns in table \ref{table:theories} can be realized by complex deformations of the Kodaira singularities.  This therefore also shows that all these deformation patterns have realizations in terms of EM duality monodromies.  Indeed, the presentations of the submaximal singularities given in \eqref{subgrouping1}, \eqref{subgrouping2}, \eqref{subgrouping3}, \eqref{subgrouping4}, and \eqref{subgrouping5} give the monodromies of the singularities in their deformation patterns explicitly.  Furthermore, since these submaximal deformations are found by coalescing singularities of the maximal deformations by implementing specific linear constraints on their linear masses, the explicit SW curves of the submaximal deformations are thereby constructed.  We will give some details in two examples below, to help make this construction more concrete, and also to illustrate how the flavor symmetry associated with a submaximal deformation appears.

Note that all these same arguments can also be applied to the non-scale-invariant $I_n$ and $I^*_n$ series of Kodaira singularities to show the existence of many commensurate-charge deformations in addition to the maximal ones.  Indeed, these deformations are predicted to occur from the realization of these singularities as IR free lagrangian theories.  For instance, the $I_n$ singularity corresponds to a $\U(1)$ gauge theory with beta function proportional to $n=\sum_a Q_a^2$ where $Q_a$ are the $\U(1)$ charges of the hypermultiplets.  The maximal deformation is the case where all $Q_a=1$, while all other choices with unequal charges give sub-maximal deformations.  A similar story holds for the $I_n^*$ series where the maximal deformation is the $\SU(2)$ gauge theory with all hypermultiplets in the fundamental representation, while other choices of representations give sub-maximal deformations.  For more details see \cite{Argyres:2015ffa} and \cite{Argyres:2016new}.

Finally, note that this string web strategy for coalescing adjacent singularities can fail to when the singularities have charges which are not mutually local.  For example, restricting the $E_8$ linear masses by setting $\ba_6(\bn)=\ba_7(\bn)=\ba_8(\bn)=0$ coalesces the five right-most singularities in figure \ref{f2} to give a single singulariy with $\SL(2,\Z)$ monodromy $K=\left(\begin{smallmatrix}-3 & 2 \\ 1 & -1\end{smallmatrix}\right)$.  Since $|\Tr(K)|>2$, this is in a hyperbolic conjugacy class, so cannot be conjugate to any of the monodromies in table \ref{Table:Kodaira}.  Thus it cannot be the case that tuning the masses in this way actually coalesces these five singularities.  (Instead it forces them to combine into two separate singularities.)

\paragraph{Two examples of the explicit construction of SW curves and determination of flavor symmetries for submaximal deformations.} 

The two examples will be the submaximal deformations with patterns $II^*\to\{{I_1}^2,{I_4}^2\}$ and $II^*\to\{{I_1}^6,I_4\}$.  We will show how the flavor group of the first is determined to be $\Sp(4)$ while that of the second is determined to be either $\Sp(10)$ or $\SO(11)$.  In section \ref{section:one-form}, when we construct the SW 1-forms for these curves, we will see that the only consistent flavor group in the second case is $\Sp(10)$.  In section \ref{RGflow}, when we apply the constraints from the safely irrelevant conjecture to \emph{all} mass deformations (not just a generic deformation), we will see that the first theory is ruled out as being unphysical (or, if physical, it would be a counter example to the safely irrelevant conjecture).

\vspace{.2cm}

\noindent\underline{Example 1:  $II^*\to\{{I_1}^2,{I_4}^2\}$}

In the presentation of the maximal deformation given in \eqref{Ftheory}, by comparing to the presentation of the submaximal deformation in the first line of \eqref{subgrouping1} we see that we need to coalesce 2 groups of 4 adjacent $\bz=(1,0)$ singularities.  From figure \ref{f2} this means we want to set to zero the linear mass coordinates, $\bn$, dual to the simple roots $\ba_i$ for $i\in\{1,2,3,5,6,7\}$.  In other words we solve the system of linear equations $\ba_i(\bn) = 0$ to find, using \eqref{E8roots}, that
\begin{align}\label{C2masspatt}
n_2 = n_3 = n_4 = n_5 &= \frac14(n_1-n_8), &
n_6 = n_7 &= n_8.
\end{align}
Plugging these values into the $II^*$ maximal deformation curve found in \cite{Minahan:1996cj}, gives an explicit curve for this deformation of the $II^*$ singularity which depends polynomially on only two mass parameters, $\{n_1,n_8\}$, and so its coefficients must be polynomials in just 2 algebraically independent homogeneous combinations of $\{n_1,n_8\}$.  They turn out\footnote{There are standard, though computationally intensive, algorithms for finding a basis of algebraically independent polynomials which generate a given set of polynomials; see, e.g., \cite{Cox:1997}.} to be
\begin{align}\label{C2Ms}
M_2 &= \frac1{16}\left(5n_1^2-2n_1n_8+13 n_8^2\right), &
M_4 &= \frac1{64}\left(n_1^2+2n_1n_8-3n_8^2\right)^2,
\end{align}
and the resulting curve is given in appendix \ref{appA1}, eqn.\  \eqref{C2sigM}.  The degrees $\{2,4\}$, of the mass invariants, $\{M_2,M_4\}$, imply by the Chevalley-Shephard-Todd theorem that the automorphism group of the SW curve acting on the linear masses is the order eight group $\Z_2 \ltimes (\Z_2)^2$.  Since this group is the same as Weyl$(\Sp(4))$, so we deduce that the flavor algebra of this deformation must be $\Sp(4)$.  (Since $\Sp(4) \simeq \SO(5)$, there is no ambiguity in deducing $F$ from Weyl$(F)$ in this case.)

We have thus constructed a SW curve $\S(u,M_a)$ in terms of the Weyl-invariant mass parameters whose discriminant has the factorization \eqref{discfac} required by the deformation pattern, and deduced the associated flavor symmetry from the discrete automorphism group of the curve as a function of any set of linear mass parameters.   For instance, the Weyl$(\Sp(4))$ action can be made more obvious for the polynomials \eqref{C2Ms} given in terms of the $E_8$ linear masses by defining new linear mass parameters linearly related to the $\bn$ by
\begin{align}\label{E8ntoC2m}
m_1 &:= \frac12\left(n_1-n_8\right), &
m_2 &:= \frac14\left(n_1+3n_8\right),
\end{align}
in terms of which
\begin{align}\label{C2M}
M_2 &= m_1^2 + m_2^2, &
M_4 &= m_1^2 m_2^2 .
\end{align}
These are then clearly a basis of invariants of Weyl$(\Sp(4))\simeq\Z_2\ltimes\Z_2^2$ since the $\Z_2$ factors can be taken to act on the $m_i$ by permutations and independent sign changes.

But it is important to note that the dependence of the $M_a$ mass invariants in terms of linear masses $\bm$ proposed in \eqref{C2M} is not unique.  For instance, the parameterization of the $M_a$ in terms of linear masses $\til m_i$ given by
\begin{align}\label{C2Mt}
M_2 &= a\, (\til m_1^2 + \til m_2^2),& 
M_4 &= b\, \til m_1^2 \til m_2^2 + c\, (\til m_1^2 + \til m_2^2)^2,
\end{align} 
for arbitrary complex constants $a$, $b$, $c$ gives $M_a$ which are invariant under  Weyl$(\Sp(4))$ (with the same action on the $\til m_i$ as on the $m_i$).  And (except for special values of $a$, $b$, $c$) the $\til m_i$ in \eqref{C2Mt} are not linearly related to the $m_i$ in \eqref{C2M}.  

Thus our construction by itself does not determine the linear mass dependence of the SW curve.  However, a specific linear mass dependence, as in \eqref{C2Ms}-\eqref{C2M}, is picked out in our construction by the linear mass dependence of the original maximally deformed curve.  It turns out that in all cases this linear mass dependence is, in fact, the physical one determined by the SW one-form.  We will discuss the reasons for this in section \ref{sec:end}.

\vspace{.2cm}

\noindent\underline{Example 2:  $II^*\to\{{I_1}^6,I_4\}$}

We can repeat the previous construction for the $II^*\to\{{I_1}^6,I_4\}$ deformation pattern.  In this case we only need to coalesce one set of four adjacent $\bz=(1,0)$ singularities in \eqref{Ftheory}.  From figure \ref{f2} we can again read off a set simple roots which will do this if they are set to zero, e.g., $\ba_i(\bn)=0$ for $i\in\{1,2,3\}$.  This implies, by \eqref{E8roots}, that
\begin{align}\label{C5masspatt}
n_2 = n_3 = n_4 = n_5.
\end{align}
Plugging these values into the $II^*$ maximal deformation curve found in \cite{Minahan:1996cj}, gives a curve for the deformation of the $II^*$ singularity which depends polynomially on only five mass parameters, $\{n_1,n_5,n_6,n_7,n_8\}$, and so its coefficients must be polynomials in 5 algebraically independent homogeneous combinations of those mass parameters.  They turn out to be
\begin{align}\label{C5Ms}
M_2 &= N_2, \qquad
M_4 =3 N_2^2-12N_4, \qquad
M_6 = -\frac{9}{2} N^3_2+18N_2N_4-108N_6,
\nonumber\\
M_8 &= \frac{45}{8}N_2^4-45N_2^2N_4+90N_4^2+216N_8, \qquad
M_{10} = -2592 N_{10},
\end{align}
where we have defined
\begin{align}\label{C5Ms2}
N_{2k} &:= \sum_{1\le i_1<...<i_k\le5} 
m_{i_1}^2 \cdots m_{i_k}^2,
\end{align}
in terms of a new basis of linear masses, $\bm$, defined by
\begin{align}\label{E8ntoC5m}
m_1 &:= \frac1{2\sqrt6} \left( n_1 + n_6 + n_7 + n_8 \right)
&
m_2 &:= \frac1{2\sqrt6} \left( n_1 + n_6 - n_7 - n_8 \right)
\nonumber\\
m_3 &:= \frac1{2\sqrt6} \left( n_1 - n_6 + n_7 - n_8 \right)
&
m_4 &:= \frac1{2\sqrt6} \left( n_1 - n_6 - n_7 + n_8 \right)
\\
m_5 &:= \frac2{\sqrt6} n_5 .
\nonumber
\end{align}
The resulting SW curve $\S(u,M_a)$ can be found in appendix \ref{appA1}, eqn.\ \eqref{C5sigM}. 

The degrees, $\{2,4,6,8,10\}$, of the $M_a$'s implies by the Chevalley-Shephard-Todd theorem that the automorphism group of the SW curve acting on the linear masses is $S_5\ltimes (\Z_2)^5$.  The definition of the $N_{2k}$'s in \eqref{C5Ms2} make it clear that this groups acts by permutations and independent sign changes of the $m_i$.   Note that in this case the curve does not uniquely determine the flavor group since both Weyl$(\Sp(10)) \simeq \text{Weyl}(\SO(11)) \simeq S_5\ltimes (\Z_2)^5$.  

Construction of the SW one-form (discussed in the next section) both determines the dependence of the SW curve on linear mass parameters, and determines the flavor group.  In this case the linear mass dependence is precisely the one given above in \eqref{C5Ms}-\eqref{E8ntoC5m}, inherited from the linear mass dependence of the maximal deformation of the $II^*$ singularity, and the flavor group turns out to be $\Sp(10)$.

The above construction might seem to be non-unique since there are many (70) different ways of coalescing 4 of the 8 charge $(0,1)$ singularities in the presentation of the $E_8$ maximal deformation shown in figure \ref{f2}.  However, these are all equivalent (up to linear redefinitions of the mass parameters) because they are all related by the action of Weyl$(E_8)$ on the simple roots of $E_8$.  In addition, by braiding the presentation shown in figure \ref{f2} in suitable ways, one can find (infinitely) many new presentations with four adjacent singularities with parallel charge vectors.  These can then be coalesced to find yet more curves realizing the $\{I_1^6,I_4\}$ deformation pattern.  In all cases that we have checked, these are equivalent to the curve constructed in appendix \ref{appA1}.  We suspect that all such braidings and coalescences must be equivalent by virtue of the Weyl$(E_8)$ action, but we do not have a proof, since the way the braid group and Weyl group actions are related seems complicated; cf., \cite{Hauer:2000xy}.

\subsection{Relation of the submaximal to the maximal flavor algebra\label{s3.3}}

This procedure can be carried out for all the other submaximal deformation patterns in table \ref{table:theories} as well, with the resulting curves as functions of Weyl-invariant mass parameters, $\S(u,M_d)$, recorded in the appendix.  In particular, this allows us to read off --- up to the ambiguity Weyl$(\SO(2n+1))=\text{Weyl}(\Sp(2n))$ --- the flavor symmetries of the CFT corresponding to the various deformations.  These flavor symmetries are listed in table \ref{table:theories}.

It is natural to ask if this construction gives a simple relation between the flavor symmetries of the submaximal deformations and that of the maximal deformation of a given scale invariant singularity.   We have seen that the construction consists of restricting the Cartan subalgebra of the maximal flavor algebra (e.g., $E_8$ in the above examples) to a linear subspace determined by setting $\bw_i(\bm)=0$ for some set of $\bw_i\in\L_{F\text{-max}}$, the root lattice of the maximal flavor algebra.  This subspace is identified with the Cartan of the resulting submaximal flavor algebra.  The root lattice of the submaximal algebra, $\L_{F\text{-submax.}}$, (which is in the dual of the Cartan subalgebra) is the quotient
\begin{align}\label{submaxL}
\L_{F\text{-submax}} \simeq \L_{F\text{-max}}/\L_{\{\bw_i\}}
\end{align}
of the maximal flavor root lattice by the sublattice $\L_{\{\bw_i\}}$ generated by the $\bw_i$.  

This construction is an unfamiliar one in the context of Lie algebra theory.  For instance it is unrelated to subalgebra constructions, in which the subalgebra root lattice is induced by orthogonal projection with respect to the Killing form of the maximal algebra: $\ba \to \ba - \sum_{i,j} \bw_i \, [\bw_i(\bw_j^*)]^{-1} \ba(\bw_j^*)$.   (The Killing form is needed to define the dual vectors $\bw_i^*$.)  Indeed, it is not too hard to see this in the above examples, where the root lattices of the $\Sp(4)$ and of the $\Sp(10)$ or $\SO(11)$ submaximal algebras can be deduced from \eqref{E8ntoC2m} and \eqref{E8ntoC5m}.   For instance, in the case of the $\Sp(4)$ submaximal deformation, since the $n_i$ $E_8$ linear masses are coordinates in an orthonormal basis (with respect to the $E_8$ Killing form), it is clear that the $\Sp(4)$ Cartan coordinates $m_i$ given by \eqref{E8ntoC2m} are not orthonormal.  However, they \emph{are} orthonormal coordinates with respect to the $\Sp(4)$ Killing form.   Thus the submaximal Killing forms are not induced from the Killing form of the maximal flavor algebra.  As a further indication of this, note for example that $E_8$ does not even have $\Sp(10)$ as a subalgebra! 

Indeed, the Killing form does not enter directly into any of the data (SW curve or one-form) specifying the special K\"ahler structure.  Instead, as we have seen --- and will be made more explicit with our construction of the SW one-form in the next section --- what enters is the root lattice (without metric) and a (linear) action of the Weyl group on it.  This data allows one to reconstruct the flavor algebra (and therefore its root system and its Killing form up to overall normalization) uniquely.\footnote{We do not know of a proof of this statement from first principles, but it is easy to check it directly for all reductive Lie algebras, e.g., by inspection of explicit descriptions of the simple root systems \cite{Argyres:2012ka}.}  The property which seems to characterize the choice of sublattices $\L_{\{\bw_i\}}$ one can mod out by is that the resulting quotient lattice has an action of a Weyl group which is not the restriction of the maximal algebra's Weyl group on the quotient lattice.

Unfortunately, it seems difficult to evaluate whether this property is satisfied for any given set $\{\bw_i\}$, so we are are uncertain whether the ``string web" procedure of this section generates all such sublattices (up to equivalences), or only a subset of them.  

\subsection{Relation between the two submaximal deformations of the $I_0^*$ singularity}

Among the submaximal deformations we have constructed two are of the $I_0^*$ singularity.  Both deformation patterns have three generic singularities and the same flavor symmetry, $\Sp(2) \simeq \SU(2)$.   Since the $I_0^*$ singularity has a marginal deformation parameter, $\t$, which can be taken to the limit $\t=i\infty$ in which the low energy theory on the CB becomes weakly coupled, it is natural to associate it with a scale-invariant lagrangian theory.  (See, however, section \ref{sec5.3} for another interpretation.)  The only such theory with a $\D(u)=2$ CB vev and $\Sp(2)$ flavor symmetry is the $\cN=2^*$ $\SU(2)$ gauge theory with one adjoint hypermultiplet.  We thus expect these two deformations to somehow be equivalent.  

We will show here that they are related by a 2-isogeny of their fibers.  That is, the elliptic curves of the two deformations are related by a 2-to-1 holomorphic map from the $\{{I_1}^2,I_4\}$ curve to the $\{{I_2}^3\}$ curve.  This rescales the low energy $\U(1)$ gauge coupling by a factor of $\sqrt2$ which rescales electric and magnetic charges by opposite factors of $\sqrt2$.  By acting with an overall $\til M=\left(\begin{smallmatrix}1&0\\-1&1\end{smallmatrix}\right)$ $SL(2,\Z)$ transformation on the $I_0^*$ presentation in \eqref{subgrouping1}, we obtain a new presentation
\beq
{\bz_a}=\{(0,2),(-1,2),(1,0)\}\quad \text{for}\ I_0^*\to\{{I_1}^2,I_4\}.
\eeq
This presentation makes it obvious that the rescaling of the electric and magnetic charges by opposite $\sqrt{2}$ factors is what is needed to change from the $\{I_1^2,I_4\}$ to the $\{I_2^3\}$ pattern \eqref{subgrouping5}. This change in normalization between the two descriptions is discussed in \cite{Seiberg:1994aj}, who also noted the existence of the two different curves.

Note that these two curves give physically equivalent CB theories only if there is a restriction on the allowed charges appearing in the BPS spectrum of the $\{{I_1}^2, I_4\}$ theory since its electric charge number 1 states get mapped to charge number $1/2$ states in the $\{{I_2}^3\}$ theory.  This means that only states with even electric charges in the $\{{I_1}^2, I_4\}$ theory can have counterparts in the $\{{I_2}^3\}$ theory.  But since hypermultiplets in the fundamental of $\SU(2)$ contribute electric charge-1 states in the $\{{I_1}^2,I_4\}$ deformation (as per the discussion in section 4.2 of \cite{Argyres:2015ffa}), this is consistent with the fact that only adjoint hypermultiplets (with twice the low energy $\U(1)$ charge of fundamentals) enter into the $\cN=2^*$ theory.

The isogeny between the two curves, \eqref{I0*C1sigM} and \eqref{I0*C1sigSW2}, can be found explicitly.  There the CB vev, mass parameter, marginal coupling, and Weierstrass curve coordinates for the $\{{I_2}^2,I_4\}$ curve are $u$, $m$, $\a$, $x$ and $y$, respectively, while for the $\{{I_2}^3\}$ curve they are $U$, $M$, $\t$, $X$ and $Y$.  The coupling $\t$ appears in the $\{{I_2}^3\}$ curve in terms of three modular forms, $e_j(\tau)$, while it appears via a different modular function, $\a(\t)$, in the $\{{I_1}^2,I_4\}$ curve.  The map between the curves is
\begin{align}\label{}
U      &= \frac{1}{3 e_3} \left( u + \frac23 m^2 \right), 
\qquad M^2 = \frac{2}{9 e_3^2} m^2, \qquad
\frac{(e_1-e_3)(e_2-e_3)}{e_3^2}     = \frac{9}{4}\a^2, 
\nonumber\\
&\qquad\qquad X = \frac{1}{6 e_3} \frac{A B_2^2}{B_1^2} ,
\qquad\qquad\qquad Y =  u (e_1-e_2) y\,\frac{A B_2}{B_1^3} ,
\end{align}
where
\begin{align}\label{}
A &:= 2u(e_1-e_3) - 3m^2 \a^2 e_3 ,
&
B_j &:= u(e_3-2e_j)+(3x+2m^2 \a^2) e_3 .
\end{align}
One then checks that $\frac{dX}{Y} = - 2 \frac{dx}{y}$, giving the factor of two rescaling of $\t$ between the two curves.   

There is thus a change of variables for any given choice of the coupling and mass parameters making the predictions of the low energy theory on the CB of these two theories the same.  Nevertheless, these two CB geometries are different since they have different global dependence on the marginal coupling.  In particular, the  $\{{I_2}^3\}$ curve's coupling takes values in a fundamental domain of $PSL(2,\Z)$ while that of the $\{{I_1}^2,I_4\}$ curve takes values in a fundamental domain of an index 3 subgroup $\G^0(2)\subset PSL(2,\Z)$.  This distinction has striking implications for the discrete gauging constructions considered in \cite{Argyres:2016yzz} (see especially section 3.4.2 of that paper).

\section{Construction of the one-form}\label{section:one-form}

Knowing that the curves exist is only half the battle, since we still need to show that a SW one-form, $\l$, also exists satisfying the rigid special K\"ahler (SK) conditions \eqref{SWform}.  The SK conditions are very powerful requirements which by themselves fix the linear mass dependencies of SW curves, but are difficult to solve.   

We will be inspired by a strategy developed by Minahan and Nemeschansky (MN) \cite{Minahan:1996cj,Minahan:1996fg} for finding a one-form satisfying the SK condition and invariant under a given Weyl group symmetry.  The MN strategy has three parts.  They start with the SW curve as a function of the Weyl-invariant deformation parameters, $\S(M_d)$, and posit an ansatz for the form of $\l$ and for the positions of its poles $x=x_\bw$ that automatically satisfies the SK condition that the residues of $\l$ form a lattice in the weight space of $F$.  Then they simultaneously solve for the pole position dependence on the linear mass parameters, $x_\bw(u,\bm)$, and for the dependence of the Weyl-invariant deformation parameters appearing in the curve on the linear masses, $M_d(\bm)$.  This step is computationally intensive since it involves solving factorization constraints on polynomials in many variables.  Finally, they fix a few remaining parameters in $\l$ by solving the differential SK condition in \eqref{SWform}.  Either of these steps might fail to have a solution, in which case we would learn only that the MN ansatz fails, but could not conclude that there is no SK geometry associated to the $\S(M_d)$.  But it so happens that we find a solution in every case with the MN ansatz.  In fact, we often find more than one solution, and we discuss the physical equivalence of these multiple solutions in the next section.

The SW curves as functions of the Weyl-invariant deformation parameters, $\S(M_d)$, are easy to write down for the maximal deformations, and are listed in table \ref{t4}.  MN computed the curve and one-form as functions of the linear masses, $\S(\bm)$ and $\l(\bm)$, for the $II^*$, $III^*$ and $IV^*$ maximal deformations in \cite{Minahan:1996cj,Minahan:1996fg} using their method, while SW computed the same for the $I_0^*$ maximal deformation in \cite{Seiberg:1994aj} using different methods.  The MN strategy (with a slight generalization of the original MN ansatz for $\l(\bm)$, presented below) also works for the maximal deformations of the $IV$, $III$ and $II$ singularities; the resulting one-forms for these cases are presented in appendix \ref{c1}, \ref{c2} and \ref{c3} for completeness, since they do not seem to have been written down in full generality elsewhere.

\subsection{The MN ansatz and factorization at poles.}

We now follow the MN strategy to compute SW one-forms for the submaximal deformations.  We start by parameterizing the possible form of the one-form on the curve as 
\begin{align}\label{MNform}
\l(\bm) := \left[ 2 \Delta(u) a\, u + 6b\m\, x + 2 W(M_d) 
+ \sum_i r_i \!\!\! \sum_{\bw_i\ \text{orbit}} 
\frac{y_{\bw_i}(u,\bm)}{\bw_i(\bm)^2\, x 
- x_{\bw_i}(u,\bm)} \right] \frac{dx}{y} .
\end{align}
Here $a$, $b$, and the $r_i$ are constants, $W$ is a Weyl invariant polynomial in the masses, $\bw_i(\bm)$ is the linear combination of the $\bm$ corresponding to a weight $\bw_i$ of $F$, the sum is over Weyl orbits of each $\bw_i$, and $x_{\bw_i}(u,\bm)$ is polynomial in $u$ and the $\bm$ such that 
\begin{align}\label{polepos}
y_{\bw_i}(u,\bm) &:=\left. \bw_i(\bm)^3\, y\right|_{x\,=\,\bw_i(\bm)^{-2}\, x_{\bw_i}(u,\bm)}
\end{align}
is also polynomial in $u$ and the $\bm$.  We will call a triple $\{\bw, x_\bw, y_\bw\}$ satisfying these constraints --- i.e., \eqref{polepos} and the conditions that $x_\bw$ and $y_\bw$ be polynomials --- a ``pole position solution".  Any $\l$ of the form \eqref{MNform} is then manifestly a Weyl-invariant meromorphic one-form on $\S(M_d)$ with pairs of poles with residues $\pm r_i \bw_i(\bm)$ at the two points on the fiber with $x=\bw_i(\bm)^{-2} x_{\bw_i}$.  The reason for the ``$i$" index on the $\bw_i$ (Weyl orbit of) weights is to accomodate cases where there are multiple distinct pole position solutions.

The $r_i$ parameters could be absorbed by a rescaling of the $\bw_i$.  In particular, if $\{\bw, x_\bw, y_\bw\}$ is a pole position solution, then $\{r\bw, r^2 x_\bw, r^3 y_\bw\}$ is also a pole position solution for any constant $r$.  However, there turn out to be curves having multiple distinct solutions of the differential SK condition in \eqref{SWform} which involve different normalizations of the pole position solutions.  It is therefore convenient to introduce explicit pole position normalizations, $r_i$, in order to describe these solutions in a uniform way.  

The condition \eqref{SWform} that the residues of $\l$ are of the form $\bw(\bm)$ with $\bw$ in the root lattice of $F$ can be justified as follows.  For generic masses, the flavor symmetry is broken to $\U(1)^{\text{rank}(F)}$, and these $\U(1)$ ``quark number" charges thus span a lattice of rank rank$(F)$.  The terms in the central charge which depend linearly on the quark numbers and are proportional to the linear masses come from the residues of $\l$.  Thus the $\bw$'s appearing in $\l$, i.e., the set $\{r_i\bw_i\}$, should span the quark number charge lattice.  Since the linear mass parameters $\bm$ transform in the adjoint of $F$, whenever $\ba(\bm)=0$ for $\ba$ a root of $F$ there should be a degeneracy in the BPS spectrum since on these subspaces the flavor symmetry is not completely abelianized (i.e., it has some unbroken non-abelian factors).  Conversely, if there were some $\bw$ not in the root lattice of $F$, then there will be additional degeneracies in the BPS mass spectrum at every point on the CB for masses satisfying $\bw(\bm)=0$ which are not due to an enhanced symmetry.  Discounting the existence of such accidental degeneracies which persist for all values of $u$, we conclude that the lattice of quark number charges must be the root lattice of $F$.

This leads to some constraints on the $r_i$.  If one chooses (as we will) to normalize the $\bw_i$ to all lie in a given normalization of the root lattice, one then has a restriction on the allowed values of the $r_i$ such that the set $\{r_i\bw_i\}$ spans a possibly rescaled root lattice.  In particular, the $r_i$ have to be real and all their ratios must be rational (for simple $F$), and there may be further constraints for them to span the whole lattice and not just a sublattice.

Note that our ansatz \eqref{MNform} for $\l$ is a slight generalization of the MN ansatz in \cite{Minahan:1996cj}, differing from it by the addition of the $\m x$ term.  Here $\m$ is the relevant deformation parameter with scaling dimension $\Delta(\m) = 2-\Delta(u)$ which exists whenever the dimension of $u$ is in the range $1 < \Delta(u) < 2$.  Note that the $\m x$ term contributes a double pole with no residue at the single point on the fiber at $x=\infty$ for $\m\neq0$.

The condition that $y_\bw$ given in \eqref{polepos} be polynomial in $u$ and $\bm$ is a stringent constraint on the curve $\S(\bm)$.  In particular, it requires that the right side of \eqref{rank1curve} be a perfect square in $u$ and $\bm$ when $x=\bw(\bm)^{-2} x_\bw(u,\bm)$.  This condition is strong enough to determine $\S(\bm)$ given $\S(M_d)$, or, equivalently, to determine the dependence of the Weyl$(F)$-invariant mass polynomials on the linear masses, $M_d(\bm)$.

Solutions to this factorization condition can be found by the following procedure \cite{Minahan:1996cj,Minahan:1996fg}.  First, pick a residue, $\bw(\bm)$, linear in the $\bm$ associated to a weight $\bw\in\ff^*$.  Second, write an ansatz for the associated pole position $x_\bw(u,\bm)$ which is either linear or quadratic in $u$ and is invariant under the subgroup of Weyl$(F)$ which fixes $\bw$.  (We limit ourselves to $x_\bw$ at most quadratic in $u$ just for computational ease.)  Next, parameterize the possible dependence of the Weyl-invariant masses $M_d$ appearing in the curve on the linear masses, $\bm$ (as in the discussion around \eqref{C2Mt} above).  Fourth, pick a convenient direction in the complexified Cartan subalgebra of $F$, $\be\in\ff_\C$, set $\bm = m \be$, and demand that the curve factorizes as a polynomial in $u$ and $m$ at $x = m^{-2} \bw(\be)^{-2} x_\bw(u, m\be)$.  If this has a solution, it will fix some linear combination of the coefficients in the $x_\bw$ and $M_d(\bm)$ polynomials.  Continue this for other choices of directions in $\ff_\C$ until all coefficients are determined.  (Convenient directions in $\ff_C$ are often proportional to weights which are fixed by a large subgroup of Weyl$(F)$.)  This is a laborious process, made more so by the fact that the factorization step can result in a tree of possibilities which needs to be exhausted.  See \cite{Minahan:1996cj,Minahan:1996fg} for detailed examples carrying out this procedure and for some tricks to simplify the factorization step.  

In the appendices we record the results of this process for all the submaximal deformations of the scale-invariant Kodaira singularities.  For each deformation, $\S(M_d)$, we find multiple solutions for Weyl orbits of $\bw$ and associated $x_\bw(u,\bm)$, but all correspond to a single solution for the deformed curve, $\S(\bm)$, in terms of the linear masses $\bm$.

\subsection{Solving the differential constraint}

So far we have constructed SW one-forms satisfying the second of the SK condtions in \eqref{SWform}.  We next need to check whether there are values of the constants $a$, $b$, $r_i$ and the Weyl-invariant polynomial $W(M_d)$ in the MN ansatz \eqref{MNform} such that the one-form satisfies the first, differential, SK constraint in \eqref{SWform}.  

It is straightforward, although slightly technical, to convert this constraint to linear algebra following and generalizing an argument in  \cite{Noguchi:1999xq}.  In particular, appendix A of \cite{Noguchi:1999xq} shows that 
\begin{align}\label{dul}
\del_u\l(b{=}0) &= 2a\frac{dx}{y}+(A_1 x+A_0)\frac{dx}{y^3}
+d\f'
\end{align}
for some meromorphic $\f'$, where
\begin{align}\label{A12}
A_1 & := a \delta f - W \del_u f + h_1, &
A_0 & := a \delta g - W \del_u g + h_0,
\end{align}
and $\delta := \m\del_\m+\sum_i m_i\del_{m_i}$.  Denote the scaling dimensions of the polynomials $f(u,\bm)$ and $g(u,\bm)$ appearing in the Weierstrass form of the curve $\S(\bm)$ \eqref{rank1curve} by $\Delta(f)$, $\Delta(g)$.  Then   the weighted homogeneity of $f$ and $g$ in $u$, $\m$, and the $m_i$ implies
\begin{align}\label{Deltafg}
\delta f &= \D(f)\, f - \D(u)\, u\del_u f , &
\delta g &= \D(g)\, g - \D(u)\, u\del_u g.
\end{align}
Finally, $h_{0,1}$ in \eqref{A12} are given by
\begin{align}\label{h12new}
h_1 & := 
\frac12 \sum_i r_i \!\!\! \sum_{\bw_i\ \text{orbit}}\!\!\! \bw_i(\bm)^{-4}
\left[2 x_{\bw_i} \del_u y_{\bw_i}
- 3 y_{\bw_i}\del_u x_{\bw_i} \right], \\
h_0 & := 
\frac16 \sum_i r_i \!\!\! \sum_{\bw_i\ \text{orbit}}\!\!\!
\Bigl\{
\bw_i(\bm)^{-6} \left[
6 x_{\bw_i}^2 \del_u y_{\bw_i}
- 9 y_{\bw_i} x_{\bw_i} \del_u x_{\bw_i} 
\right] +
\bw_i(\bm)^{-2} \left[
4 f \del_u y_{\bw_i}
- 3 y_{\bw_i} \del_u f 
\right]\Bigr\}. \nonumber
\end{align}

This can be generalized to include the $\l_b := 6b \m x dx/y$ term in \eqref{MNform} using the identity
\begin{align}\label{btermident}
\del_u\l_b
&= \left(-3b \m \del_u g\, x+b \m f\del_u f\right)
\frac{dx}{y^3}+d\phi'' ,
\end{align}
which implies that \eqref{dul} holds for $\l$ with $b\neq0$ with the replacements
\begin{align}\label{btermA12}
A_1 &\to A_1 -3b\m\del_ug, &
A_0 &\to A_0 +b\m f\del_uf,
\end{align}
in \eqref{A12}.  However we cannot yet conclude that the differential constraint in \eqref{SWform} is satisfied if $A_0=A_1=0$ since $dx/y$, $dx/y^3$, and $xdx/y^3$ are related up to a total derivative by the identity
\begin{align}\label{(1,0)cohomol}
0=\frac{1}{y} - (2 f x+ 3 g)\frac{1}{y^3} + 2 \del_x\left(\frac{x}{y}\right).
\end{align}
Multiplying this equation by $-c\, dx$ and adding this to  \eqref{dul} (for $b\neq0$) gives
\begin{align}\label{dul2}
\del_u\l &= (2a-c)\frac{dx}{y}+(B_1 x +B_0)\frac{dx}{y^3}+d\f,  \\
B_1 & :=a \delta f - 3b \m \del_u g + 2c f - W \del_u f 
+ h_1, \nonumber\\
B_0 & := a \delta g + b \m f \del_u f + 3c g - W \del_u g 
+ h_0.  \nonumber
\end{align}
Thus the SW condition is fulfilled if $a$, $b$, $c$, $r_i$, and $W(M_d)$ exist such that $B_1=B_0=0$ and $c \neq2 a$.  Such a solution for $\l$ then satisfies the differential constraint in \eqref{SWform} with normalization $\k=(2a-c)$.

Since $B_1$ and $B_0$ are relatively high-order polynomials in $u$, their vanishing greatly over-constrains $a$, $b$, $c$, $r_i$ and $W(M_d)$.  Nevertheless, for every set of $x_{\bw_i}$ solving the factorization condition described in the previous paragraph, we find  at least one solution, and typically many solutions to the differential condition.   The resulting values of $a$, $b$, $c$, $r_i$, and $W(M_d)$ are recorded in the appendices.  

\subsection{Ambiguities in the one form and the flavor symmetry}

We have described how the Weyl group of the flavor symmetry, $F$, is encoded in the curve.  But this cannot distinguish between $\SO(2r+1)$ and $\Sp(2r)$ flavor factors since their Weyl groups are equal.  We have argued that the residues of the one form span the root lattice of $F$, and a root lattice with an action of the Weyl group on it determines $F$ uniquely.

For instance, there is a basis $\{\be^1,\ldots,\be^r\}$ of $\R^r$ such that every element of the root lattices of $\SO(2r+1)$ and $\Sp(2r)$ can be written as $n_i \be^i$ for some integers $n_i$, and such that the action of their Weyl groups is by permutations and independent sign flips of the $n_i$.  However in this case the $\SO(2r+1)$ root lattice is the lattice with all $n_i$ allowed, while the $\Sp(2r)$ root lattice is the sublattice of elements such that $\sum_i n_i$ is even.  In fact, in this presentation, the $\be^i$ are orthonormal with respect to the Killing forms of either  algebra.   Given a lattice and an action of the Weyl group on it, one can reconstruct (up to normalization) the Killing form on the lattice by demanding that the Weyl group is generated by hypersurface-orthogonal reflections.

It would thus seem that there is enough information in the curve and one-form to uniquely determine the flavor symmetry.  However, in practice we do not find that this is always true because we do not find unique solutions for the one-form from the MN ansatz.  Indeed, this non-uniqueness of the SW one form was already pointed out in \cite{Minahan:1996cj} and discussed further in \cite{Noguchi:1999xq}.  

The non-uniqueness of the one form comes about because there can be many distinct solutions for the pole positions.   Each pole position is labelled by a residue in its Weyl orbit, $r_i \bw_i(\bm)$, and the index $i=1,\ldots,P$ thus runs over the distinct Weyl orbits of pole positions.  The complex parameters, $r_i$, are only constrained by the differential constraint --- the first of the SK conditions in \eqref{SWform}.  If there are fewer than $P$ of these constaints then there will be a multi-parameter family of SW one forms for the curve. 

Different one-forms in such a multi-parameter family generically give rise to different physical predictions for the BPS spectrum, so are not physically equivalent.  For even though the differential constraint in \eqref{SWform} ensures that the $u$-derivative of the central charge, $Z$, will be independent of the pole positions, this does not determine the linear mass dependence of $Z$, as these effectively appear as constants of integration of the differential constraint.  

As an obvious illustration of this, note that if the set of $r_i\bw_i$ are not commensurate (as elements of the dual Cartan algebra of the flavor symmetry) or have different complex phases, then they will not span a real lattice in $\ff_\C^*$, and so cannot describe the quark number charges of BPS states at all.  

This physical requirement is addressed in the second of the SK conditions in \eqref{SWform}, which states that the residues of $\l$ should span the root lattice of the flavor symmetry.  To span a real rank$(F)$ lattice only requires that the $r_i$ all be real and commensurate.  Demanding that the lattice also be a root lattice of a Lie algebra which has as its Weyl group the Weyl group of the curve adds a further restriction.  This is because different lattices can now be distinguished by the action of the Weyl group, as the above example of the $\SO(2r+1)$ and $\Sp(2r)$ root lattices illustrates.

If the root lattice is uniquely specified by the Weyl group, then the multi-parameter family of SW forms which satisfy the above restrictions will by definition give identical predictions for the central charges of the theory.  But precisely in the case of the $\SO(2r+1)$/$\Sp(2r)$ ambiguity of root lattices associated to a single Weyl group, we can potentially get two different SW one-forms for a given curve, one realizing the $\SO(2r+1)$ flavor group, and the other realizing $\Sp(2r)$.  Depending on the number of pole positions, the constraints among the $r_i$ coming from solving the differential constraint, and the conditions of commensurability and of being in a root lattice of the Weyl group, there can be one of two outcomes:  (1) only one root lattice, $\SO(2r+1)$ or $\Sp(2r)$, satisfies the constraints, and the one-form and flavor symmetry are uniquely determined; or (2) there are solutions for both symmetries and there are two distinct theories.  A third logical possibility is that no values of the $r_i$ satisfy all the constraints, and there is no physical one form satisfying the MN ansatz; however, this does not occur for any of our deformation patterns.

Among the submaximal curves and one-forms we have constructed, there are four that have this potential $\SO(2r+1)$/$\Sp(2r)$ ambiguity.   (Recall that $\SO(5)=\Sp(4)$ so there is only a potential ambiguity for $r>2$.)  There is an $r=5$ ambiguity for the $II^*\to\{{I_1}^6,I_4\}$ theory, and $r=3$ ambiguities for the $II^*\to\{{I_1}^3,I_1^*\}$, $III^* \to \{{I_1}^5,I_4\}$, and $III^*\to\{{I_1}^3,I_0^*\}$ theories.
\begin{itemize}
\item $II^*\to\{{I_1}^6,I_4\}$:  In this case there is only a single pole orbit, and the integral span of its resiues fills our the $\Sp(10)$ root lattice.  This identification of the flavor symmetry agrees with the identification based on S-dualities \cite{Argyres:2007tq}.
\item $II^*\to\{{I_1}^3,I_1^*\}$: Here there are three pole orbits and one relation among their residues imposed by the differential constraint.  The resulting 2-parameter family of solutions has members which generate both the $\SO(7)$ and the $\Sp(6)$ root lattices, as described in appendix \ref{appA1}.  Thus we find two distinct SW geometries for this deformation.  However, this distinction turns out to be moot since neither of these flavor symmetry assignments are self-consistent under RG flows, as described in the next section.
\item $III^* \to \{{I_1}^5,I_4\}$:  In this case there are five pole orbits and one relation.  But only two of the pole positions transform under the $\SO(7)$/$\Sp(6)$ Weyl factor, and the solutions only generate the $\Sp(6)$ root lattice, as described in appendix \ref{appA2}.  Again, this identification of the flavor symmetry agrees with the identification based on S-dualities \cite{Argyres:2007tq}.
\item $III^*\to\{{I_1}^3,I_0^*\}$: This example has four pole orbits and one relation giving a 3-parameter family of solutions which realizes both the $\SO(7)$ and the $\Sp(6)$ root lattices.  But, unlike the previous ambiguous case, one can flow to this theory by tuning masses in the $II^*\to\{{I_1}^4,I_0^*\}$ theory.  This theory has flavor symmetry $F_4$, and the masses which flow to the $III^*\to\{{I_1}^3,I_0^*\}$ theory are the ones which implement the adjoint breaking $F_4 \to \SO(7)\oplus \U(1)$.  (These and other RG flows are discussed in section \ref{RGflow}.)  We take this as evidence that the physical flavor symmetry of this theory is $\SO(7)$ and not $\Sp(6)$.
\end{itemize}

\subsection{Relation to the one forms of the maximally deformed curve}

Finally, we address the extent to which the one-forms we have (laboriously) constructed are simply restrictions of the known one-forms of the maximal deformations.  After all, in the last section we constructed the submaximal deformation curves by imposing restrictions on the linear masses of the maximal deformation curves.  

This construction ensures that the restricted curves will have the property that they will factorize at some pole positions.  This is simply because the restriction of a pole position solution will be a pole position solution of the restricted curve.  But this does not mean that the set of poles occuring in the restricted one form are the restriction of the poles in the maximal deformation one-form.  The reason is that, as we have seen, the Weyl group of the restricted curve is not the same as (and is generally larger than) the restriction of the Weyl group of the maximal deformation curve.  By the Weyl invariance of the restricted curve, it will necessarily also factorize on the Weyl orbit of any pole position solution.  Since the Weyl group of the restricted curve is different from that of the maximal one, some of the pole positions in the orbit of a pole position found by restriction will not themselves be the restriction of any maximal curve pole position.  Thus we see that although the existence of pole positions for which the restricted curve factorizes is ensured by the restriction, the one form resulting from summing over the Weyl orbits of these poles will not be the restriction of the maximal deformation one form.

This also means that it is not the case that the differential constraint on the one form will be automatically satisfied by virtue of the restriction construction:  the ``new" poles generated by the Weyl group of the submaximal deformation curve add new terms to the differential constraint equation.  An example is the $II^*\to\{I_2, IV^*\}$ deformation, where there is a pole position $x_{\bw_1}$ with residue $r_1 \bw_1(\bm)$ found by restriction from the $II^*$ maximal deformation, but whose coefficient $r_1$ is set to zero by the differential constraint, and so does not appear in the one-form for this submaximal deformation.

\section{Constraints from RG flows and gauged rank 0 SCFTs}\label{RGflow}

We now have constructed SW geometries for all 28 deformation patterns shown in table \ref{table:theories}.   As discussed in section \ref{sec:math}, these deformation patterns were the only ones which satisfy the constraints coming from the safely irrelevant conjecture, from the Dirac quantization condition, and from having consistent EM duality monodromies.  These constraints were applied only to the case of \emph{generic} mass deformations, for which an initial UV singularity on the Coulomb branch is split into a collection of distinct ``frozen" singularities corresponding to massless IR theories which admitted no further (splitting) deformations.  

However, the safely irrelevant conjecture together with the Dirac quantization condition should also apply to these theories at non-generic values of their mass parameters $\bm\in\ff_\C$.  These correspond to subspaces of the flavor Cartan algebra, $\ff_\C$, at which some of the above-mentioned frozen singularities coalesce to form higher singularities.  These higher singularities then must have the correct flavor symmetries to account for the part of the UV flavor symmetry left unbroken by the non-generic masses.

For example, we have seen that the $III^*\to\{I_1^*,I_1^2\}$ deformation pattern corresponds to a consistent SW geometry with flavor algebra $\SU(2)\oplus\SU(2)$ (the UV flavor symmetry).  Now suppose we turn on a special mass, $\bm\in\ff_\C$ that performs the adjoint flavor breaking
\begin{align}\label{IIIs-split1}
\SU(2)\oplus\SU(2) \xrightarrow{\ \bm\ } 
\SU(2) \oplus \U(1),
\end{align}
and find that the $III^*$ singularity splits as
\begin{align}\label{IIIs-split2}
III^* \xrightarrow{\ \ \ } \{I_2^* , I_1 \}.
\end{align}
(This in fact occurs, and will be discussed below.)  Since the $I_1$ singularity can only have a $\U(1)$ flavor symmetry, this splitting can only be consistent with \eqref{IIIs-split1} if the (IR free) theory at the $I_2^*$ singularity has flavor symmetry $\SU(2)$.  (It may also have an extra $\U(1)$ flavor factor if the massless hypermultiplets at the $I_2^*$ and $I_1$ singularities have non-mutually local charges;  see \cite{Argyres:2015ffa} for  a discussion of the counting of the low energy $\U(1)$ flavor factors.)

If one finds that the $I_2^*$ theory has a larger-rank flavor symmetry, it is not a priori clear that there is a contradiction.  For instance, it is possible that the symmetry is accidentally enlarged in the IR, or that symmetry is not actually the apparent larger symmetry by virtue of a discretely gauged subgroup.   As argued in \cite{Argyres:2015ffa}, these possibilities are not consonant with the way deformations of rank-1 SW geometries actually occur, and we have conjectured that they do not.  As will be discussed in examples below, when there is such an enlargement in the rank of the flavor symmetry, there is no way of reconciling the low energy degrees of freedom on the Coulomb branch at the special mass values with those at generic masses.  So in these cases, discussed in section \ref{sec5.1}, we conclude that the geometries fail to have consistent RG flows.

If, on the other hand, one found that the $I_2^*$ theory had an enlargement of the flavor symmetry that did not change its rank, or even a smaller-than-expected flavor symmetry, then there is no sharp contradiction with the spectrum of light states on the Coulomb branch.   In these cases, discussed in section \ref{sec5.2}, we cannot conclude that the RG flows are inconsistent.

We call this test {\it RG flow consistency}.  This check is not trivial, as evidenced by the existence of deformations for which it fails.  It depends crucially on determining the possible flavor symmetries of fixed point theories, like the $I^*_2$ theory in the above example.  A careful analysis of the various Kodaira singularities and their flavor symmetries and charge normalizations is given in section 4.2 of \cite{Argyres:2015ffa}.  We will use this analysis heavily in this section.

There is an interesting way in which theories which fail the RG flow consistency check could still be consistent as long as one is willing to posit the existence of special new rank-0 interacting $\cN=2$ SCFTs (i.e., ones without a Coulomb branch).  We will discuss these in more detail below.  Though there is no direct evidence for the existence of such theories, we will describe an indirect way of looking for such rank-0 theories.  Such a search, however, is computationally intensive, and will not be carried out here.  Frozen $I_n^*$ singularities can be interpreted as such theories (instead of as IR free $\SU(2)$ gauge theories).  RG flows in the $I_n^*$ series of singularities will be analyzed from this perspective in section \ref{sec5.3}.

In the rest of this section we will perform such checks to conclude that the three geometries whose maximal flavor symmetries are shaded red in table \ref{table:theories} are not consistent field theories with these flavor symmetry assignments, modulo the existence of these new rank-0 SCFTs.\footnote{An analysis of the possible sub-maximal flavor symmetry assignments for all the geometries in table \ref{table:theories} and their consistency under RG flows is carried out in \cite{Argyres:2016xua}.}  Carrying this out involves determining all the special relations among the mass parameters at which some of the generic singularities on the CB collide.  This is equivalent to mapping out the web of RG flows connecting the various rank-1 theories discussed here.  It is a daunting task to do this when there are many relevant deformation parameters.  The most complicated cases are the maximal deformations, whose RG flows have already been studied \cite{Minahan:1996cj, Noguchi:1999xq}.  In that case one finds that maximal deformation CFTs only flow to other maximal deformation CFTs.  For example, a simple part of the ``web" of flows is
\begin{align}\label{E-RGflows}
[II^*,E_8] \to
[III^*,E_7] \to
[IV^*,E_6] \to 
[I_0^*,D_4] \to 
[IV,A_2] \to
[III,A_1] \to
[II,-].
\end{align}
Here we are denoting a CFT by its Kodaira singularity and its flavor symmetry using Dynkin notation for their Lie algebras.  The arrows in \eqref{E-RGflows} denote turning on a particular relevant deformation and flowing to a specific IR fixed point on the CB.  The flows shown in \eqref{E-RGflows} are ``minimal" in the sense that they correspond to turning on a relevant operator which splits off the smallest possible (often $I_1$) singularity from the original singularity.  There are many other fixed points on the CB and other special deformation directions which flow these theories to the IR free $[I_n, A_{n-1}\oplus U_1]$ and $[I_n^*, D_{n+4}]$ theories. These are the ``standard'' $\U(1)$ gauge theories with $n$ charge-1 hypermultiplets and the $\SU(2)$ gauge theory with $(n+4)$ fundamental hypermultplets, respectively. (We are denoting $\U(1)$ factors by $U_1$.)

The other submaximal deformations have more complicated 
RG flow patterns.  They can be organized as in the maximal case \eqref{E-RGflows} in terms of sequences of ``minimal" flows. For instance, the ones whose deformation patterns involve a single frozen $I_4$ obey
\begin{align}\label{C-RGflows}
[II^*,C_4] \to
[III^*,C_3\oplus C_1] \to
[IV^*,C_2\oplus U_1] \to 
[I_0^*,C_1],
\end{align}
together with flows to the IRF theories $[I_n, A_{n-5}\oplus U_1\oplus U_1]$, $[I^*_n, C_{n+1}]$, and $[I^*_n, C_1\oplus D_n]$.  These latter are the $\U(1)$ gauge theories with $n{-}4$ charge-1 and 1 charge-2 hypermultiplets, the $\SU(2)$ gauge theory with $(n+1)$ adjoint hypermultplets, and the $\SU(2)$ gauge theory with 1 adjoint and $n$ fundamental hypermultplets, respectively, as we will show below.

The ones with a frozen $I_0^*$ obey
\begin{align}\label{I0-RGflows}
[II^*,F_4] \to
[III^*,B_3] \to
[IV^*,A_2] \to
[II, -]
\end{align}
together with flows to $[I_n, A_{n-1}\oplus U_1]$ singularities and $I_n^*$ singularities for which we will be able to find a consistent interpretation in terms of novel gauged rank-0 SCFTs.  Similarly, the ones with a frozen $I_1^*$ obey
\begin{align}\label{I1-RGflows}
\begin{array}{ccccc}
[II^*,BC_3] &\to&
[III^*,A_1\oplus A_1] &\to&
[IV^*,U_1] \\
&&\downarrow \\
&&[II,-]
\end{array}
\end{align}
together with flows to $[I_n, A_{n-1}\oplus U_1]$ singularities and $I_n^*$ singularities for which we will not be able to find a consistent IR free gauge theory interpretation, but will be able to find a consistent interpretation in terms of gauged rank-0 SCFTs.  

Finally, the ones with frozen $III^*$ or $IV^*$ singularities have minimal flows
\begin{align}\label{III-IVs-RGflows}
[II^*,G_2] & \to
[III^*,A_1] \to
[IV^*,-]\nonumber\\
[II^*,A_1] & \to
[IV^*,-]\\
[II^*, A_1]' & \to
[III^*,-].\nonumber
\end{align}
Here $[II^*,A_1]$ refers to the geometry with generic deformation pattern $\{{I_1}^2,IV^*\}$, while $[II^*,A_1]'$ refers to the one with pattern $\{I_1,III^*\}$.

In general, the full web of possible RG flows is very complicated and difficult to compute.   But in cases where there are only two relevant parameters, it is not too hard to do so and we will carry out the full analysis.  For example, from the form of the $[II^*,C_2]$ curve given in \eqref{C2sigM} -- \eqref{C2MtoN} it follows that there are four  directions (up to equivalences under the Weyl$(C_2)$ action) in the space of $\bm$'s for which the $II^*$ singularity is not fully split into the four singularities $\{{I_1}^2,{I_4}^2\}$ of its deformation pattern.  These are
\begin{align}\label{IIsC2-RG}
[II^*, C_2] \ 
\begin{cases}
\to \{ I_1, I_3^*\} &\text{for}\ m_1=0,\\
\to \{ {I_1}^2, I_8\} &\text{for}\ m_1=m_2,\\
\to \{ II, {I_4}^2\} &\text{for}\ m_1=\sqrt{2-\sqrt3}\,m_2,\\
\to \{ I_1 ,I_4, I_5\} &\text{for}\ m_1=2\,m_2.
\end{cases}
\end{align}
Any other direction results in the generic deformation pattern.  Thus for this curve the complete set of possible IR fixed points one can flow to is $\{I_1, I_4, I_5, I_8, I_3^*, II\}$.  We will discuss this example in detail below, and will determine whether there is an identification of IR free gauge theories with the $I_4$, $I_5$, $I_8$, and $I_3^*$ fixed points that is consistent with the flavor symmetry and with Dirac quantization. 

For theories with higher rank a complete analysis is computationally intensive and will not be carried out here.  An easier task is to study the web of their minimal adjoint breaking flows. We turn to this discussion now.

\subsection{Adjoint flavor breakings}\label{sec5.1}

For generic values of the mass deformation parameters, the flavor symmetry, $F$, is broken to rank$(F)$ $\U(1)$ factors since the masses transform in the adjoint representation of $F$.   There exist specific patterns of mass parameters (e.g., if $\ba(\bm)=0$ for any root $\ba$) for which the flavor group $F$ is not completely broken to abelian factors but some nonabelian factors are left unbroken.  For these mass patterns, groups of the undeformable singularities must merge to form new singularities on the CB, as we argued in section \ref{StrWeb}.  Turning on a mass $\bm$ such that $\ba(\bm)=0$ for all but one of the simple roots of $F$ thus picks out directions in the flavor Cartan along which $F$ is minimally broken, and we will call them the {\it minimal adjoint breakings}.  These breakings are all of the form $F \to \U(1) \oplus \text{semi-simple}$, where the semi-simple factors can be read off from the $F$ Dynkin diagram by crossing out one node.

We will now use these minimal adjoint breakings to locate some RG flows of the submaximal deformation theories.  We will then check their RG flow consistency, i.e., check that the set of resulting merged singularities accounts for --- in terms of corresponding IR free or conformal theories --- the expected simple unbroken flavor symmetry factors, and is consistent with charge normalizations.  

First we discuss two examples in detail: the $[II^*,C_2]$ special K\"ahler geometry describing the generic deformation pattern $II^*\to\{{I_1}^2,{I_4}^2\}$ with $\Sp(4)$ flavor symmetry which fails this test and thus does not give rise to a consistent CFT; and the $[II^*,C_5]$ geometry with $II^*\to\{{I_1}^6,I_4\}$ deformation pattern and flavor symmetry $\Sp(10)$, which instead satisfies this condition in a interesting way.  We will then describe the result of similar considerations for the rest of the submaximal deformations.

\vspace{.2cm}

\noindent\underline{Example 1:  $II^*\to\{{I_1}^2,{I_4}^2\}$}

This theory has flavor symmetry $\Sp(4)$ which has Dynkin diagram
\begin{align}\nonumber
\begin{tikzpicture}[decoration={markings,mark=at position 0.7 with {\arrow{stealth}}}]
\node[wc] (r1) at (0,.5) {};
\node at (0,.15) {$\ba_1$};
\node[wc] (r2) at (1,.5) {};
\node at (1,.15) {$\ba_2$};
\draw[double, postaction={decorate},double distance=2pt] (r2) -- (r1);
\end{tikzpicture}
\end{align}
where the simple roots can be written in terms of a set of two orthonormal vectors, $\be_i$, $i=1,2$, as
\beq
\ba_1=\frac{1}{\sqrt{2}}(\be_1-\be_{2}), \qquad\&\qquad\ba_2=\sqrt{2}\be_2 .
\eeq
It is straightforward to list the two adjoint breakings obtained by choosing a particular mass configuration such that $\ba(\bm)\neq0$ for only one of the simple roots. 

\emph{Minimal adjoint breaking 1:}
\begin{align}\label{C2mab1}
\begin{tikzpicture}
\draw[thick] (0.9,.4) -- (1.1,0.6);
\draw[thick] (0.9,.6) -- (1.1,0.4);
\node[wc] (r2) at (0,.5) {};
\node at (0,.15) {$\ba_1$};
\node at (1,.15) {$\ba_2\neq0$};
\node[right] at (2.5,.35) {$\Longleftrightarrow \qquad\Sp(4)\to\SU(2)\oplus\U(1)$.};
\end{tikzpicture}%
\end{align}
Imposing the condition $\ba_1(\bm)=0$, $\ba_2(\bm)\neq0$, implies that we have to set $m_1=m_2$ in the generic $II^*\to\{{I_1}^2,{I_4}^2\}$ curve \eqref{C2sigM}--\eqref{C2MtoN}. From the explicit expression of the curve we can study the behavior of the geometry near the zero of its $x$ discriminant and determine that for this particular mass deformation
\beq\label{C2Spl2}
II^*\to\{{I_1}^2,I_8\} .
\eeq
This is the second line reported in \eqref{IIsC2-RG}.
The $I_1$'s in \eqref{C2Spl2} can only contribute abelian flavor factors since their global symmetry is $\U(1)$.  The $\SU(2)$ flavor factor must therefore come from the $I_8$ singularity.  The $I_8$ singularity can arise from an IR free $\U(1)$ gauge theories with hypermultiplets of charges $Q_i$ satisfying $\sum_i Q_i^2=8$ where, as reviewed in section \ref{sec2.2}.  The $Q_i$ are normalized so that their squares are integers.  Since, by \eqref{C2Spl2}, these charges inhabit a moduli space with an $I_1$ singularity (which can only be given by a $\U(1)$ gauge theory with a charge-1 hypermultiplet), Dirac quantization implies that the $Q_i$'s must, in fact, be integers.  There are then just 3 possible $\U(1)$ theories giving the $I_8$ singularity:  one with 8 charge-1 massless hypermultiplets and flavor symmetry $\U(8)$; one with 1 charge-2 and 4 charge-1 hypermultiplets and flavor symmetry $\U(1)\oplus\U(4)$; and one with 2 charge-2 hypermultiplets and flavor symmetry $\U(2)$.  The last one has the expected $\SU(2)$ flavor symmetry.  Furthermore, the two charge-2 hypermultiplets are just what one expects from the generic deformation pattern of this theory, $II^*\to\{{I_2}^2,{I_4}^2\}$, since the two $I_4$ singularities, in order to be ``frozen", must each have a charge-2 massless hyper.
We have thus found that the splitting \eqref{C2Spl2} associated to this minimal adjoint flavor breaking gives a consistent account of the expected flavor symmetries.

\emph{Minimal adjoint breaking 2:}
\begin{align}\label{C2mab2}
\begin{tikzpicture}
\draw[thick] (-0.1,.4) -- (.1,0.6);
\draw[thick] (-0.1,.6) -- (.1,0.4);
\node at (0,.15) {$\ba_1\neq0$};
\node[wc] (r2) at (1,.5) {};
\node at (1,.15) {$\ba_2$};
\node[right] at (2.5,.35) {$\Longleftrightarrow \qquad\Sp(4)\to\U(1)\oplus\SU(2)$.};
\end{tikzpicture}
\end{align}
Imposing the condition $\ba_1(\bm)\neq0$, $\ba_2(\bm)=0$, implies that we have to set $m_2=0$ in the generic $II^*\to\{{I_1}^2,{I_4}^2\}$ curve. From the explicit expression of the curve, studying the behavior of geometry near the zero of its $x$ discriminant we obtain
\beq\label{C2Spl1}
II^*\to\{I_1,I_3^*\}.
\eeq
This is the first line reported in \eqref{IIsC2-RG}.
The $I_1$ singularity in \eqref{C2Spl1} has a $\U(1)$ flavor symmetry, so the expectation is that the $I^*_3$ singularity will carry the $\SU(2)$ symmetry.  The $I^*_3$ singularity is an IR free $\SU(2)$ gauge theory with one-loop beta function coefficient satisfying 
\begin{align}\label{beta-a}
b_0 = \frac{3}{a^2}
\end{align}
for some charge normalization factor $a$.  If the massless half-hypermultiplets are in $\SU(2)$ irreps $\bf R$ of dimension $R$, then $b_0$ is
\begin{align}\label{beta-0}
b_0=-T({\bf 3})+\frac{1}{2}\sum_i T(\bR_i),\qquad 
T(\bR):=2{\rm tr}_\bR(t_3^2)
=\frac{1}{6}(R-1)R(R+1).
\end{align}
Here $t_3$ is the $\U(1)$ generator of $\SU(2)$ unbroken on the Coulomb branch, $t_3(\bR) = \frac12 \text{diag}\{ R-1, R-3, \ldots, -R+1\}$, and the CB charges of its components are
\begin{align}\label{u1su2norm}
Q=2a t_3(\bR) .
\end{align}
Since, by \eqref{C2Spl1}, the $I^*_3$ singularity cohabitates with an $I_1$ singularity, $Q$ in \eqref{u1su2norm} must be an integer.  This implies that $a \in \Z/2$ if all $R_i$ are odd, while $a\in \Z$ if some $R_i$ are even.  Furthermore, $\cN=2$ supersymmetry and the absence of global $\SU(2)$ anomalies imply that there must be an even number of half-hypermultiplets in each $R$-odd irrep, and the total number of half-hypermultiplets in irreps with $R=2 \mod 4$ must be even.  (See, e.g., section 4.2 of \cite{Argyres:2015ffa} for a review of these constraints.)  

With these constraints, one finds only four solutions to \eqref{beta-a}--\eqref{u1su2norm} for the IR free $\SU(2)$ gauge theory giving the $I^*_3$ singularity:  it can have half-hypermultiplets in one of the four representations
\begin{align}\label{possthrs}
8\cdot {\bf 3} 
\quad &\text{with}\quad a= \frac12, & 
\Rightarrow\quad & Q =1
 &\text{and}\quad 
F &= \Sp(8) ,
\nonumber\\
14\cdot {\bf 2} 
\quad &\text{with}\quad a= 1, & 
\Rightarrow\quad & Q =1
 &\text{and}\quad 
F &= \SO(14) ,
\\
6\cdot {\bf 2}\oplus 2\cdot{\bf 3} 
\quad &\text{with}\quad a= 1, & 
\Rightarrow\quad & Q \in\{1,2\}
 &\text{and}\quad 
F &= \SO(6)\oplus\Sp(2) ,
\nonumber\\
4\cdot {\bf 2}\oplus1\cdot{\bf 4} 
\quad &\text{with}\quad a= 1, & 
\Rightarrow\quad
& Q \in \{1,3\}
 &\text{and}\quad F &= \SO(4) .
\nonumber
\end{align}
None of these has the expected $\SU(2)$ flavor symmetry, so this can only be consistent if we posit that there is an accidentally enhanced flavor symmetry at the $I^*_3$ fixed point.  

But even this assumption is not sufficient to provide a consistent picture of the RG flow.  This is because deforming the $[II^*,C_2]$ theory by its additional relevant operator --- the mass not turned on in \eqref{C2mab2} --- causes the splitting $I^*_3 \to \{ I_1, {I_4}^2\}$.  But this splitting has no consistent interpretation in terms of the IR free field theories in \eqref{possthrs}.  In other words, the special mass deformation which splits the $I^*_3$ in this way in these theories does not have the interpretation of three frozen singularities (i.e., one with a single $Q=1$ hypermultiplet, and two each with a single $Q=2$ hypermultiplet).  This is just a reflection of the fact that since all the theories in \eqref{possthrs} have more than one relevant deformation, this special mass deformation can always be followed by further deformations which split the singularities, and under generic mass deformations these IR free theories split into a number of (electric) $I_{Q^2}$ singularities, together with a monopole $I_1$ singularity and a dyon $I_1$ singularity in the first three cases in \eqref{possthrs}, or a frozen $I_1^*$ singularity in the fourth.  Thus in these theories, the $\{ I_1, {I_4}^2\}$ pattern, however arrived at, can always be split further.

The discussion above does not exhaust all possibilities. If we posit that in this case the $I^*_3$ singularity corresponds to some novel (non-lagrangian) field theory the RG-flow has a consistent low energy interpretation. Because the low energy $\U(1)$ coupling on the CB of the $I_3^*$ geometry becomes free at the singularity (i.e., $\lim_{u\to 0} \t(u) = i\infty$), it would seem natural to assume that the field theory is IR free.  But if this were the case, we have just seen that it cannot be any known (lagrangian) IR free theory.  An alternative to an IR free theory is one for which the low energy modes on the Coulomb branch become free at the singularity, but the theory as a whole is not free. 

An example of such a situation --- an interacting field theory with a free Coulomb branch sector --- could arise if there existed an interacting rank-0 $\cN=2$ SCFT with a flavor symmetry, $F$, with flavor central charge $k_F$.  ``Rank 0" means simply that this SCFT has no Coulomb branch of its moduli space.  Then the $I^*_n$ geometry could arise coupling this theory to a vector multiplet by gauging an $\SU(2)\subset F$ \cite{Argyres:2007cn,Argyres:2007tq}.  The resulting theory would have a rank-1 CB with $\D(u)=2$ and Kodaira singularity $I_n^*$ with $n = a^2 (\frac12 I_{\SU(2)\hookrightarrow F} k_F-4)$, as long as $n\ge0$.\footnote{Here $a$ is the charge normalization factor, as in \eqref{beta-a}, and $I_{\SU(2)\hookrightarrow F}$ is the Dynkin index of embedding of $\SU(2)$ in $F$; see \cite{Argyres:2007cn} for details.} We will discuss this possibility in more detail below.

\vspace{.2cm}

\noindent\underline{Example 2:  $II^*\to\{{I_1}^6,I_4\}$}

Consider now the $[II^*,C_5]$ theory whose generic deformation pattern is $II^*\to\{{I_1}^6, I_4\}$, and which --- along with the other theories in the series \eqref{C-RGflows} --- was constructed by S-duality techniques \cite{Argyres:2007cn, Argyres:2007tq, Chacaltana:2014nya}.  If an analysis of the RG flows of this theory, along the lines of the above analysis of the $[II^*,C_2]$ theory, found a violation of the RG flow condition, this could then be taken as strong evidence for the existence of new rank-0 SCFTs.  We will perform the RG flow test for the minimal adjoint breakings of this theory.  We will find no violations, and will instead find that these flows are consistent in an intricate and interesting way. 

The Dynkin diagram describing the flavor symmetry of this theory is:
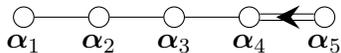
\begin{align}\nonumber
\begin{tikzpicture}[decoration={markings,mark=at position 0.7 with {\arrow{stealth}}}]
\node[wc] (r2) at (0,.5) {};
\node at (0,.15) {$\ba_1$};
\node[wc] (r2) at (1,.5) {};
\node at (1,.15) {$\ba_2$};
\node[wc] (r3) at (2,.5) {};
\node at (2,.15) {$\ba_3$};
\node[wc] (r4) at (3,.5) {};
\node at (3,.15) {$\ba_4$};
\node[wc] (r5) at (4,.5) {};
\node at (4,.15) {$\ba_5$};
\draw (r1) -- (r2);
\draw (r2) -- (r3);
\draw (r3) -- (r4);
\draw[double,postaction={decorate},double distance=2pt] (r5) -- (r4);
\end{tikzpicture}
\end{align}
where the simple roots can be written in terms of a set of five orthonormal vectors, $\be_i$, $i=1,...,5$, as
\beq
\ba_i=\frac{1}{\sqrt{2}}(\be_i-\be_{i+1}), \quad i=1,...,4\qquad\&\qquad\ba_5=\sqrt{2}\,\be_5
\eeq
It is straightforward to list all possible adjoint breakings obtained by choosing a particular mass configuration such that $\ba(\bm)\neq0$ for only one of the five simple roots at each time.

\emph{Minimal adjoint breaking 1:}
\begin{align}\label{C5mab1}
\begin{tikzpicture}[decoration={markings,mark=at position 0.7 with {\arrow{stealth}}}]
\draw[thick] (-0.1,.4) -- (.1,0.6);
\draw[thick] (-0.1,.6) -- (.1,0.4);
\node at (0,.15) {$\ba_1\neq0$};
\node[wc] (r2) at (1,.5) {};
\node at (1,.15) {$\ba_2$};
\node[wc] (r3) at (2,.5) {};
\node at (2,.15) {$\ba_3$};
\node[wc] (r4) at (3,.5) {};
\node at (3,.15) {$\ba_4$};
\node[wc] (r5) at (4,.5) {};
\node at (4,.15) {$\ba_5$};
\draw (r2) -- (r3);
\draw (r3) -- (r4);
\draw[double,postaction={decorate},double distance=2pt] (r5) -- (r4);
\node[right] at (5,.35) {$\Longleftrightarrow \qquad \Sp(10)\to \U(1)\oplus \Sp(8)$.};
\end{tikzpicture}
\end{align}
Imposing the condition $\ba_1(\bm)\neq0$, $\ba_2(\bm)=\ba_3(\bm)=\ba_4(\bm)=\ba_5(\bm)=0$, implies that we have to set $m_2=m_3=m_4=m_5=0$ in the generic $[II^*,C_5]$ curve \eqref{C5sigM}--\eqref{C5MtoN}.  Studying the behavior of geometry near the zero of its $x$ discriminant we obtain
\beq\label{C5Spl1}
II^*\to\{I_1,I_3^*\}.
\eeq
Now we have to check that the splitting \eqref{C5Spl1} is consistent with the expected flavor symmetry. The $\Sp(8)$ factor comes from the IR free $\SU(2)$ gauge theory giving the $I^*_3$ singularity. This is the one with half-hypermultiplet representation content $8\cdot{\bf 3}$ and with charge rescaling factor $a=1/2$. Using \eqref{u1su2norm}, since all the fields are in the $\bf 3$, we obtain that the CB charge for these fields is $Q=1$ and it is compatible with Dirac quantization and the existence of the $I_1$ singularity. The $I_1$ provides the remaining $\U(1)$ factor.

\emph{Minimal adjoint breaking 2:}
\begin{align}\label{C5mab2}
\begin{tikzpicture}[decoration={markings,mark=at position 0.7 with {\arrow{stealth}}}]
\node[wc] (r1) at (0,.5) {};
\node at (0,.15) {$\ba_1$};
\draw[thick] (0.9,.4) -- (1.1,0.6);
\draw[thick] (0.9,.6) -- (1.1,0.4);
\node at (1,.15) {$\ba_2\neq0$};
\node[wc] (r3) at (2,.5) {};
\node at (2,.15) {$\ba_3$};
\node[wc] (r4) at (3,.5) {};
\node at (3,.15) {$\ba_4$};
\node[wc] (r5) at (4,.5) {};
\node at (4,.15) {$\ba_5$};
\draw (r3) -- (r4);
\draw[double,postaction={decorate},double distance=2pt] (r5) -- (r4);
\node[right] at (5,.35) {$\Longleftrightarrow \qquad \Sp(10)\to \SU(2)\oplus \U(1)\oplus \Sp(6)$.};
\end{tikzpicture}
\end{align}
Imposing the condition $\ba_2(\bm)\neq0$, $\ba_1(\bm)=\ba_3(\bm)=\ba_4(\bm)=\ba_5(\bm)=0$, implies that we have to set $m_1=m_2$ and $m_3=m_4=m_5=0$. The splitting associated to this particular value of the masses is:
\beq\label{C5Spl2}
II^*\to\{I_1,III^*\}.
\eeq
This is one of the ``minimal" RG flows recorded in \eqref{C-RGflows}.  In this case it is easy to see that the splitting \eqref{C5Spl2} is consistent with the expected flavor symmetry: the $\U(1)$ factor comes as usual from the $I_1$ singularity, while the $III^*$ singularity is the isolated CFT with flavor group $\Sp(6)\oplus\Sp(2)\simeq\Sp(6)\oplus\SU(2)$ matching exactly the remaining non-abelian component of the flavor algebra.

\emph{Minimal adjoint breaking 3:}
\begin{align}\label{C5mab3}
\begin{tikzpicture}[decoration={markings,mark=at position 0.7 with {\arrow{stealth}}}]
\node[wc] (r1) at (0,.5) {};
\node at (0,.15) {$\ba_1$};
\node[wc] (r2) at (1,.5) {};
\node at (1,.15) {$\ba_2$};
\draw[thick] (1.9,.4) -- (2.1,0.6);
\draw[thick] (1.9,.6) -- (2.1,0.4);
\node at (2,.15) {$\ba_3\neq0$};
\node[wc] (r4) at (3,.5) {};
\node at (3,.15) {$\ba_4$};
\node[wc] (r5) at (4,.5) {};
\node at (4,.15) {$\ba_5$};
\draw (r1) -- (r2);
\draw[double,postaction={decorate},double distance=2pt] (r5) -- (r4);
\node[right] at (5,.35) {$\Longleftrightarrow \qquad \Sp(10)\to \SU(3)\oplus \U(1)\oplus \Sp(4)$.};
\end{tikzpicture}
\end{align}
Imposing the condition $\ba_3(\bm)\neq0$, $\ba_1(\bm)=\ba_2(\bm)=\ba_4(\bm)=\ba_5(\bm)=0$, implies that we have to set $m_1=m_2=m_3$ and $m_4=m_5=0$. The splitting associated to this particular value of the masses is:
\beq\label{C5Spl3}
II^*\to\{I_3,I_1^*\}.
\eeq
The IR free theory at the singularity $I_3$ in \eqref{C5Spl3} is a $\U(1)$ gauge theory with 3 hypermultiplets with charge 1. This theory has a $\U(3)$ flavor symmetry. To match the rest of the flavor symmetry, we need the $I_1^*$ to be an IR free theory with $\Sp(4)$ flavor symmetry.  This is in fact the case, as the $I_1^*$ arises as a singularity for the IR free $\SU(2)$ with half-hypermultiplet representation content $4\cdot{\bf 3}$ and with charge rescaling factor $a=1/2$.  Again, since all the fields are in the $\bf 3$, from \eqref{u1su2norm} the CB charge normalization is $Q=1$.  This is compatible with Dirac quantization and the existence of the $I_3$ singularity.

\emph{Minimal adjoint breaking 4:}
\begin{align}\label{C5mab4}
\begin{tikzpicture}[decoration={markings,mark=at position 0.7 with {\arrow{stealth}}}]
\node[wc] (r1) at (0,.5) {};
\node at (0,.15) {$\ba_1$};
\node[wc] (r2) at (1,.5) {};
\node at (1,.15) {$\ba_2$};
\node[wc] (r3) at (2,.5) {};
\node at (2,.15) {$\ba_3$};
\draw[thick] (2.9,.4) -- (3.1,0.6);
\draw[thick] (2.9,.6) -- (3.1,0.4);
\node at (3,.15) {$\ba_4\neq0$};
\node[wc] (r5) at (4,.5) {};
\node at (4,.15) {$\ba_5$};
\draw (r1) -- (r2);
\draw (r2) -- (r3);
\node[right] at (5,.35) {$\Longleftrightarrow \qquad \Sp(10)\to \SU(4)\oplus \U(1)\oplus \SU(2)$.};
\end{tikzpicture}
\end{align}
Imposing the condition $\ba_4(\bm)\neq0$, $\ba_1(\bm)=\ba_2(\bm)=\ba_3(\bm)=\ba_5(\bm)=0$, implies that we have to set $m_1=m_2=m_3=m_4$ and $m_5=0$. The splitting associated to this particular value of the masses is
\beq\label{C5Spl4}
II^*\to\{I_1,I_3^*\}.
\eeq
The $I_1$ singularity in \eqref{C5Spl4} provides the usual $\U(1)$ factor, so the semi-simple component of the flavor symmetry should arise from the $I_3^*$.  In fact this is the IR free $\SU(2)$ theory with charge normalization $a=1$ and with half-hypermultiplet representation content $2\cdot{\bf 3}\oplus 6\cdot{\bf2}$ which has flavor group $\Sp(2)\oplus\SO(6)\simeq\SU(2)\oplus\SU(4)$ as expected.  The Dirac quantization condition is trivially satisfied.  Notice that in this particular case, because of the presence of half-hypermultiplets in the fundamental representation of the $\SU(2)$, an overall rescaling of the charges for the $I_3^*$ while allowed by the theory itself, would be incompatible with the presence of the $I_1$ singularity.

\emph{Minimal adjoint breaking 5:}
\begin{align}\label{C5mab5}
\begin{tikzpicture}[decoration={markings,mark=at position 0.7 with {\arrow{stealth}}}]
\node[wc] (r1) at (0,.5) {};
\node at (0,.15) {$\ba_1$};
\node[wc] (r2) at (1,.5) {};
\node at (1,.15) {$\ba_2$};
\node[wc] (r3) at (2,.5) {};
\node at (2,.15) {$\ba_3$};
\node[wc] (r4) at (3,.5) {};
\node at (3,.15) {$\ba_4$};
\draw[thick] (3.9,.4) -- (4.1,0.6);
\draw[thick] (3.9,.6) -- (4.1,0.4);
\node at (4,.15) {$\ba_5\neq0$};
\draw (r1) -- (r2);
\draw (r2) -- (r3);
\draw (r3) -- (r4);
\node[right] at (5.5,.35) {$\Longleftrightarrow \qquad \Sp(10)\to \SU(5)\oplus \U(1)$.};
\end{tikzpicture}
\end{align}
Imposing the condition $\ba_5(\bm)\neq0$, $\ba_1(\bm)=\ba_2(\bm)=\ba_3(\bm)=\ba_4(\bm)=0$, implies that we have to set $m_1=m_2=m_3=m_4=m_5$. The splitting associated to this particular value of the masses is
\beq\label{C5Spl5}
II^*\to\{I_1,I_4,I_5\}.
\eeq
The simple factor of the flavor symmetry comes from the $I_5$ singularity in \eqref{C5Spl5} if it represents an IR-free $\U(1)$ theory with 5 hypermultiplets with charge 1.  Both the $I_1$ and the $I_4$ only provide $\U(1)$ factors being, respectively, the IR-free $\U(1)$ theory with a single charge 1 hypermultiplet and the IR-free $\U(1)$ with a single charge 2 hypermultiplet. Notice that Dirac quantization is again satisfied. 

\paragraph{Minimal adjoint breaking RG flows for other submaximal deformations.}

We will now simply list the results of similar analyses for the minimal adjoint breaking flows for the remaining submaximal deformation CB geometries.  We organize them according to the three series shown in \eqref{C-RGflows}, \eqref{I0-RGflows}, \eqref{I1-RGflows}, and \eqref{III-IVs-RGflows}, which we will call the $I_4$, $I_0^*$, $I_1^*$, and $III^*/IV^*$ series, respectively.  We will postpone the discussion of the $I_0^*$ series until section \ref{sec5.3} since it requires special consideration.  For economy of presentation, we will use the Dynkin names for simple Lie algebras, and will denote $\U(1)$ factors by $U_1$.  

\vspace{.2cm}

\noindent\underline{$I_4$ series:}

The minimal adjoint breakings of the $II^*\to\{I_4,{I_1}^6\}$ geometry with $C_5$ flavor symmetry was analysed above. 

The $III^*\to\{I_4,{I_1}^5\}$ generic deformation pattern has flavor symmetry $C_3\oplus C_1$, and gives rise to minimal adjoint breaking flows 
\begin{align}\label{}
C_3  \oplus C_1 \ 
\begin{cases}
\to C_3                 &\to \{I^*_2,I_1\} \ \cmark\\
\to A_2\oplus A_1  &\to \{I_3,I_2,I_4\} \ \cmark\\
\to A_1\oplus A_1\oplus A_1  &\to \{I_2^*,I_1\} \ \cmark\\
\to C_2\oplus A_1                 &\to \{I_1^*,I_2\} \ \cmark
\end{cases}
\end{align}
(Here we have not written the $\U(1)$ factors in the adjoint breakings.)  The checks or crosses record whether each flow passes the RG flow test.  In the first line it is consistent to identify $I_2^* \simeq \SU(2) \ \text{with}\ 6\cdot{\bf 3}$ having $C_3$ symmetry, and $I_1 \simeq \U(1) \ \text{w/}\ 1\cdot1$ having $U_1$ symmetry.     In the second line it is consistent to identify $I_3 \simeq \U(1) \ \text{w/}\ 3\cdot1$ having $U_1\oplus A_2$ symmetry, $I_2 \simeq \U(1) \ \text{w/}\ 2\cdot1$ having $U_1\oplus A_1$ symmetry, and $I_4 \simeq \U(1) \ \text{w/}\ 1\cdot2$ having $U_1$ symmetry.   In the third line it is consistent to identify $I_2^* \simeq \SU(2) \ \text{with}\ 4\cdot{\bf 2}\oplus2\cdot{\bf 3}$ having $D_2\oplus C_1 \simeq A_1\oplus A_1\oplus A_1$ symmetry, and $I_1 \simeq \U(1) \ \text{w/}\ 1\cdot1$ having $U_1$ symmetry.  In the fourth line it is consistent to identify $I_1^* \simeq \SU(2) \ \text{with}\ 4\cdot{\bf 3}$ having $C_2$ symmetry, and $I_2 \simeq \U(1) \ \text{w/}\ 2\cdot1$ having $A_1\oplus U_1$ symmetry.  

The $IV^*\to\{I_4,{I_1}^4\}$ generic deformation pattern has flavor symmetry $C_2\oplus U_1$, and gives rise to minimal adjoint breaking flows 
\begin{align}\label{}
C_2 \oplus U_1 \ 
\begin{cases}
\to A_1^S  &\to \{I_0^*,{I_1}^2\} \ \cmark\\
\to A_1^L                 &\to \{I_6,{I_1}^2\} \ \cmark\\
\to C_2                 &\to \{I_1^*,I_1\} \ \cmark
\end{cases}
\end{align}
Where the $L$ ($S$) super-script indicates that the unbroken $A_1$ is associated to the long (short) $C_2$ root. The RG-flow above gives consistent result. In the first line is consistent to identify the $I_0^* \simeq \SU(2) \ \text{with}\ 2\cdot{\bf 3}$ having $A_1$ symmetry, while each $I_1 \simeq \U(1) \ \text{w/}\ 1\cdot1$ having $U_1$ symmetry. In the second line it is consistent to identify $I_6 \simeq \U(1) \ \text{w/}\ 2\cdot1\oplus1\cdot2$ having $U_1\oplus U_1\oplus A_1$ symmetry, and each $I_1 \simeq \U(1) \ \text{w/}\ 1\cdot1$ having $U_1$ symmetry. In the third line it is consistent to identify the $I_1^* \simeq \SU(2) \ \text{with}\ 4\cdot{\bf 3}$ with charge normalization $a=1/2$ having $C_2$ symmetry while the $I_1 \simeq \U(1) \ \text{w/}\ 1\cdot1$ having $U_1$ symmetry. 

The $I_0^* \to\{I_4, {I_1}^2\}$ or $\{{I_2}^3\}$ generic deformation patterns have as their flavor symmetry $A_1$, and no minimal adjoint breaking flows.  In fact their only flow is precisely the generic deformation pattern, so there is no nontrivial RG flow test for this theory. 

\vspace{.2cm}

\noindent\underline{$I^*_1$ series:}\footnote{A similar analysis applies to the $I_2^*$ and $I_3^*$ ``series" of deformations as well --- i.e., entries numbered 7, 8 and 17 in tabel \ref{table:theories}.  But these all have only rank 1 or 2 flavor symmetries, so the analysis of their RG flow consistency gives no interesting constraints.}  

The $II^*\to\{I_1^*,I_1^3\}$ generic deformation pattern has as flavor symmetry $BC_3$, by which we mean either $B_3$ or $C_3$.  It turns out that both $B_3$ and $C_3$ have the same pattern of adjoint breakings and the curve gives rise to minimal adjoint breaking flows 
\begin{align}\label{}
BC_3 \ 
\begin{cases}
\to A_2                 &\to \{I^*_1,I_3\} \ \cmark\\
\to A_1\oplus A_1  &\to \{III^*,I_1\} \ \cmark\\
\to C_2                 &\to \{I_3^*,I_1\} \ \xmark
\end{cases}
\end{align}
In the first line it is consistent to identify $I_1^* \simeq \SU(2) \ \text{w/}\ 1\cdot{\bf 4}$ having no symmetry, and $I_3 \simeq \U(1) \ \text{w/}\ 3\cdot1$ having $A_2\oplus U_1$ symmetry.
In the second line it is consistent to identify $III^* \simeq (III^*,A_1\oplus A_1)$ with $A_1\oplus A_1$ symmetry, and $I_1 \simeq \U(1) \ \text{w/}\ 1\cdot1$ having $U_1$ symmetry.   
In the third line there is no IR free $\SU(2)$ theory with flavor symmetry $C_2$ and consistent with the $I^*_3$ singularity and the $Q=1$ charge normalization forced by the $I_1$ factor, and it thus fails the test.   

The $III^*\to\{I_1^*,I_1^2\}$ generic deformation pattern has as flavor symmetry $A_1\oplus A_1$, and minimal adjoint breaking flows 
\begin{align}\label{}
A_1\oplus A_1 \ 
\begin{cases}
\to A_1\oplus U_1  &\to \{I_1^*,I_2\} \ \cmark\\
\to U_1\oplus A_1  &\to \{I^*_2,I_1\} \ \xmark
\end{cases}
\end{align}
In the first line it is consistent to identify $I_1^* \simeq \SU(2) \ \text{w/}\ 1\cdot{\bf 4}$ having no symmetry, and $I_2 \simeq \U(1) \ \text{w/}\ 2\cdot1$ having $A_1\oplus U_1$ symmetry.
In the second line there is no IR free $\SU(2)$ theory with flavor symmetry $A_1$ and consistent with the $I^*_2$ singularity and the $Q=1$ charge normalization forced by the $I_1$ factor, and it thus fails the test.  

The $II^*\to\{I_0^*,I_1\}$ generic deformation pattern has as its flavor symmetry $U_1$, and no adjoint breaking flows.  In fact its only flow is precisely the generic deformation pattern, so there is no nontrivial RG flow test for this theory.

\vspace{.2cm}

\noindent\underline{$III^*/IV^*$ series:}

The $II^*\to\{III^*,I_1\}$, $II^*\to\{IV^*,I_2\}$, and $III^*\to\{IV^*,I_1\}$ generic deformation patterns all have flavor symmetry $A_1$.  Since this is rank 1, their only flows are precisely the generic deformation patterns, so there are no nontrivial RG flow tests for these theories.

The $II^*\to\{IV^*,{I_1}^2\}$ generic deformation pattern has as flavor symmetry the exceptional $G_2$ Lie algebra whose adjoint breakings give rise to the flows 
\begin{align}\label{}
G_2 \ 
\begin{cases}
\to A_1\oplus U_1  &\to \{III^*,I_1\} \ \cmark\\
\to U_1\oplus A_1  &\to \{IV^*,I_2\} \ \cmark
\end{cases}
\end{align}
In the first line it is consistent to identify $III^* \simeq (III^*,A_1)$ with flavor symmetry $A_1$, and $I_1 \simeq \U(1) \ \text{w/}\ 1\cdot1$ having $U_1$ symmetry.    
In the second line it is consistent to identify $IV^* \simeq [IV^*,-]$ as the frozen $IV^*$ singularity with no flavor symmetry, and $I_2 \simeq \U(1) \ \text{w/}\ 2\cdot1$ having $U_1\oplus A_1$ symmetry.   

Note that in this series there are actually two different assumed frozen $IV^*$ singularities:  one with charges coming in multiples of $Q=1$ and one with charges multiples of $Q=\sqrt2$.   The first is the one that appears in the $[II^*,G_2]$ and $[III^*,A_1]$ deformations, while the second is the one appearing in the $[II^*,A_1]$ deformation.  So, if these deformations all correspond to SCFTs, then there must be two distinct frozen rank-1 SCFTs corresponding to the $IV^*$ singularity.

\subsection{Non-adjoint breaking RG flows}\label{sec5.2}

There are further conditions that consistency under RG flow imposes.  There can be special patterns of masses for which some of the undeformable singularities will merge into other singularities on the CB, even though there is no enhanced unbroken flavor symmetry.   In this case consistency requires that the new singularities correspond to IR SCFTs which are compatible with the $\U(1)^{\text{rank}(F)}$ unbroken flavor symmetry along these flows.   In every case we have checked, we find that this consistency condition is satisfied.

These non-adjoint breaking special RG flows correspond to additional ``accidental" collisions of roots of the curve discriminant, and can be algebraically complicated to locate.  We have searched for such flows for geometries with just 2 relevant deformations:

\vspace{.2cm}

\noindent\underline{$[II^*,C_2]$ geometry:}

In this case there are just two non-adjoint breaking special flows.  They are the ones recorded in the last two lines of \eqref{IIsC2-RG}.  In each both cases they satisfy the RG flow test to reproduce the $\U(1)\oplus\U(1)$ flavor symmetry.  The $\{II,{I_4}^2\}$ breaking is consistent with the interpretation of the $II$ singularity as the $[II,-]$ CFT with no flavor symmetry, and each $I_4$ as the IR free $\U(1)$ gauge theory with one charge-4 hypermultiplet with flavor symmetry $U_1$.  Thus the total flavor symmetry is the expected $U_1\oplus U_1$.  The $\{I_1,I_4,I_5\}$ breaking is consistent to identify: $I_1 \simeq \U(1) \ \text{w/}\ 1\cdot1$ having $U_1$ symmetry, $I_4 \simeq \U(1) \ \text{w/}\ 1\cdot2$ having $U_1$ symmetry, and $I_5 \simeq \U(1) \ \text{w/}\ 1\cdot1\oplus1\cdot2$ having $U_1\oplus U_1$ symmetry.  Thus the total flavor symmetry is $U_1\oplus U_1 \oplus U_1 \oplus U_1$.  This may seem like too many $U_1$'s, but we recall that some of them can be identified with global parts of the electric and magnetic charge $\U(1)$'s.

\vspace{.2cm}

\noindent\underline{$[II^*,G_2]$ geometry:}

The non-adjoint special RG flows for this theory is:
\begin{align}\label{IIsG2-RG}
[II^*, G_2] \ 
\to \{ IV^*, II  \} \ \text{for}\ m_1= e^{2\pi i/3}\,m_2. & \cmark
\end{align}
The $IV^*$ contributes no flavor symmetry, while the $II$ can only be the $[II,-]$ CFT, which also has no flavor symmetry. This flow does not present an inconsistency in the low energy action on the CB even though it predicts that there is no flavor symmetry in the IR.  This is possible if all the states which become massless on the CB are neutral under the flavor symmetry.  A sufficient condition for this to be the case is if, at  the $IV^*$ or $II$ singularity on the CB, none of the poles of the SW one form are located at the same point on the SW curve \eqref{rank1curve} as the point where the branch cuts in the $x$-plane collide.  For if this condition is satisfied, it is easy to show that the vanishing cycles at the singularities (whose homology class determines the EM and flavor charges of the light states near the singularities) can all be taken to have vanishing flavor charges.  This check is easy to carry out using the explicit curve and one-form recorded in the appendix, and is satisfied.

\vspace{.2cm}

\noindent\underline{$[III^*,A_1\oplus A_1]$ geometry:}

The non-adjoint special RG flows for this theory are:
\begin{align}\label{}
[III^*, A_1\oplus A_1] \ 
\begin{cases}
\to \{ I_1^*, II\} \ \text{for}\ m_1=\sqrt{8/3}\,m_2, &\cmark\\
\to \{ IV^*, I_1  \} \ \text{for}\ m_1= i/\sqrt{3}\,m_2. & \cmark
\end{cases}
\end{align}
In the first line, the $I_1^*$ singularity must correspond to the frozen IR free $\SU(2)$ w/ $1\cdot{\bf 4}$ theory and no flavor symmetry, while the $II$ is the $[II,-]$ CFT.  This is consistent with the $U_1\oplus U_1$ flavor symmetry not acting on the light states on the CB, and can be checked as in the last example from the curve and one-form.

In the second line, the $IV^*$ singularity must be identified with the  $[IV^*,U_1]$ CFT since this is the only $IV^*$ that deforms to a terminal $I_1^*$.  The $I_1$ is the $\U(1)$ gauge theory with 1 charge-1 hypermultiplet1, with flavor symmetry $U_1$.  Thus this flow has the expected $U_1\oplus U_1$ symmetry.

\subsection{Flows to frozen gauged rank-0 SCFTs}\label{sec5.3}

We now wish to carry out the RG flow test for minimal adjoint and special RG flows of the $I_0^*$ series of deformations.  However, the $I_0^*$ series requires a separate treatment.   

Recall that the $I_0^*$ Kodaira singularity is actually a one-parameter set of geometries, parameterized by $\t$ taking values in a fundamental domain of the $\SL(2,\Z)$ action on the upper half plane.  This singularity also has $\D(u)=2$, so, by the general discussion of $\cN=2$ deformations of superconformal field theories given in \cite{Argyres:2015ffa}, these theories will have a corresponding exactly marginal deformation parameter, $f$.  As also argued in \cite{Argyres:2015ffa}, by holomorphy, $\t(f)$ can either be a fixed value, or it will have to cover a fundamental $\t$ domain as $f$ varies.  In fact, a closer examination of the above flows shows that they can lead to different values of $\t$ for the $I_0^*$ singularity, so $\t(f)$ cannot be constant.  As a result it must be that the frozen $I_0^*$ singularity includes the whole $\t$ family, and, in particular, the $\t=i\infty$ value for which the CB degrees of freedom are free.

If this weak coupling limit of the CB were due to the whole $I^*_0$ theory becoming free, then it would have to be a lagrangian theory.  Such theories are the familiar $N_f=4$ and $\cN=2^*$ $\SU(2)$ SCFTs (which are $[I^*_0,D_4]$ and $[I_0^*,C_1]$ in our current notation).  These theories, however, have mass deformations, so are not frozen.

It was pointed out in \cite{Argyres:2016yzz} that an alternative lagrangian possibility is a free $\U(1)$ theory (a massless vector multiplet) for which a certain $\Z_2$ symmetry has been gauged.  These, in fact, give an interpretation of the $I_0^*$ series of CB geometries consistent under RG flows.

We can now ask whether there is an alternative consistent intepretation of the frozen $I^*_0$ singularity as a non-lagrangian theory (i.e., not described purely in terms of weakly coupled degrees of freedom)?
Such a frozen non-lagrangian $I_0^*$ singularity but with weakly coupled CB degrees of freedom could arise as an example of an interacting rank-0 SCFT, ``$X_0$", coupled to a vector multiplet through gauging an $\SU(2)$ subgroup of its flavor group.  In order for the singularity to be scale-invariant, the beta function of this $\SU(2)$ must vanish.  In order for it to be frozen, there must be no commutant of the $\SU(2)$ in the rank-0 SCFT's flavor group \cite{Argyres:2007cn,Argyres:2007tq}.

The RG flow test for the $I_0^*$ series is changed by this assumption, because the $I^*_n$ singularities need not all correspond to lagrangian IR free $\SU(2)$ gauge theories, but rather, we must include the new rank-0 SCFT as a new form of interacting ``matter" that IR free $\SU(2)$ gauge fields can couple to.  With this interpretation, the $I_0^*$ series passes the RG flow test, as we will see below.

Note that the same interpretation could be made of the $I_n^*$ series of theories.  E.g., instead of interpreting the frozen $I^*_1$ singularity as the IR free $\SU(2)$ gauge theory with a half-hypermultiplet in the $\bf 4$ representation, we could try to interpret it as a rank-0 CFT, $X_1$, with flavor symmetry $F$, coupled to an $\SU(2)$ vector multiplet which gauges an $\SU(2)\subset F$ such that the coefficient of the beta function of the $\SU(2)$ is 1 and such that there is no commutant of the $\SU(2)$ in $F$.  This new rank-0 SCFT can then also play the role of interacting matter in IR free $\SU(2)$ w/ $X_1 \oplus \ldots$ gauge theories, thus giving rise to new field theory interpretations of the $I^*_n$ singularities. If we interpret the $I_3^*$ arising in \eqref{C2Spl1} in this way we can find a consistent interpretation of the RG-flow (more below). Thus this could provide the sought $I_3^*$ non-langrangian theory mentioned above.

There is at present no direct evidence for the existence of rank-0 $\cN=2$ SCFTs.  It is possible that $\cN=2$ conformal bootstrap methods \cite{Beem:2014zpa,Lemos:2015awa,Lemos:2015orc,Liendo:2015ofa} could conceivably provide such evidence in the future.  Also, it may be possible to adduce indirect evidence in favor of their existence by the following strategy.  

Start by assuming that deformations which fail the RG flow test --- such as the $[II^*,C_2]$ deformation discussed above --- in fact correspond to consistent theories with the assumption of the existence of new rank-0 SCFTs.  One would then interpret some of the singularities they flow to --- such as the $I^*_3$ discussed above --- as weakly-gauged rank-0 SCFTs and deduce some of their properties.  For example, in the case of the $I^*_3$ singularity, we would learn that its rank-0 SCFT has a flavor group $F$ with a maximal subgroup $\SU(2)\oplus\SU(2)'$ and central charge $k_F$ satisfying $I_{\SU(2)\hookrightarrow F} k_F = 10$.  If similar matching arguments for the $\U(2)_R$ symmetry are also possible, then constraints on the $c$ and perhaps $a$ central charges of the rank-0 SCFT may also be deduced.  Presumably many assignments of $(F,k_F,c,a)$ will be consistent with these constraints.  However, by repeating this procedure for many theories, one might find that a small set of rank-0 CFT flavor and central charge assignments suffices to explain many RG flows of higher-rank SCFTs.  For instance, a given rank-0 SCFT might show up in different guises as $I_n^*$ or $I_n$ singularities according to whether different $\SU(2)$ or $\U(1)$ subalgebras, respectively, of its flavor algebra are weakly gauged.  

To carry out such a program more generally would involve examining the RG flows implied by all otherwise consistent deformations of scale-invariant Coulomb branch geometries.  In the rest of this section we analyse the RG flows of the $I_0^*$ and $I_1^*$ series in terms of $\SU(2)$ gauge theories coupled to rank-0 CFTs $X_0$ and $X_1$, respectively.

\subsubsection{Minimal adjoint breaking flows}

\vspace{.2cm}

\noindent\underline{$I^*_0$ series:}

The $II^*\to\{I_0^*,{I_1}^4\}$ generic deformation pattern has as flavor symmetry the exceptional simple algebra $F_4$, and gives rise to minimal adjoint breaking flows 
\begin{align}\label{I0sF4-mab}
F_4 \ 
\begin{cases}
\to C_3                 &\to \{I^*_3,I_1\} \ \cmark\\
\to A_1\oplus A_2  &\to \{I^*_1,I_3\} \ \cmark\\
\to A_2\oplus A_1  &\to \{IV^*,I_2\} \ \cmark\\
\to B_3                 &\to \{III^*,I_1\} \ \cmark
\end{cases}
\end{align}
In the first line, there is no IR free $\SU(2)$ theory with flavor symmetry $C_3$ and consistent with the $I^*_3$ singularity and the $Q=1$ charge normalization forced by the $I_1$ factor, and it would fail the RG test if the $I^*_3$ had to be interpreted as a lagrangian theory.  However, the $I^*_3$ singularity must eventually split to the frozen $I^*_0$ and other singularities.  Thus we should interpret the $I^*_3$ singularity as being an IR free $\SU(2)$ gauge theory coupled to a rank-0 SCFT, $X_0$, as well as to some massless hypermultiplets.  In particular, it can be interpreted as the IR free theory: $\SU(2)$ w/ $X_0\oplus 6\cdot{\bf 3}$.  This has flavor group $C_3$ since, by assumption, the $X_0$ matter contributes no flavor factor.  With charge normalization factor $a=1/2$ as in \eqref{u1su2norm}, the 6 adjoint half-hypermultiplets then contribute $12$ to $b_0$, giving an $I_3^*$ singularity, and contribute light charge $Q=1$ states on the CB, consistent with Dirac quantization and the existence of an $I_1$ singularity.

In the second line in \eqref{I0sF4-mab} it is similarly consistent to identify $I_1^* \simeq \SU(2) \ \text{w/}\ X_0 \oplus 2\cdot{\bf 3}$ having $C_1 \simeq A_1$ symmetry, and $I_3 \simeq \U(1) \ \text{w/}\ 3\cdot1$ having $A_2\oplus U_1$ symmetry.   In the third line it is consistent to identify $IV^* \simeq [IV^*,A_2]$ with $A_2$ symmetry, and $I_2 \simeq \U(1) \ \text{w/}\ 2\cdot1$ having $A_1\oplus U_1$ symmetry.  In the fourth line it is consistent to identify $III^* \simeq [III^*,B_3]$ with $B_3$ symmetry, and $I_1 \simeq \U(1) \ \text{w/}\ 1\cdot1$ having $U_1$ symmetry.  

The $III^*\to\{I_0^*,{I_1}^3\}$ generic deformation pattern has as flavor symmetry $B_3$, and minimal adjoint breaking flows 
\begin{align}\label{}
B_3 \ 
\begin{cases}
\to A_2                 &\to \{IV^*,I_1\} \ \cmark\\
\to A_1\oplus A_1  &\to \{I^*_1,I_2\} \ \cmark\\
\to C_2                 &\to \{I^*_2,I_1\} \ \cmark
\end{cases}
\end{align}
In the first line it is consistent to identify $IV^* \simeq [IV^*,A_2]$ with $A_2$ symmetry, and $I_1 \simeq \U(1) \ \text{w/}\ 1\cdot1$ having $U_1$ symmetry.   In the second  line it is consistent to identify $I_1^* \simeq \SU(2) \ \text{w/}\ X_0\oplus2\cdot{\bf 3}$ having $A_1$ symmetry, and $I_2 \simeq \U(1) \ \text{w/}\ 2\cdot1$ having $A_1\oplus U_1$ symmetry.  In the third line it is consistent to identify $I_2^* \simeq \SU(2) \ \text{w/}\ X_0\oplus4\cdot{\bf 3}$ having $C_2$ symmetry, and $I_1 \simeq \U(1) \ \text{w/}\ 1\cdot1$ having $U_1$ symmetry.  

The $IV^*\to\{I_0^*,{I_1}^2\}$ generic deformation pattern has as flavor symmetry $A_2$, and minimal adjoint breaking flow 
\begin{align}\label{}
A_2 \ 
&\to A_1     \quad \to \{I^*_1,I_1\} \ \cmark
\end{align}
It is consistent to identify $I_1^* \simeq \SU(2) \ \text{w/}\ X_0\oplus2\cdot{\bf 3}$ having $A_1$ symmetry, and $I_1 \simeq \U(1) \ \text{w/}\ 1\cdot1$ having $U_1$ symmetry.  

Thus, at least as far as adjoint breaking flows are concerned, the $I_0^*$ series of theories passes the RG flow test.

\vspace{.2cm}

\noindent\underline{$I^*_1$ series:}

Now we will revisit the flows for the $I^*_1$ series positing the existence of a non-lagrangian frozen $I_1^* \simeq \SU(2) \ \text{w/}\ X_1$ singularity. The $II^*\to\{I_1^*,I_1^3\}$ generic deformation pattern has as flavor symmetry $BC_3$, and the curve gives rise to minimal adjoint breaking flows 
\begin{align}\label{}
BC_3 \ 
\begin{cases}
\to A_2                 &\to \{I^*_1,I_3\} \ \cmark\\
\to A_1\oplus A_1  &\to \{III^*,I_1\} \ \cmark\\
\to C_2                 &\to \{I_3^*,I_1\} \ \cmark
\end{cases}
\end{align}
In the first line we must identify $I_1^* \simeq \SU(2) \ \text{w/}\ X_1$ having no symmetry, and $I_3 \simeq \U(1) \ \text{w/}\ 3\cdot1$ having $A_2\oplus U_1$ symmetry.
In the second line it is consistent to identify $III^* \simeq [III^*,A_1\oplus A_1]$ with $A_1\oplus A_1$ symmetry, and $I_1 \simeq \U(1) \ \text{w/}\ 1\cdot1$ having $U_1$ symmetry.   
In the third line it is consistent to identify $I_3^* \simeq \SU(2) \ \text{w/}\ X_1\oplus4\cdot{\bf 3}$ having $C_2$ symmetry, and $I_1 \simeq \U(1) \ \text{w/}\ 1\cdot1$ having $U_1$ symmetry.  

The $III^*\to\{I_1^*,I_1^2\}$ generic deformation pattern has as flavor symmetry $A_1\oplus A_1$, and minimal adjoint breaking flows 
\begin{align}\label{}
A_1\oplus A_1 \ 
\begin{cases}
\to A_1\oplus U_1  &\to \{I_1^*,I_2\} \ \cmark\\
\to U_1\oplus A_1  &\to \{I^*_2,I_1\} \ \cmark
\end{cases}
\end{align}
In the first line it is consistent to identify $I_1^* \simeq \SU(2) \ \text{w/}\ X_1$ having no symmetry, and $I_2 \simeq \U(1) \ \text{w/}\ 2\cdot1$ having $A_1\oplus U_1$ symmetry.
In the second line it is consistent to identify $I_2^* \simeq \SU(2) \ \text{w/}\ X_1\oplus2\cdot{\bf 3}$ having $A_1$ symmetry, and $I_1 \simeq \U(1) \ \text{w/}\ 1\cdot1$ having $U_1$ symmetry.  

The $IV^*\to\{I_1^*,I_1\}$ generic deformation pattern has as its flavor symmetry $U_1$, and no adjoint breaking flows.  In fact its only flow is precisely the generic deformation pattern, so there is no nontrivial RG flow test for this theory.

Thus, at least as far as adjoint breaking flows are concerned, the $I_1^*$ series of theories passes the RG flow test if $I^*_1$ is interpreted as $\SU(2)$ w/ $X_1$ instead of as $\SU(2)$ w/ $1\cdot{\bf 4}$.  

\subsubsection{Non-adjoint breaking flows}

We again only check the non-adjoint special flows for the theories with rank 2 flavor groups.

\vspace{.2cm}

\noindent\underline{$[IV^*,A_2]$ geometry:}

The only non-adjoint special flow for this theory is
\begin{align}\label{IVsA2-RG}
[IV^*, A_2] \ 
\to \{ I_0^*, II\} \ \text{for}\ m_1=e^{2\pi i/3}\,m_2. 
\quad \cmark
\end{align}
In the first line, the $I_0^*$ singularity must be identified with the frozen $I_0^*$ CFT of the generic deformation, so contributes no flavor symmetry, while the $II$ can only be the $[II,-]$ CFT, which also has no flavor symmetry.  This is consistent with the $U_1\oplus U_1$ flavor symmetry not acting on the light states on the CB, and can be checked as in the last example from the curve and one-form.

\vspace{.2cm}

\noindent\underline{$[III^*,A_1\oplus A_1]$ geometry:}

The non-adjoint special RG flows for this theory are:
\begin{align}\label{}
[III^*, A_1\oplus A_1] \ 
\begin{cases}
\to \{ I_1^*, II\} \ \text{for}\ m_1=\sqrt{8/3}\,m_2, &\cmark\\
\to \{ IV^*, I_1  \} \ \text{for}\ m_1= i/\sqrt{3}\,m_2. & \cmark
\end{cases}
\end{align}
In the first line, the $I_1^*$ singularity must correspond to the frozen IR free $\SU(2)$ w/ $X_1$ theory and no flavor symmetry, while the $II$ is the $[II,-]$ CFT.  This is consistent with the $U_1\oplus U_1$ flavor symmetry not acting on the light states on the CB, and can be checked as in the last example from the curve and one-form.  In the second line, the $IV^*$ singularity must be identified with the frozen $[IV^*,U_1]$ CFT since this is the only $IV^*$ that deforms to a terminal $I_1^*$.  The $I_1$ is the $\U(1)$ gauge theory with 1 charge-1 hypermultiplet, with flavor symmetry $U_1$.  Thus this flow has the expected $U_1\oplus U_1$ symmetry.

Thus the $I_0^*$ and $I_1^*$ series interpreted in terms of gauged rank-0 CFTs pass all RG flow tests we have checked.

\section{Summary and open questions\label{sec:end}}

This is the second paper in a series of three.  In the first we developed a strategy for classifying physical rank-1 CB geometries of $\cN=2$ SCFTs.  In this paper we have shown how to carry out this strategy computationally.  We find that each geometry can be uniquely labeled (with one exception) by its ``deformation pattern" which simply lists the set of Kodaira singularities the given initial scale invariant singularity splits into under generic deformation.  The full list is reported in table \ref{table:theories}.   

We developed a method for explicitly constructing the SW curves of the submaximal deformations using the known curves for the maximal deformations.   We also constructed SW one-forms for each curve each by slightly generalizing a technique presented in \cite{Minahan:1996cj,Minahan:1996fg}.  The explicit curves and one-forms determine the flavor symmetry for each entry in table \ref{table:theories}.

We discussed at length physical consistency of the geometries under arbitrary RG flows.  This is an extra, quite restrictive, condition that each geometry must satisfy.  The geometries which pass it do so in intricate and non-trivial ways, while those that do not are shown in blue in table \ref{table:theories}.

Our analysis and results highlighted some questions specific to the construction of special K\"ahler geometries:
\begin{itemize}
\item {\bf Uniqueness of curves with a given deformation pattern.} 
It is not obvious that the pattern of a deformation uniquely characterizes the monodromies of the deformation, let alone the analytic form of the curve.  
There are three obvious braid and $\SL(2,\Z)$ invariants of a set of monodromies $\{K_a\}$:
\begin{itemize}
\item the deformation pattern,
\item the $\SL(2,\Z)$ conjugacy class $[K_0]$ of the total monodromy,
\item and the number $\ell := \gcd\{\vev{\bz_a,\bz_b}, \forall a,b\}$.
\end{itemize}
The ``asymptotic charge invariant" $\ell$ was introduced in \cite{DeWolfe:1998eu}, who also conjectured that the last two invariants uniquely characterize the orbit of the monodromy set $\{K_a\}$ under the combined action of the braid group and overall $\SL(2,\Z)$ conjugation.\footnote{In \cite{DeWolfe:1998eu} only maximal deformation patterns were considered.}  In this paper we have ignored the asymptotic charge invariant. While we have no proof that $\ell$ can be neglected in general, we have found no examples of deformations of Kodaira singularities belonging to the same monodromy class [$K_0$] with the same deformation patterns but different $\ell$.

Furthermore, we find only a single curve up to analytic equivalence for each deformation pattern although we have no proof that this needs to be the case.  In many cases we performed lengthy direct searches for curves with appropriate discriminant factorization patterns, and in each case only found a single solution.

\item {\bf Theory of the SW one form.} 
Our strategy for showing the existence of a SW one-form using the MN ansatz was highly over-constrained, but in each case yielded solutions. Furthermore, in most cases we find a multi-parameter family of solutions which are nevertheless physically equivalent.  This suggests we are missing a more efficient global and possibly geometric argument for the existence of the one form. 

\end{itemize}
More general questions about the space of $\cN=2$ SCFTs also raised by our analysis were discussed in the first paper in this series \cite{Argyres:2015ffa}.   The third paper \cite{Argyres:2016ccharges} will focus on how to extract other $\cN=2$ SCFT data, such as certain Higgs branch dimensions, and conformal and flavor central charges, from their CB geometries.

\acknowledgments

It is a pleasure to thank C. Beem, M. Del Zotto, T. Dumitrescu, P. Esposito, S. Gukov, K. Intriligator, C. Long, D. Morrison, L. Rastelli, V. Schomerus, N. Seiberg, A. Shapere, Y. Tachikawa, R. Wijewardhana, J. Wittig, and D. Xie for helpful comments and discussions. This work was supported in part by DOE grant DE-SC0011784.  MM was also partially supported by NSF grant PHY-1151392.

\appendix
\section{Curves and one forms\label{app:curves}}

In this section we report the curves and the one-forms of all the deformations listed in table \ref{table:theories} except for those of the maximal deformations of the $I^*_0$, $II^*$, $III^*$ and $IV^*$ singularities which can be found in \cite{Seiberg:1994aj,Minahan:1996cj,Minahan:1996fg}. 

Recall that in section \ref{section:one-form} we introduced $(x_\bw, y_\bw)$ which are proportional to the coordinates $(x,y)$ of the one-form pole, and are defined by
\begin{align}\label{}
x_\bw &:=\bw(\bm)^2\, x,&
y_\bw &:=\bw(\bm)^3\, y ,
\end{align}
where $r\, \bw(\bm)$ is the residue associated to the pole and depends linearly on the $\bm$.  We find explicit expressions for the one-form by solving for pole positions $(x_\bw, y_\bw)$ depending polynomially on the linear masses $\bm$ and vev $u$.  We restrict ourselves to searching for $x$ pole positions at most quadratic in $u$.   If there is more than one solution for the pole position, we label them by subscripts on their residues, as $r_i \bw_i$.   The coefficients $r_i$ can be reabsorbed into the definition of $\bw_i$, as discussed in section \ref{section:one-form}, but it is more convenient to keep them explicit in cases with multiple pole position solutions.

\subsection{Deformations of the $II^*$ singularity\label{appA1}}

The undeformed $II^*$ singularity is given in table \ref{Table:Kodaira}.  It is easy to see that a linear $u$-dependence for $x$ cannot work for the $II^*$ singularity: if the pole location $x$ depended on $u$ linearly, we would need to find a solution for $y^2=P(u)$ with $P(u)$ a fifth order polynomial in $u$, with $y$ polynomial in $u$, which is not possible.  Throughout this subsection we thus consider only poles of the form
\begin{align}\label{}
x(u,\bm)&=\bw(\bm)^{-2}\,  \left(u^2+Ru+S\right),& 
y(u,\bm)&=\bw(\bm)^{-3}\, \left(u^3+Ju^2+Ku+L\right),
\end{align}
where $\bw(\bm)$, $R$, $S$, $J$, $K$ and $L$ are polynomials of the linear masses we need to solve for.  Their mass dimensions are fixed by the $u$ mass dimension: $\Delta(u)=6$ implies $\Delta(R)=6$, $\Delta(S)=12$, $\Delta(J)=6$, $\Delta(K)=12$, $\Delta(L)=18$, and $\Delta(\bw(\bm))=1$.

In the $II^*$ case $\Delta(u)>2$ so there is no $\mu$ parameter in the MN-ansatz (\ref{MNform}).

\subsubsection{$\{ I_1^6,I_4\}$ with $\Sp(10)$ flavor symmetry}

The $\{I_1^6,I_4\}$ deformation of the $II^*$ singularity has curve
\begin{align}\label{C5sigM}
y^2 &= x^3+3 x \bigl[2 u^3 M_2+u^2 \left(M_4^2-2 M_8\right)+2 u M_4 M_{10}-M_{10}^2\bigr]
\nonumber\\
&\qquad \ \text{}+2 \bigl[u^5+u^4 M_6
+u^3 \left(2 M_4^3-3 M_4 M_8-3 M_2 M_{10}\right)
\\
&\qquad\qquad\text{}+3 u^2 M_8 M_{10}
-3 u M_4 M_{10}^2+M_{10}^3\bigr].
\nonumber
\end{align}
Its spectrum of dimensions of mass invariants implies it has a discrete $\text{Weyl}(B_5)\simeq\text{Weyl}(C_5)$ group of symmetries acting on the linear masses $\bm$.  

Choose a basis, $\bm=m_i \be^i$, of the linear masses so that Weyl$(BC_5) \simeq S_5 \ltimes \Z_2^5$ acts by permutations and independent sign flips of the five $m_i$, and define a standard basis of Weyl$(BC_5)$ invariant polynomials by
\begin{align}\label{BCinv}
N_{2k} := \sum_{i_1<\cdots<i_k} 
m_{i_1}^2\cdots m_{i_k}^2 
\end{align}
for $k=1,\ldots,5$.  Then either by restriction from the maximal deformation of the $II^*$ singularity (described in section \ref{s3.2}), or by solving the factorization condition for the MN ansatz for the SW one-form, or by demanding increased zero multiplicities of the curve's discriminant when $\ba(\bm)=0$ for $\ba$'s fixed by the Weyl group, we find the same dependence, $M_d(\bm)$, of the invariant masses appearing in the curve \eqref{C5sigM} on the linear masses:
\begin{align}\label{C5MtoN}
M_{10} &= - 2592 N_{10},
\nonumber\\
M_8 &= \frac{45}{8} N_2^4 - 45 N_2^2 N_4 
+ 90 N_4^2 + 216 N_8,
\nonumber\\
M_6 &= -\frac{9}{2} N_2^3 + 18 N_2 N_4 - 108 N_6,
\\
M_4 &= 3 N_2^2 - 12 N_4,
\nonumber\\
M_2 &= N_2.
\nonumber
\end{align}

We solve for the one-form using the MN ansastz \eqref{MNform}. We tried two choices\footnote{For all other deformations considered in this paper, the algebra is simple enough that we can take a general linear ansatz for $\bw(\bm)$ and solve for it simultaneously with the curve factorization, and so do not need to make additional choices restricting the possible residues.} for $\bw(\bm)$, 
\begin{align}\label{BC5wt}
\bw(\bm) = m_1 + m_2,
\end{align}
and $\bw(\bm) = m_1$.  
(Note that the normalizations of these $\bw$'s are arbitrary in the sense that they can be absorbed in a rescaling of the coefficient $r$ in the ansatz for $x_\bw$.)  
There is no solution to the curve factorization condition for the second choice.  A single solution for the first choice \eqref{BC5wt} is found by parameterizing the possible mass polynomials appearing in $x_\bw$ and $y_\bw$ in terms of a basis of invariant polynomials for the stabilizer subgroup of $\bw(\bm)$ in Weyl$(BC_5)$.   This subgroup is $\Z_2\times \text{Weyl}(BC_3)$ where the $\Z_2$ interchanges $m_1$ and $m_2$, while Weyl$(BC_3)$ acts on $m_{3,4,5}$ in the usual way.  So define a basis of its invariant polynomials by
\begin{align}\label{Z2BC3invts}
S_1 &:= m_1 + m_2, &
S_2 &:= m_1 m_2, &
T_{2k} &:=\!\!\! \sum_{{i_1 < \cdots < i_k}\atop{3\le i_j\le 5}}
\!\!\! m^2_{i_1} \cdots m_{i_k}^2, 
\qquad k=\{1,2,3\}.
\end{align}
Then the solution for the pole position $(x_\bw,y_\bw)$ for which the curve is a perfect square is
\begin{align}\label{C5pole}
(i \sqrt{6})^2 x_\bw &= u^2 
+ 18 u \bigl[ S_1^6 +S_1^4 \left(5 S_2-2 T_2\right)
-S_1^2 \left(2 S_2 T_2-T_2^2+4 T_4\right)
-3 S_2 \left(T_2^2-4 T_4\right)
\bigr]
\nonumber\\
&\qquad\text{} + 243 S_2^2 \bigl[
3 S_1^8-12 S_1^6 T_2
+6 S_1^4 \left(3 T_2^2-4 T_4\right)
\nonumber\\
&\qquad\qquad\qquad\text{}
-4 S_1^2 \left(3 T_2^3-12 T_2 T_4+32 T_6\right)
+3 \left(T_2^2-4 T_4\right)^2
\bigr],
\nonumber\\
(i \sqrt{6})^3 y_\bw &=
u^3-27 u^2 \left[3 S_1^6+S_1^4 \left(3 S_2-2 T_2\right)+3 S_2 \left(T_2^2-4 T_4\right)+S_1^2 \left(2 S_2 T_2-T_2^2+4 T_4\right)\right]
\nonumber\\
&\qquad\text{}-729 u S_2 \biggl[S_1^{10}+S_1^8 \left(5 S_2-4 T_2\right)-2 S_1^6 \left(6 S_2 T_2-3 T_2^2+4 T_4\right)
\nonumber\\
&\qquad\qquad\qquad\text{}+2 S_1^4 \left(S_2 \left[3 T_2^2-4 T_4\right]-2 \left[T_2^3-4 T_2 T_4+16 T_6\right]\right)\nonumber\\
&\qquad\qquad\qquad\text{}+S_1^2 \bigl(\left[T_2^2-4 T_4\right]^2-3 S_2 \left[T_2^2-4 T_4\right]^2+4 S_2 \left[T_2^3-4 T_2 T_4+16 T_6\right]\bigr)\biggr]
\nonumber\\
& \qquad \text{}-19683 S_2^3 \left[S_1^4-2 S_1^2 T_2+T_2^2-4 T_4\right] 
\bigl[S_1^8-4 S_1^6 T_2+S_1^4 \left(6 T_2^2-8 T_4\right)
\nonumber\\
&\qquad\qquad\qquad\text{}+\left(T_2^2-4 T_4\right)^2-4 S_1^2 \left(T_2^3-4 T_2 T_4+16 T_6\right)\bigr] .
\end{align}

Weyl invariance of the curve implies that it factorizes for all $\bw$'s in the orbit of \eqref{BC5wt}.  Summing over these 40 poles in the $x$-plane in the MN ansatz \eqref{MNform} for the SW 1-form and imposing the differential condition \eqref{SWform} (over-)determines the remaining parameters of the 1-form to be
\begin{align}\label{C5-1form}
a &= \frac{7}{9} i \sqrt{6}, &
c &= -\frac{10}{9} i \sqrt{6}, &
W &= 36 i \sqrt{6} \left(N_2 N_4 - 6 N_6\right),
\end{align}
where we have fixed the overall normalization by choosing the 1-form normalization constant, $\k:=2a-c = \frac{8}{3} i \sqrt{6}$.  

The residues at the poles, $\bw(\bm)$, are
\begin{align}\label{BC5-40orbit}
\bw(\bm) &= \pm m_i \pm m_j, &
\text{for} \quad & 1 \le i\neq j \le 5. 
\end{align}
The integral span of \eqref{BC5-40orbit} is the root lattice of $C_5$, and is not the $B_5$ root lattice.  Thus we conclude that \eqref{C5sigM} describes a CFT with $C_5 \simeq \Sp(10)$ symmetry group. This conclusion agrees with the identification of the flavor symmetry of the rank-5 deformation of the $II^*$ singularity based on S-dualities \cite{Argyres:2007tq}.

\subsubsection{$\{ I_1^2,I_4^2\}$ with $\Sp(4)$ flavor symmetry}

The $\{I_1^2,I_4^2\}$ deformation of the $II^*$ singularity has curve
\begin{align}\label{C2sigM}
y^2 &= x^3+x \bigl[u^3 L_2 + u^2 L_8 + u L_{14} + L_{20}\bigr]
+\bigl[ 2 u^5 + u^3 L_{12} + u^2 L_{18}
+ u L_{24} + L_{30}\bigr]
\end{align}
where
\begin{align}
L_{30} &= 
-\frac{1024}{2278125} M_2^{15}
+\frac{8192}{151875} M_2^{13} M_4
-\frac{38912}{50625} M_2^{11} M_4^2
+\frac{2601472}{455625} M_2^9 M_4^3
-\frac{3891968}{151875} M_2^7 M_4^4
\nonumber\\
&\qquad\text{}
+\frac{6143488}{84375} M_2^5 M_4^5
-\frac{2227712}{18225} M_2^3 M_4^6
+\frac{896}{9} M_2 M_4^7,
\nonumber\\
L_{24} &= 
-\frac{256}{10125} M_2^{12}
-\frac{1024}{10125} M_2^{10} M_4
+\frac{24704}{10125} M_2^8 M_4^2
-\frac{172544}{10125} M_2^6 M_4^3
+\frac{568384}{10125} M_2^4 M_4^4
\nonumber\\
&\qquad\text{}
-\frac{40064}{405} M_2^2 M_4^5
+\frac{160}{3} M_4^6,
\nonumber\\
L_{20} &= 
\frac{64}{3375} M_2^{10}
-\frac{2048}{3375} M_2^8 M_4
+\frac{55042}{1125} M_2^6 M_4^2
-\frac{64576}{3375} M_2^4 M_4^3
+\frac{5072}{135} M_2^2 M_4^4
-32 M_4^5,
\nonumber\\
L_{18} &= 
\frac{64}{225} M_2^9
-\frac{1024}{675} M_2^7 M_4
+\frac{1472}{225} M_2^5 M_4^2
-\frac{2656}{225} M_2^3 M_4^3
+\frac{544}{27} M_2 M_4^4,
\nonumber\\
L_{14} &= 
\frac{32}{225} M_2^7
+\frac{128}{75} M_2^5 M_4
-\frac{272}{25} M_2^3 M_4^2
+\frac{208}{9} M_2 M_4^3,
\nonumber\\
L_{12} &= 
-\frac{112}{135} M_2^6
+\frac{64}{15} M_2^4 M_4
-\frac{824}{45} M_2^2 M_4^2
+\frac{560}{27} M_4^3,
\nonumber\\
L_8 &= -\frac{28}{15} M_2^4
+\frac{32}{15} M_2^2 M_4
-\frac{20 M_4^2}{3},
\nonumber\\
L_2 &= 4 M_2.
\end{align}
Its spectrum of dimensions of mass invariants implies it has a discrete Weyl($C_2$) group of symmetries acting on the linear masses $\bm$.  

Choose a basis, $\bm=m_i \be^i$, of the linear masses so that Weyl$(C_2) \simeq S_2 \ltimes \Z_2^2$ acts by permutations and independent sign flips of the two $m_i$, and define a standard basis of Weyl$(C_2)$ invariant polynomials 
\begin{align}\label{C2Ndef}
N_2 &:= m_1^2+m_2^2, &
N_4 &:= m_1^2 m_2^2.
\end{align}
Then either by solving the factorization condition for the MN ansatz for the SW one-form, or by demanding increased zero multiplicities of the curve's discriminant when $\ba(\bm)=0$ for $\ba$'s fixed by the Weyl group, we find the same dependence, $M_d(\bm)$, of the invariant masses appearing in the curve \eqref{C2sigM} on the linear masses:
\begin{align}\label{C2MtoN}
M_4 &= N_4, &
M_2 &= N_2.
\end{align}

We find three solutions for pole positions for which the curve factorizes:
\begin{align}\label{C2pole}
r_1\,\bw_1(\bm) &:= r_1\,m_1, 
\nonumber\\
(15i)^2 x_{\bw_1} 
&= 225 u^2
+u \left(30 m_1^6-180 m_1^4 m_2^2-180 m_1^2 m_2^4-120 m_2^6\right)
-24 m_1^{12}+88 m_1^{10} m_2^2
\nonumber\\
&\qquad\text{}
-276 m_1^8 m_2^4+464 m_1^6 m_2^6-16 m_1^4 m_2^8+48 m_1^2 m_2^{10}+16 m_2^{12} ,
\nonumber\\
(15i)^3 y_{\bw_1} 
&= \left(15 u-4 m_1^6-26 m_1^4 m_2^2
+4 m_1^2 m_2^4-4 m_2^6\right)^2
\nonumber\\
&\qquad\text{}
\cdot \left(15 u-4 m_1^6+4 m_1^4 m_2^2
-26 m_1^2 m_2^4-4 m_2^6\right);
\nonumber\\[2mm]
r_2\, \bw_2(\bm) &:= r_2\, m_1,
\nonumber\\
(30i)^2 x_{\bw_2} 
&= 225 u^2
+u \left(-1320 m_1^6+2520 m_1^4 m_2^2
-180 m_1^2 m_2^4-120 m_2^6\right)
+336 m_1^{12}
\nonumber\\
&\qquad\text{}
-992 m_1^{10} m_2^2+2784 m_1^8 m_2^4-3856 m_1^6 m_2^6+164 m_1^4 m_2^8
+48 m_1^2 m_2^{10}+16 m_2^{12},
\nonumber\\
(30i)^3 y_{\bw_2} 
&= \left(15 u-4 m_1^6+4 m_1^4 m_2^2
-26 m_1^2 m_2^4-4 m_2^6\right)^2
\\
&\qquad\text{}
\cdot\left(15 u+356 m_1^6-236 m_1^4 m_2^2+34 m_1^2 m_2^4-4 m_2^6\right) ;
\nonumber\\[2mm]
r_3\,\bw_3(\bm) &:= r_3\,(m_1 + m_2),
\nonumber\\
(15i)^2 x_{\bw_3} 
&= 225 u^2
+u \bigl(30 m_1^6+300 m_1^5 m_2-30 m_1^4 m_2^2
+300 m_1^3 m_2^3-30 m_1^2 m_2^4+300 m_1 m_2^5
\nonumber\\
&\qquad\text{}
+30 m_2^6\bigr)
-24 m_1^{12}-80 m_1^{11} m_2+48 m_1^{10} m_2^2-276 m_1^8 m_2^4-520 m_1^7 m_2^5 
\nonumber\\
&\qquad\text{}
+204 m_1^6 m_2^6-520 m_1^5 m_2^7-276 m_1^4 m_2^8+48 m_1^2 m_2^{10}-80 m_1 m_2^{11}-24 m_2^{12} ,
\nonumber\\
(15i)^3 y_{\bw_3} 
&= \left(15 u-4 m_1^6+4 m_1^4 m_2^2
-26 m_1^2 m_2^4-4 m_2^6\right) 
\nonumber\\
&\qquad\text{}
\cdot\left(15 u-4 m_1^6-26 m_1^4 m_2^2
+4 m_1^2 m_2^4-4 m_2^6\right)
\nonumber\\
&\qquad\text{}
\cdot \left(15 u-4 m_1^6+34 m_1^4 m_2^2+90 m_1^3 m_2^3+34 m_1^2 m_2^4-4 m_2^6\right).
\nonumber
\end{align}

Summing over the Weyl$(C_2)$ orbits of these poles in the $x$-plane in the MN ansatz \eqref{MNform} for the SW 1-form, imposing the differential condition \eqref{SWform} and choosing the normalization $2a-c=i/2$ gives a two-parameter family of solutions:
\begin{align}
a &=\frac{i}{12}\left(r_1+4r_2+2r_3\right),
\qquad \quad
c =\frac{i}{3}\left(2r_1-r_2-2r_3\right),&
1 &= -r_1 + 2r_2+2r_3,
\nonumber\\
W=&-\frac{i}{15}N_2 \left\{
\left(2r_1+8r_2+4r_3\right) N_2^2
-\left(8r_1+17r_2+16r_3\right)N_4 \right\}.
\end{align}

\subsubsection{$\{ I_1^4,I_0^*\}$ with $F_4$ flavor symmetry\label{appA1.3}}

The $\{I_1^4,I_0^*\}$ deformation of the $II^*$ singularity has curve
\begin{equation}
y^2=x^3+3 x[2u^3 M_2+u^2 M_8]+ 2[u^5+u^4 M_6+u^3 M_{12}]
\end{equation}
The spectrum of dimensions of the masses is $\{2,6,8,12\}$ which implies that the curve has a discrete Weyl($F_4$) group of symmetries acting on the linear masses $\bm$.

To construct a standard basis of Weyl($F_4$) invariant polynomials, first choose a basis, $\bm=m_i \be^i$, of the linear masses so that Weyl($BC_4$)$\simeq S_4 \ltimes \Z_2^4$ acts by permutations and independent sign flips of the four $m_i$, and define the standard basis of Weyl($BC_4$) invariant polynomials:
\begin{equation}\label{GenBC4}
P_{2k}:=\sum_{i_1<..<i_k} m_{i_1}^2...m_{i_k}^2
\end{equation}
for $k=1,...,4$.  Now exploit the fact that Weyl($F_4$) is Weyl($BC_4$) plus two extra generators $\{\mathcal{O}_1,\mathcal{O}_2\}$ which act on the $m_i$ as
\begin{equation}\label{extraF4}
\mathcal{O}_1:=\left\{
\begin{array}{l}
m_1\to\frac{1}{2}(\phantom{-}m_1-m_2-m_3-m_4)\\
m_2\to\frac{1}{2}(-m_1+m_2-m_3-m_4)\\
m_3\to\frac{1}{2}(-m_1-m_2+m_3-m_4)\\
m_4\to\frac{1}{2}(-m_1-m_2-m_3+m_4)
\end{array}\right.,\quad
\mathcal{O}_2:=\left\{
\begin{array}{l}
m_1\to\frac{1}{2}(m_1+m_2+m_3+m_4)\\
m_2\to\frac{1}{2}(m_1+m_2-m_3-m_4)\\
m_3\to\frac{1}{2}(m_1-m_2+m_3-m_4)\\
m_4\to\frac{1}{2}(m_1-m_2-m_3+m_4)
\end{array}\right..
\end{equation}
A basis of Weyl($F_4$) invariant polynomials is then independent combinations of the $P_{2k}$'s in (\ref{GenBC4}) which are invariant under the (\ref{extraF4}) action.  One such basis is
\begin{align}
N_2&=P_2,& 
N_8&=12 P_8-3 P_2P_6+P_4^2,\\
N_6&=6 P_6-P_2P_4,& 
N_{12}&=288P_4P_8-27P_2^3P_6
-8P_4^3+12P_2^2P_4^2 .
\nonumber
\end{align}

Then either by solving the factorization condition for the MN ansatz for the the SW one-form, or by demanding increased zero multiplicities of the curve's discriminant when $\ba(\bm)=0$ for $\ba$'s fixed by the Weyl group, we find the following dependence of the $M_d$ on the linear masses:
\begin{align}
M_2&=N_2,& 
M_8&=-36 N_8,
\\
M_6&=-18N_6,& 
M_{12}&=-81 N_6^2-243 N_2^2N_8+27N_{12}.
\nonumber
\end{align}

We find two solutions for pole positions for which the curve factorizes:
\begin{align}\label{F4pole}
r_1\,\bw_1(\bm) &:=r_1\,(m_1+m_2), 
\nonumber\\
(i\sqrt{3})^2 x_{\bw_1} 
&= u^2+
18u (m_1+m_2)^2\left(2m_1^2 m_2^2 -(m_1^2+m_2^2)(m_3^2+m_4^2)+2 m_3^2m_4^2\right),
\nonumber\\
(i\sqrt{3})^3 y_{\bw_1} 
&= u^3
-54u^2 (m_1+m_2)^2 \left(m_1m_2(m_1^2+m_2^2+m_1m_2-m_3^2-m_4^2)-m_3^2m_4^2\right),
\nonumber\\[2mm]
r_2\,\bw_2(\bm) &:= r_2\, (2 m_1),
\\
(i\sqrt{3})^2 x_{\bw_2} &=u^2
+36 u (9 m_1^6 - 5 m_1^4 S_2 + m_1^2 S_4 + 3 S_6)
+2916 T_6^2,
\nonumber\\
(i\sqrt{3})^3 y_{\bw_2} &= u^3
-54 u^2 (15 m_1^6 - 3 m_1^4 S_2 - m_1^2 S_4 - 3 S_6)
\nonumber\\
&\qquad\ \mbox{}-2916 u (9 m_1^6 - 5 m_1^4 S_2 + m_1^2 S_4 + 3 S_6) T_6
-157464 T_6^3,
\nonumber
\end{align}
where
\begin{align}\label{}
S_{2k} &:= \sum_{2\le j_1<..<j_k\le 4} m_{j_1}^2...m_{j_k}^2, &
T_6 &:= \prod_{2\le j \le4}(m_1^2-m_j^2).
\end{align}

Summing over the Weyl$(F_4)$ orbits of these poles in the $x$-plane in the MN ansatz \eqref{MNform} for the SW 1-form, imposing the differential condition \eqref{SWform} and choosing the normalization $2a-c=-i 4\sqrt3$ gives a unique solution for the SW one-form:
\begin{align}\label{}
a &=0, &
c &=4i\sqrt{3}, &
W &= 18i\sqrt{3}(4 N_6+N_2^3), &
r_1 &= -4, &
r_2 &= 1.
\end{align}
Note that the residues fill out the $F_4$ root lattice.

\subsubsection{$\{ I_1^3,I_1^*\}$ with $\Sp(6)$ or $\SO(7)$ flavor symmetry\label{appA1.4}}

The $\{I_1^3,I_1^*\}$ deformation of the $II^*$ singularity has curve
\begin{align}\label{BC3curve}
y^2 = x^3 + 3 x u^2 (2 u M_2 - M_4^2) + 2 u^3 (u^2 + M_4^3 + u M_6).
\end{align}
From the spectrum of the $M$'s, $\{2,4,6\}$, we infer that the curve is invariant under the reflection group $S_3 \ltimes \Z_2^3 \simeq \text{Weyl}(B_3)\simeq\text{Weyl}(C_3)$ and thus the curve corresponds to a conformal field theory with flavor symmetry group either $B_3\simeq\SO(7)$ or $C_3\simeq\Sp(6)$.  Introducing a basis of homogeneous Weyl($BC_3$) invariant polynomials,
\begin{align}
N_2 = m_1^2 + m_2^2 + m_3^2, \qquad
N_4 = m_1^2 m_2^2 + m_2^2 m_3^2 + m_1^2 m_3^2, \qquad 
N_6 = m_1^2 m_2^2 m_3^2,
\end{align}
we find that
\begin{align}
M_2 = N_2,\qquad
M_4 = \frac{3}{2}N_2^2 - 6 N_4,\qquad 
M_6 =  -\frac{9}{2}N_2^3 + 18 N_2 N_4 - 108 N_6.
\end{align}

The following pole positions solve the curve factorization problem: 
\begin{align}\label{BC3root}
r_1\,\bw_1(m)&:=r_1\, m_1 ,
\nonumber\\
\left(i\sqrt{6}\right)^2\, x_{\bw_1}&=
u^2
+18 u\ m_1^2 \Big(m_1^4
-2m_1^2(m_2^2+m_3^2)+(m_2^2-m_3^2)^2\Big) ,
\nonumber\\
\left(i\sqrt{6}\right)^3\, y_{\bw_1}&=
u^3
-27 u^2 m_1^2 \Big(3m_1^4
+2m_1^2(m_2^2+m_3^2)-(m_2^2-m_3^2)^2\Big) ,
\nonumber\\[2mm]
r_2\,\bw_2(m)&:=r_2\,(m_1+m_2+m_3) ,
\nonumber\\
\left(i\sqrt{3/2}\right)^2\,x_{\bw_2}&=
u^2
-\frac{9}{4}u(m_1+m_2+m_3)^2
\Big(m_1^4 -2m_1^2(m_2^2+m_3^2)
+(m_2^2-m_3^2)^2\Big) ,
\nonumber\\
\left(i\sqrt{3/2}\right)^3\,y_{\bw_2}&=
u^3
-27 u^2 (m_1+m_2+m_3)^3 m_1m_2m_3 ,
\nonumber\\[2mm]
r_3\,\bw_3(m)&:=r_3\,(m_1+m_2),
\\
\left(i\sqrt{6}\right)^2\,x_{\bw_3}&=
u^2
+18 u (S_2-m_3^2)
\Big(
(S_2-2m_1m_2) (S_2+5m_1m_2)
- (S_2-3m_1m_2) m_3^2
\Big)
\nonumber\\
&\qquad\ \mbox{}+729m_1^2m_2^2 (S_2-m_3^2)^4,
\nonumber\\
\left(i\sqrt{6}\right)^3\,y_{\bw_3}&=
u^3
-27 u^2 \Big( 
3 S_2^2 (S_2+m_1m_2)
-2 S_2 (S_2-m_1m_2) m_3^2 
-(S_2-3m_1m_2) m_3^4 
\Big)
\nonumber\\
&\qquad\ \mbox{}-729 u m_1m_2
(S_2-m_3^2)^3
\Big(S_2 (S_2+5m_1m_2)
-(S_2-3m_1m_2) m_3^2 \Big)
\nonumber\\
&\qquad\ \mbox{}-19683m_1^3m_2^3
(S_2-m_3^2)^6,
\nonumber
\end{align}
where $S_2:=(m_1+m_2)^2$. 

After solving the differential equation, we obtain a two-parameter family of solutions for the SW one-form:
\begin{align}\label{}
a&=\frac{i}{6\sqrt{6}} (-3r_1+4r_2+18r_3),&
c&=\frac{i}{3\sqrt{6}}(3r_1-8r_2-6r_3),
\nonumber\\
W&=-i36\sqrt{6}(N_4N_2+6 r_3 N_6),&
1&=r_1-2r_2-4r_3,
\end{align}
where we imposed the 1-form normalization $2a-c=-i\sqrt{2/3}$. 

Choosing particular values of the $r_i$'s, we find solutions of the residues which generate both the $C_3$ and $B_3$ root lattices.  For example, for $(r_1,r_2,r_3)=(1,-1,1/2)$ the residues in \eqref{BC3root} generate a face-centered cubic lattice ($C_3$ root lattice), while for $(r_1,r_2,r_3)=(1,0,0)$  a simple cubic lattice ($B_3$ root lattice).  Thus in this case, there are two distinct physical CB geometries, one corresponding to flavor symmetry $B_3$ and the other to $C_3$.

\subsubsection{$\{ I_3,I_1^*\}$, $\{{I_1}^2,I_2^*\}$, and $\{I_1,I_3^*\}$}

In this section we simply list the SW curves (in terms of the linear mass paramters) for the above three geometries without giving their SW one-forms.  The reason is that these one-forms can all be easily constructed by restriction of masses of the $\{{I_1}^4,I_0^*\}$ one-form given in section \ref{appA1.3} above, and their (lengthy) expressions do not illuminate any interesting points since all the resulting flavor symmetries are of low rank.

\paragraph{The $\{I_3,I_1^*\}$ curve.}

The $II^* \to \{ I_3,I_1^*\}$ deformation is described by the curve
\begin{align}\label{}
y^2& = x^3 + 12 u^2 x M_2 (u - 9 M_2^3) - 
 2 u^3 (u^2 - 36 u M_2^3 + 216 M_2^6).
\end{align}
The dimension of the mass, 2, indicates that the curve is invariant under the action of Weyl$(A_1)$, indicating that this curve describes a CFT with $A_1$ flavor group, and so is written in terms of a linear mass $m$ as $M_2 = m^2$.

\paragraph{The $\{{I_1}^2,I_2^*\}$ curve.}

The $II^* \to \{ {I_1}^2,I_2^*\}$ deformation is described by the curve
\begin{align}\label{}
y^2& = x^3 + 3 u^2 x (2 u M_2 - M_4^2) + 
 2 u^3 (u^2 - 3 u M_2 M_4 + M_4^3).
\end{align}
The spectrum of dimensions of the masses, $\{2,4\}$, indicates that the curve is invariant under the action of Weyl$(BC_2)$, indicating that this curve describes a CFT with $B_2 \simeq C_2$ flavor group.  Choose a basis of the linear masses and a basis $\{N_2,N_4\}$ of Weyl$(BC_2)$-invariant polynomials as in \eqref{C2Ndef}.  Then we find the dependence of the $M_d$ on the linear masses to be 
\begin{align}\label{}
M_2&= N_2,& 
M_4&=\frac{3}{2} (N_2^2-4 N_4).
\end{align}

\paragraph{The $\{I_1,I_3^*\}$ curve.}

The $II^* \to \{ I_1,I_3^*\}$ deformation is described by the curve
\begin{align}\label{}
y^2& = x^3 + 12 u^2 x M_2 (u - M_2^3) + 
 2 u^3 (u^2 - 20 u M_2^3 - 8 M_2^6).
\end{align}
The dimension of the mass, 2, indicates that the curve is invariant under the action of Weyl$(A_1)$, indicating that this curve describes a CFT with $A_1$ flavor group, and so is written in terms of a linear mass $m$ as $M_2 = m^2$.

\subsubsection{$\{ {I_1}^2,IV^*\}$ with $G_2$ flavor symmetry}

The $\{ I_1^2,IV^*\}$ deformation of the $II^*$ singularity has the form:
\begin{align}\label{G2Curve}
y^2& = x^3 - \frac{1}{8} x (2u - M_6)^3 M_2
-\frac{1}{8} (2u - M_6)^4 (u + 2M_6).
\end{align}
In this case the spectrum of dimensions of the masses, $\{2,6\}$, indicates that the curve is invariant under the action of the Weyl$(G_2)\simeq D_6$, where $D_6$ is the dihedral group of order 6. Thus this curve describes a CFT with $G_2$ flavor group.

Choose a basis of the linear masses, $\bm=\sum_{i=1}^3 m_i \be^i$ with $\sum_{i=1}^3 m_i =0$, so that Weyl$(G_2)$ acts as permutations of the three $m_i$ and by an overall sign change.   Then a basis of Weyl$(G_2)$-invariant polynomials is 
\begin{align}\label{G2poly}
N_2&:=m_1^2-m_1m_2+m_2^2,& 
N_6&:=m_1^2(m_1-m_2)^2m_2^2.
\end{align}
Then we find the same dependence of the $M_d$ on the linear masses to be simply
\begin{align}\label{}
M_2&=N_2,& 
M_6&=N_6.
\end{align}

We find two solutions for pole positions for which the curve factorizes:
\begin{align}
r_1\, \bw_1(\bm) &:= r_1\, m_1,
\nonumber\\
(10i)^2\, x_{\bw_1}&=(10 u-M_6)^2,
\nonumber\\
(10i)^3\, y_{\bw_1}&=(10u-M_6)^2
\left(10 u +m_1^2(m_1-m_2)m_2(9m_1^2+M_2)
\right),
\nonumber\\[2mm]
r_2\, \bw_2(\bm)&:=r_2\ (m_1+m_2),
\nonumber\\
(10i)^2\, x_{\bw_2}&=
100u^2
-60u\, m_1^2 m_2^2 (7m_1^2+6m_1m_2 +7m_2^2)
\\
&\qquad\qquad \mbox{}+m_1^4 m_2^4
(41 m_1^4 - 44 m_1^3 m_2 + 406 m_1^2 m_2^2 - 44 m_1 m_2^3 + 41 m_2^4),
\nonumber\\
(10i)^3\, y_{\bw_2}&=
1000u^3
+300 u^2 m_1 m_2 (10 m_1^4 + 19 m_1^3 m_2 + 42 m_1^2 m_2^2+ 19 m_1 m_2^3 + 10 m_2^4)
\nonumber\\
&\qquad\qquad\ \mbox{}- 30 u\, m_1^3 m_2^3
(20 m_1^6 - m_1^5 m_2 + 424 m_1^4 m_2^2 + 314 m_1^3 m_2^3 + 424 m_1^2 m_2^4 
\nonumber\\
&\qquad\qquad\qquad\qquad\qquad\quad \mbox{}- m_1 m_2^5 + 20 m_2^6) 
\nonumber\\
&\qquad\qquad\ \mbox{}+ m_1^5 m_2^5 
(30 m_1^8 - 61 m_1^7 m_2 + 1266 m_1^6 m_2^2 
- 1395 m_1^5 m_2^3 + 8320 m_1^4 m_2^4
\nonumber\\
&\qquad\qquad\qquad\qquad\qquad \mbox{} 
- 1395 m_1^3 m_2^5 + 1266 m_1^2 m_2^6 
- 61 m_1 m_2^7 + 30 m_2^8).
\nonumber
\end{align} 

Summing over the Weyl$(G_2)$ orbits of these poles in the $x$-plane in the MN ansatz \eqref{MNform} for the SW 1-form, imposing the differential condition \eqref{SWform} and (arbitrarily) choosing the normalization $2a-c=2i$ gives a unique solution for the SW one-form: 
\begin{align}
a &= \frac{i}{2}, &
c &= -i ,&
W &=\frac{3 i}{10}M_6,&
r_1 &= 1, &
r_2 &= 0.
\end{align}

\subsubsection{$\{ I_2,IV^*\}$ with $\SU(2)$ flavor symmetry}

The $\{ I_2,IV^*\}$ deformation of the $II^*$ singularity has the form:
\begin{align}\label{A1IVsCurve}
y^2& = x^3 - 6 x M_2 \left(5 u- \frac{2}{5} M_2^3 \right)^3
- 2 \left(u - \frac{2}{5} M_2^3\right)^4 \left(u + \frac{8}{5} M_2^3 \right).
\end{align}
From the spectrum of its mass invariants, namely $\{2\}$, it follows that this curve is invariant under the action of Weyl$(A_1)\simeq \Z_2$.  In this case the root lattice is only one dimensional and thus there is no ambiguity in defining the invariant polynomial $M_2$ in terms of a linear mass (the overall normalization is irrelevant),
\beq
M_2=m^2,
\eeq
where Weyl$(A_1)$ acts on $m$ by a sign change.

The factorization solutions for the curve are:
\begin{align}
r_1\, \bw_1(\bm)&:=r_1 m,
\nonumber\\
(i)^2\, x_{\bw_1}&= 
\left(u  - \frac{1}{160} m^6\right)^2 ,
\nonumber\\
(i)^3\, y_{\bw_1}&=
\left(u +\frac{39}{160} m^6\right)
\left(u - \frac{1}{160} m^6\right)^2,
\nonumber\\[2mm]
r_2\, \bw_2(\bm)&:=r_2 m,
\\
(i/2)^2\, x_{\bw_2}&=
\left(u  - \frac{1}{160} m^6\right)^2 ,
\nonumber\\
(i/2)^3\, y_{\bw_2}&=
\left(u -\frac{3}{80} m^6\right)
\left(u - \frac{1}{160} m^6\right)^2,
\nonumber\\[2mm]
r_3\, \bw_3(\bm)&:=r_3 m,
\nonumber\\
(3i/2)^2\, x_{\bw_3}&=
u^2
-\frac{141}{80} u m^6
+\frac{1881}{25600} m^{12},
\nonumber\\
(3i/2)^3\, y_{\bw_3}&=
\left(u -\frac{3}{80} m^6\right)
\left(u^2 - \frac{399}{80} u m^6 
- \frac{13599}{25600} m^{12} \right).
\nonumber
\end{align}

Summing over the Weyl orbits of these pole solutions in the MN ansatz \eqref{MNform} for the SW 1-form and imposing the differential condition \eqref{SWform} we find a two-parameter family of solutions for the one-form:
\begin{align}\label{}
a &= \frac{i}{18} (9r_1+36r_2+20r_3), &
c &= \frac{i}{9} (-3r_1-18r_2-10r_3),
\nonumber\\
W &= \frac{i}{160} (-3r_1-12r_2+20r_3) M_2^3,&
1 &= 2r_1 + 9 r_2 + 5 r_3,
\end{align}
where we have arbitrarily normalized the 1-form so that $2a-c=2i/3$.

\subsubsection{$\{ I_1,III^*\}$ with $\SU(2)$ flavor symmetry}

The $\{ I_1,III^*\}$ deformation of the $II^*$ singularity has the form
\begin{align}\label{A1IIIsCurve}
y^2& = x^3 - 2 x u^3 M_2 -2 u^5.
\end{align}
The spectrum of the mass invariants implies that this curve is invariant under the action of the Weyl$(A_1)\simeq \Z_2$ group.  So we can take
\beq
M_2=m^2,
\eeq
where Weyl$(A_1)$ acts on $m$ as a sign change.

We find only two inequivalent solutions for the pole positions,
\begin{align}
r_1\, \bw_1(\bm)&:=r_1 m,
\nonumber\\
(i)^2\, x_{\bw_1}&=u^2,
\nonumber\\
(i)^3\, y_{\bw_1}&=u^3,
\nonumber\\[2mm]
r_2\, \bw_2(\bm)&:=r_2 m,
\nonumber\\
(2i)^2\, x_{\bw_2}&=
u^2-12 u m^6+4 m^{12},
\nonumber\\
(2i)^3\, y_{\bw_2}&=
u^3+30 u^2 m^6-36 u m^{12}+8 m^{18} .
\nonumber
\end{align}

Summing over the Weyl orbits of these pole solutions in the MN ansatz \eqref{MNform} for the SW 1-form and imposing the differential condition \eqref{SWform} we find a one-parameter family of solutions for the one-form:
\begin{align}\label{A1IIIs1form}
a &= \frac{i}{12} (4r_1-11r_2), &
c &= \frac{i}{6} (-2r_1+r_2), &
W &= i r_2 M_2^3,&
1&=r_1-2r_2,
\end{align}
where we have normalized the one form by choosing $2a-c=i$. 

\subsection{Deformations of the $III^*$ singularity}\label{appA2}

Both the quadratic and linear ansatz are consistent for the $x$ pole position for the $III^*$ curve.  In this subsection we will thus consider poles of the form
\begin{align}\label{LinAnsatz}
x &=\bw(\bm)^{-2}\, (u^2+R u+S),& 
y &=\bw(\bm)^{-3}\, (u^3+J u^2+K u+L),
\end{align}
for the quadratic ansatz, and
\begin{align}\label{QuadAnsatz}
x &= R' u+S',&
y &= T' u^2+K' u+L',
\end{align}
for the linear ansatz.  Here the coefficients are polynomials of the linear masses we will solve for.  $\Delta(u)=4$ implies $\Delta(R)=4$, $\Delta(S)=8$, $\Delta(J)=4$, $\Delta(K)=8$, $\Delta(L)=12$, $\Delta(R')=2$, $\Delta(S')=6$, $\Delta(T')=1$, $\Delta(K')=5$ and $\Delta(L')=9$.  Furthermore, from the form of the $III^*$ curve (see table \ref{Table:Kodaira}), it easily follows that $T'\propto \bw(\bm)$ and $R'\propto \bw(\bm)^2$.

Also in the $III^*$ case $\Delta(u)>2$ so there is no $\mu$ parameter in the MN-ansatz (\ref{MNform}).

\subsubsection{$\{ I_1^5,I_4\}$ with $\Sp(6)\oplus\SU(2)$ flavor symmetry}

The $\{I_1^5,I_4\}$ deformation of the $III^*$ singularity has curve
\begin{align}\label{C3C1sigM}
y^2 &= x^3
+x [12 u^3
+u^2 (-4 M_2^2-M_4)
+12 u M_2 M_6
-3 M_6^2]
\\
&\qquad\text{}
-12 u^4 (2 M_2+3 \til M_2)
+2 u^3 (M_2 M_4+6 M_6)
-u^2 (16 M_2^2+M_4) M_6
+12 u M_2 M_6^2
-2 M_6^3 .
\nonumber
\end{align}
Its spectrum of dimensions, $\{2,2,4,6\}$, of mass invariants implies it has a discrete Weyl$(BC_3\oplus A_1)$ group of symmetries acting on the linear masses $\bm$.  

Choose a basis, $\bm=\til m \til\be + \sum_{i=1}^3 m_i \be^i$, of the linear masses so that Weyl$(BC_3) \simeq S_3 \ltimes \Z_2^3$ acts by permutations and independent sign flips of the three $m_i$, and Weyl$(A_1) \simeq \Z_2$ acts by sign flip of the $\til m$ mass.  Then a standard basis of Weyl$(BC_3\oplus A_1)$ invariant polynomials is
\begin{align}\label{C3C1Ndef}
\til N_2 &:= \til m^2, &
N_2 &:= \sum_{i} m_i^2, &
N_4 &:= \sum_{i>j} m_i^2 m_j^2, &
N_6 &:= m_1^2 m_2^2 m_3^2 .
\end{align}
Then either by solving the factorization condition for the MN ansatz for the SW one-form, or by demanding increased zero multiplicities of the curve's discriminant when $\ba(\bm)=0$ for $\ba$'s fixed by the Weyl group, we find the same dependence, $M_d(\bm)$, of the invariant masses appearing in the curve \eqref{C3C1sigM} on the linear masses:
\begin{align}\label{C3C1MtoN}
\til M_2 &= 24 \til N_2, &
M_2 &= 6 ( N_2 - 4 \til N_2 ), &
M_4 &= 1296 N_4 - 36 (N_2 - 4 \til N_2)^2, &
M_6 &= 2592 N_6.
\end{align}

Solving for the one-form using the MN ansatz \eqref{MNform} trying both a quadratic and a linear ansatz for the pole position and we find four solutions for pole positions for which the curve factorizes,
\begin{align}\label{C3C1pole}\nonumber
r_1\,\bw_1(\bm) &:= r_1\, \til m,\\\nonumber
(i\sqrt6\, \til m)^{-2} x_{\bw_1} 
&= 2u (4 \til m^2 - N_2)+432 N_6,
\nonumber\\
(i\sqrt6\, \til m)^{-3} y_{\bw_1} 
&= 2 u^2 \til m;  
\nonumber\\[2mm]\nonumber
\bw_2(\bm) & :=  r_2\, \til m + r_2'\, m_1,\\\nonumber
(i\sqrt6\, \w_2)^{-2} x_{\bw_2} 
&= 
2 u \bigl[ U_1 (U_1 + 4 m_1) - T_2 \bigr]
+324 m_1^2 \bigl[ 3 U_1^4 - 6 U_1^2 T_2 
+ 3 T_2^2 - 8 T_4 \bigr],
\nonumber\\
(i\sqrt6\, \w_2)^{-3} y_{\bw_2} 
&= 
u^2 \bigl[ U_1 + m_1 \bigr]
+54 u \, m_1 \bigl[
U_1^3 (U_1 + 4 m_1)
- 2 U_1 (U_1 + 2 m_1) T_2
\nonumber\\
&\qquad\text{}
+ T_2^2 - 4 T_4
\bigr]
+5832 m_1^3 \bigl[ U_1^2-T_2 \bigr]
\cdot \bigl[ U_1^4 - 2 U_1^2 T_2
+ T_2^2 - 4 T_4 \bigr],
\nonumber\\[2mm]
r_3\, \bw_3(\bm) & := r_3\,  \til m,\\
(i\sqrt6)^2 x_{\bw_3} 
&= 
u^2 
- 18 u \bigl[ 80 \til m^4 - 8 \til m^2 N_2 
- 3 N_2^2 + 12 N_4 \bigr]
+243 \bigl[ 768 \til m^8 - 768 \til m^6 N_2 
\nonumber\\
&\qquad\text{}
+ 96 \til m^4 (3 N_2^2 - 4 N_4)
     - 16 \til m^2 (3 N_2^3 - 12 N_2 N_4 + 32 N_6)
     + 3 (N_2^2 - 4 N_4)^2
     \bigr],
\nonumber\\
(i\sqrt6)^3 y_{\bw_3} 
&= 
u^3 + 27 u^2 \bigl[ 48 \til m^4
+ 8 \til m^2 N_2 + 3 N_2^2 - 12 N_4 \bigr]
-729 u \bigl[ 1280 \til m^8
-768 \til m^6 N_2
\nonumber\\
&\qquad\text{}
+32 \til m^4 (3 N_2^2 - 4 N_4)
+16 \til m^2 (N_2^3 - 4 N_2 N_4 + 16 N_6)
-3 (N_2^2 - 4 N_4)^2  \bigr]
\nonumber\\
&\qquad\text{}
+19683 (16 \til m^4-8 \til m^2 N_2+N_2^2-4 N_4) 
\bigl[ 256 \til m^8
-256 \til m^6 N_2
\nonumber\\
&\qquad\text{}
+32 \til m^4 (3 N_2^2 - 4 N_4)
-16 \til m^2 (N_2^3 - 4 N_2 N_4 + 16 N_6)
+ (N_2^2 - 4 N_4)^2 \bigr] ;
\nonumber\\[2mm]
r_4\,\bw_4(\bm)& := r_4\, (m_1+m_2),
\nonumber\\
(i\sqrt6)^2 x_{\bw_4} 
&= 
u^2
- 72 u \bigl[ 
S_1^2 (S_2 + S_1^2)
+ (3 S_2 - S_1^2)  (4 \til m^2 - m_3^2)
\bigr]
\nonumber\\
&\qquad\text{}
+3888 S_2^2 
\bigl[ 
3 (4 \til m^2 -  m_3^2)^2 
- 2 S_1^2 (12 \til m^2 + m_3^2)
+ 3 S_1^4
\bigr] ,
\nonumber\\
(i\sqrt6)^3 y_{\bw_4} 
&=
u^3
- 108 u^2 
\Bigl[
( 4 \til m^2 - m_3^2) ( 3 S_2 - S_1^2 ) 
+ S_1^2 ( S_2 - S_1^2 ) 
\Bigr]-11664 u S_2 
\Bigl[ S_1^4 (S_1^2 + S_2) 
\nonumber\\
&\qquad\text{}
(4 \til m^2 - m_3^2)^2 (S_1^2 - 3 S_2)
-2 (4 \til m^2 + m_3^2) S_1^2 (S_1^2 - S_2)\Bigr]
\nonumber\\
&\qquad\text{}
-1259712 S_2^3 
\bigl[ 4 \til m^2 - m_3^2 - S_1^2  \bigr] 
\cdot \bigl[ (4 \til m^2 - m_3^2)^2
-2 S_1^2 (4 \til m^2+ m_3^2) + S_1^4 \bigr],
\nonumber
\end{align}
where $U_1:= 2\til m +m_1$, $T_2:= m_2^2+m_3^2$, $T_4:= m_2^2 m_3^2$, $S_1:= m_1+m_2$ and$S_2 := m_1 m_2$.  Since the second pole solution has two independent normalizations for its residues, $r_2$ and $r_2'$, we have included them in the definition of $\bw_2$, unlike in the other cases.  The ``$\w_2$" appearing in the prefactors of $x_{\bw_2}$ and $y_{\bw_2}$ stands for $\bw_2(\bm)$.

Summing over the Weyl orbits of these poles in the $x$-plane in the MN ansatz \eqref{MNform} for the SW 1-form and imposing the differential condition \eqref{SWform}, we find a 3-parameter family of solutions
\begin{align}\label{}
a &= 8 \sqrt{\frac32}(4 r_1 - 16 r_2 - 5 r_3 - 5 r_4), &
c &= \frac34 \sqrt{\frac32} (r_3 + 4 r_4),
\nonumber\\
W &= \sqrt{\frac23}\left(r_3 N_2^2 - (16 r_2 - 2 r_3) N_4\right), &
3 & = r_1 - 4 r_2 - 2 r_3 - 4 r_4 ,
\end{align}
where we have arbitrarily chosen to normalize the one-form by setting $2a-c = 3\sqrt{3/2}$.

\subsubsection{$\{ I_1^3,I_0^*\}$ with $\SO(7)$ flavor symmetry}\label{A.2.2}

The $\{I_1^3,I_0^*\}$ deformation of the $III^*$ singularity has curve
\begin{equation}\label{IIIsSo}
y^2=x^3+3x(u^3+u^2 M_4)+2(u^4 M_2+u^3 M_6)
\end{equation}
The spectrum of dimensions of invariant masses is $\{2,4,6\}$ which implies that the curve has a discrete Weyl$(BC_3)$ group of symmetries acting on the linear masses $\bm$. 

Choose a basis, $\bm=m_i \be^i$, of the linear masses so that the Weyl$(BC_3)\simeq S_3 \ltimes \Z_2^3$ acts by permutations and sign flips of the three $m_i$.  Then a basis of invariant polynomials is
\begin{equation}
N_{2k}=\sum_{i_1<...<i_k}m_{i_1}^2...m_{i_k}^2
\end{equation}
for $k=1,2,3$.  Then either by solving the factorization condition for the MN ansatz for the the SW one-form, or by demanding increased zero multiplicities of the curve's discriminant when $\ba(\bm)=0$ for $\ba$'s fixed by the Weyl group, we find the following dependence of the invariant mass polynomials $M_d(\bm)$
\begin{align}
M_2&=N_2,& 
M_4&=-\frac{4}{9}N_2^2+\frac{4}{3}N_4,&
M_6&=-\frac{8}{27}N_2^3+\frac{4}{3}N_2N_4-4N_6.
\end{align}

We find pole positions for which the curve factorizes:
\begin{align}
r_1\, \bw_1 &= r_1\, m_1,
\nonumber\\
(\sqrt{2/3}\, m_1)^{-2}\, x_{\bw_1} &= 
u\, (2m_1^2-m_2^2-m_3^2) ,
\nonumber\\
(\sqrt{2/3}\, m_1)^{-3}\, y_{\bw_1} &= 
\frac{9}{2} u^2 m_1 ,
\nonumber\\[2mm]
r_2\, \bw_2(\bm) &= r_2\, (m_1+m_2+m_3),
\nonumber\\
(\sqrt{2/3}\, S_1)^{-2} x_{\bw_2} &= 
2 u\, (S_1^2+S_2) + 4 S_3^2,
\nonumber\\
(\sqrt{2/3}\, S_1)^{-3} y_{\bw_2} &= 
\frac{9}{2} u^2 S_1 
+ 6 u\, S_3 (S_1^2 + S_2)
+ 8 S_3^3,  
\nonumber\\[2mm]
r_3\, \bw_3(\bm) &= r_3\, m_1,\nonumber\\
\left(2\sqrt{2/3}\right)^2 x_{\bw_3} &= 
u^2
- \frac{8}{9} u\, (5 m_1^4 - m_1^2 T_2 - 3 T_4)
+\frac{16}{9} (m_1^4 -m_1^2 T_2 + T_4)^2,
\\
\left(2\sqrt{2/3}\right)^3 y_{\bw_3} &=
u^3
 -\frac{16}{9}u\,
(m_1^4 -m_1^2 T_2 + T_4) 
(5 m_1^4 - m_1^2 T_2 - 3T_4)
\nonumber\\
&\qquad\ \, \mbox{}
+ \frac{4}{3} u^2 (3 m_1^4 + m_1^2 T_2 + 3 T_4)
+\frac{64}{27} (m_1^4 -m_1^2 T_2 + T_4)^3,
\nonumber\\[2mm]
r_4\,\bw_4(\bm) &= r_4\, (m_1+m_2) ,
\nonumber\\
\left(\sqrt{2/3}\right)^2 x_{\bw_4} &=
u^2
-\frac{4}{9}u\,(m_1+m_2)^2 (m_1^2+m_2^2-2 m_3^2),
\nonumber\\
\left(\sqrt{2/3}\right)^3 y_{\bw_4} &= 
u^3+
\frac{4}{3}u^2\,(m_1+m_2)^2 (m_1 m_2+m_3^2),
\nonumber
\end{align}
where $S_1 := m_1+m_2+m_3$, $S_2:= m_1m_2+m_2m_3+m_3m_1$, $S_3:= (m_1+m_2)(m_2+m_3)(m_3+m_1)$, $T_2:= m_2^2+m_3^2$, and $T_4:=m_2^2m_3^2$.

Summing over the Weyl orbits of these poles in the MN ansatz for the one-form and solving the differential constraint gives a 3-parameter family of solutions,
\begin{align}
a &= \frac{1}{8}\sqrt{\frac{3}{2}} 
(4r_1-16r_2-5r_3-4r_4), &
c &=  \frac{3}{4} \sqrt{\frac{3}{2}} (r_3+4r_4),
\\ 
W &=\sqrt{\frac23}\left( r_3 N_2^2 -(16r_2+3r_3) N_4\right), &
3 &=r_1-4r_2-2r_3-4r_4,
\nonumber
\end{align}
where we have chosen the convenient normalization $2a-c = 3\sqrt{3/2}$.
Here we have a situation similar to the $II^*$ deformation with either $\SO(7)$ or $\Sp(4)$ flavor symmetry, described in appendix \ref{appA1.4}.  In fact for particular choices of the $r_i$'s the lattice of residues can be either face-centered cubic ($\simeq C_3$ root lattice) or simple cubic ($\simeq B_3$ root lattice).  Yet we can still make a case that the theory described by \eqref{IIIsSo} has a $B_3\simeq\SO(7)$ flavor symmetry by studying the minimal adjoint breaking flows \eqref{I0sF4-mab} from the $II^*$ deformation with $F_4$ flavor symmetry described in appendix \ref{appA1.3}.

\subsubsection{$\{ I_1^2,I_1^*\}$ with $\SU(2)\oplus\SU(2)$ flavor symmetry}

The $\{I_1^2,I_1^*\}$ deformation of the $III^*$ singularity has curve
\begin{equation}
y^2=x^3+3x(u^3-u^2 {\til M}_2^2)
+2(u^4 M_2+u^3 {\til M}_2^3)
\end{equation}
The spectrum of dimensions of the invariant masses is $\{2,2\}$ which implies that the curve has a discrete Weyl$(A_1\oplus A_1)$ group of symmetries acting on the linear masses $\bm$. 

Choose a basis, $\bm=m \be + \til m \til\be$, of the linear masses so that the Weyl$(A_1\oplus A_1)\simeq \Z_2^2$ acts by independent sign flips of $m$ and $\til m$.  Then a basis of invariant polynomials is given by
\begin{align}\label{}
N_2 &:=m^2,& 
{\til N}_2 &:={\til m}^2.
\end{align}
Then by solving the factorization condition for the MN ansatz for the the SW one-form, we find the invariant masses in terms of the linear masses to be
\begin{align}\label{A1A1masses}
M_2&=N_2+\tN_2, &
\tM_2&=\frac{1}{3} (N_2-2\tN_2).
\end{align}
Note that in this case, if instead one demands increased zero multiplicities of the curve's discriminant when $\ba(\bm) = 0$ for $\ba$'s fixed by the Weyl group, instead of finding only \eqref{A1A1masses}, one also finds eight other possible linear mass dependencies.   It turns out that none of these other eight curves have a SW one form.

We find six solutions for pole positions:
\begin{align}
r_1\, \bw_1(\bm) &= r_1\, m,
\nonumber\\
m^{-2}\, x_{\bw_1} &= 
\frac13 u\, (m^2 - 2 \tm^2),
\nonumber\\
m^{-3}\, y_{\bw_1} &=
\sqrt{3}\, u^2\, m,
\nonumber\\[2mm]
\tr_2\, \bw_2(\bm) &= \tr_2\, \tm/\sqrt2,
\nonumber\\
(\sqrt2/\tm)^2\, x_{\bw_2} &= 
-\frac{2}{3} u\, (m^2 - 2 \tm^2),
\nonumber\\
(\sqrt2/\tm)^3\, y_{\bw_2} &=
\sqrt{6}\, u^2\, \tm,
\nonumber\\[2mm]
\bw_3(\bm) &= r_3 m + \tr_3 \tm/\sqrt2 ,
\nonumber\\
(S_1/\w_3)^2\, x_{\bw_3} &= 
2u\, (10m^4 + 22\sqrt2 m^3 \tm + 33 m^2 \tm^2
+10\sqrt2 m \tm^3 + 2\tm^4)
\nonumber\\
& \quad \mbox{}
+4 m^2 (2 m^6 + 10 \sqrt2 m^5 \tm + 41 m^4 \tm^2
+44\sqrt2 m^3\tm^3 + 52 m^2 \tm^4 
\nonumber\\
& \qquad\qquad\quad \mbox{}
+ 16\sqrt2 m \tm^5 + 4 \tm^6),
\nonumber\\
(S_1/\w_3)^3\, y_{\bw_3} &= 
9 u^2\, (4\sqrt2\, m^4+16m^3\tm+12\sqrt2\,m^2\tm^2
+8m\tm^3+\sqrt2\,\tm^4)
\nonumber\\
&\quad \mbox{}
+6u\, m (10\sqrt2\,m^7+94m^6\tm+183\sqrt2\,m^5\tm^2
+381m^4\tm^3 +228\sqrt2\,m^3\tm^4
\nonumber\\
& \qquad\qquad\quad \mbox{}
+156m^2\tm^5+28\sqrt2\,m\tm^6 +4\tm^7)
\nonumber\\
&\quad \mbox{}
+8m^3(2\sqrt2\, m^9 + 30 m^8 \tm 
+ 99\sqrt2\, m^7\tm^2 + 377m^6\tm^3
+456\sqrt2\, m^5\tm^4 
\nonumber\\
& \qquad\qquad\quad \mbox{}
+ 726 m^4 \tm^5 +380\sqrt2\, m^3\tm^6 
+ 252 m^2 \tm^7 +48\sqrt2\, m \tm^8 + 8 \tm^9),
\nonumber\\[2mm]
r_4\, \bw_4(\bm) &= r_4\, m,
\nonumber\\
(2/\sqrt3)^2\, x_{\bw_4} &=
u^2 -\frac{8}{9} u\, m^2 (m^2-2\tm^2),
\\
(2/\sqrt3)^3\, y_{\bw_4} &= 
u^3 +\frac43 u^2\, m^2 (m^2+2\tm^2),
\nonumber\\[2mm]
\tr_5\, \bw_5(\bm) &= \tr_5\, \tm/\sqrt2,
\nonumber\\
(4/\sqrt3)^2\, x_{\bw_5} &= 
u^2 + \frac{2}{9} u\,
(3m^4+4m^2\tm^2-20\tm^4)
+\frac19 (m^2-2\tm^2)^2 ,
\nonumber\\
(4/\sqrt3)^3\, y_{\bw_5} &= 
u^3+\frac{1}{3} u^2\, 
(3m^4+4m^2\tm^2+12\tm^4)
+\frac19 u\, (m^2-2\tm^2)^3(3m^2+10\tm^2)
\nonumber\\
&\qquad\ \mbox{}
+\frac{1}{27} (m^2-2\tm^2)^6 ,
\nonumber\\[2mm]
\bw_6(\bm)& := r_6\, m + \tr_6\,\tm/\sqrt2,
\nonumber\\
(T_1/\w_6)^2\, x_{\bw_6} &= 
u^2
+\frac19 u\,(m^2-2\tm^2) (m^2+2\sqrt2\,\tm+2\tm^2),
\nonumber\\
(T_1/\w_6)^3\, y_{\bw_6} &= 
u^3+\frac13 u^2\, m
(m^3+3\sqrt2\,m^2\tm+6m\tm^2+2\sqrt2\,\tm^3),
\nonumber
\end{align}
where $S_1:= \sqrt3\,(\sqrt2\,m+\tm)$ and $T_1:= (m + \sqrt2\,\tm)/\sqrt3$.

Summing over the Weyl orbits of these poles in the MN ansatz for the one-form, and solving the differential condition, we find
\begin{align}
a &= \frac{\sqrt3}{8}
(4 r_1 + 8 r_3 - 6 r_4 + \tr_5), 
\qquad\qquad\qquad\qquad\qquad\quad
c = \frac{\sqrt3}{4}
(2 r_4 + \tr_5 + 4 r_6)
\nonumber\\
W &= \frac{1}{2\sqrt3} 
\bigl[ (8r_3 +\tr_5)m^4 + 
4(4r_3+4\tr_3-\tr_5)m^2\tm^2
+4\tr_5\tm^4
\bigr]
\nonumber\\
0&= r_1-\tr_2+2r_3-\tr_3-2r_4+2\tr_5+r_6
= 2r_6 - \tr_6,
\qquad
1 = r_1 + 2r_3 - 2r_4 - r_6,
\end{align}
where the last relation comes from choosing the normalization $2a-c = \sqrt3$.  This is thus a 5-parameter family of solutions.

\subsubsection{$\{ I_2,I_1^*\}$ and $\{I_1,I_2^*\}$}

In this section we simply list the SW curves (in terms of the linear mass paramters) for the above three geometries without giving their SW one-forms.  The reason is that these one-forms can all be easily constructed by restriction of masses of the $\{{I_1}^3,I_0^*\}$ one-form given in section \ref{A.2.2} above, and their (lengthy) expressions do not illuminate any interesting points since the resulting flavor symmetries are of rank 1.

\paragraph{The $\{I_2,I_1^*\}$ curve.}

The $II^* \to \{ I_2,I_1^*\}$ deformation is described by the curve
\begin{align}\label{}
y^2& = x^3 + 3 u^2 x (u - M_2^2) + 
 2 u^3 M_2 (3 u + 2 M_2^2).
\end{align}
The dimension of the mass, 2, indicates that the curve is invariant under the action of Weyl$(A_1)$, indicating that this curve describes a CFT with $A_1$ flavor group, and so is written in terms of a linear mass $m$ as $M_2 = m^2$.

\paragraph{The $\{I_1,I_2^*\}$ curve.}

The $II^* \to \{ I_1,I_2^*\}$ deformation is described by the curve
\begin{align}\label{}
y^2& = x^3 + 3 u^2 x (u - 4 M_2^2) + 
 2 u^3 M_2 (3 u - 8 M_2^2).
\end{align}
The dimension of the mass, 2, indicates that the curve is invariant under the action of Weyl$(A_1)$, indicating that this curve describes a CFT with $A_1$ flavor group, and so is written in terms of a linear mass $m$ as $M_2 = m^2$.

\subsubsection{$\{ I_1,IV^*\}$ with $\SU(2)$ flavor symmetry}

The $\{I_1,IV^*\}$ deformation of the $III^*$ singularity has curve
\begin{align}
y^2 = x^3 - 2 u^3 x - u^4 M_2 .
\end{align}
The spectrum of the mass invariants implies that this curve is invariant under the action of Weyl$(A_1)\simeq \Z_2$.  So we write, without loss of generality,
\beq
M_2=m^2,
\eeq
where Weyl$(A_1)$ acts on $m$ by a sign change.

We find two pole solutions,
\begin{align}
r_1\, \bw_1(\bm)&:=r_1\, m,
\nonumber\\
x_{\bw_1}& = 4 m^4 (u - m^2),
\nonumber\\
y_{\bw_1}&= i m^4 (3 u^2 - 12 m^4 u + 8 m^8),
\\[2mm]
r_2\,\bw_2(\bm)&:= r_2\, m,
\nonumber\\
x_{\bw_2}&= -u^2,
\nonumber\\
y_{\bw_2}&= i u^2\,(u - m^4).
\nonumber
\end{align}
Summing over the Weyl orbits of these poles in the MN ansatz for the 1-form and imposing the differential condition, we find a 1-parameter family of solutions for the one-form:
\begin{align}\label{}
a &=\frac{i}{12} (12 r_1 + 5 r_2), &
c &=- \frac{i}{2} r_2, &
W &= m^4, &
1 &= 3 r_1 + 2 r_2,
\end{align}
where we have chosen the normalization $2a-c=2i/3$.

\subsection{Deformations of the $IV^*$ singularity}

Both the quadratic and linear ansatz are consistent for the $x$ pole position for the $IV^*$ curve, so we will consider pole positions of the form \eqref{LinAnsatz} and \eqref{QuadAnsatz}.  $\Delta(u)=3$ implies $\Delta(R)=3$, $\Delta(S)=6$, $\Delta(J)=3$, $\Delta(K)=6$, $\Delta(L)=9$, $\Delta(R')=1$, $\Delta(S')=4$, $\Delta(T')=0$, $\Delta(K')=3$ and $\Delta(L')=6$.  Also in the $IV^*$ case $\Delta(u)>2$ so there is no $\mu$ parameter in the MN-ansatz (\ref{MNform}).

\subsubsection{$\{ I_1^{4},I_4\}$ with $\Sp(4)\oplus\U(1)$ flavor symmetry}

The $\{I_1^4,I_4\}$ deformation of the $IV^*$ singularity has curve
\begin{align}\label{C2U1sigM}
y^2 &= x^3
+x \left[-3 u^2 (M_1^2+M_2)-12 u M_1 M_4-3 M_4^2
\right]
\\
&\qquad\text{}
-864 u^4
+2 u^3 M_1 (M_1^2-3 M_2)
-3 u^2 (5 M_1^2+M_2) M_4
-12 u M_1 M_4^2
-2 M_4^3 .
\nonumber
\end{align}
Its spectrum of mass dimensions, $\{1,2,4\}$, implies it has a discrete Weyl$(C_2\oplus U_1)$ group of symmetries acting on the linear masses $\bm$.   $M_2$ and $M_4$ are a basis of invariant polynomials for Weyl$(C_2)$ and $M_1$ is just the linear mass parameter associated to a $U(1)$ flavor group (with Lie algebra $\R$) which has trivial Weyl group.

Choose a basis, $\bm=\til m \til\be + \sum_{i=1}^2 m_i \be^i$, of the linear masses so that Weyl$(C_2) \simeq S_2 \ltimes \Z_2^2$ acts by permutations and independent sign flips of the two $m_i$, and $\til m$ is invariant.  Then a standard basis of invariant polynomials is
\begin{align}\label{C2U1Ndef}
N_1 &:= \til m, &
N_2 &:= m_1^2+m_2^2, &
N_4 &:= m_1^2 m_2^2 .
\end{align}
Then either by solving the factorization condition for the MN ansatz for the SW one-form, or by demanding increased zero multiplicities of the curve's discriminant when $\ba(\bm)=0$ for $\ba$'s fixed by the Weyl group, we find the same dependence, $M_d(\bm)$, of the invariant masses on the linear masses:
\begin{align}\label{C2U1MtoN}
M_1 &= 48 N_1, &
M_2 &= 1728 N_2, &
M_4 &= 10368 N_4.
\end{align}
(The coefficients 48 and 1728 appearing in $M_1$ and $M_2$ are arbitrary normalization factors chosen to simplify some coefficients in the following formulas.)

We find four solutions for pole positions for which the curve factorizes,
\begin{align}\label{C2U1pole}
r_1\, \bw_1(\bm) &:= \tr_1\, \tm,
\nonumber\\
(i \sqrt{6}\,\tm)^{-2} x_{\bw_1} 
&= 16 u\, \tm + 1728 N_4,
\nonumber\\
(i \sqrt{6}\,\tm)^{-3} y_{\bw_1} 
&= 2 u^2 ;
\nonumber\\[2mm]
r_2\,\bw_2(\bm) &:= \tr_2\, \tm,
\nonumber\\
(i \sqrt{6}\,\tm)^{-2} x_{\bw_2} 
&= -32 u\, \tm
+20736 \tm^4
-10368 \tm^2 N_2 
+432 (3 N_2^2 - 8 N_4) ,
\nonumber\\
(i \sqrt{6}\,\tm)^{-3} y_{\bw_2} 
&= 
2u^2 -1728 u\, \tm (4 \tm^2 - N_2)
\nonumber\\
&\qquad\quad \mbox{}
+46656 (4 \tm^2-N_2)
(4 \tm^2-[m_1-m_2]^2)(4 \tm^2-[m_1+m_2]^2) ;
\nonumber\\[2mm]
\bw_3(\bm) &:= \tr_3\, \tm + r_3\, m_1,
\nonumber\\
(i \sqrt{6}\,\w_3)^{-2} x_{\bw_3} &=
8u\, (2 \tm - 3 m_1)
+1728 m_1^2 
(12 \tm^2-12 \tm m_1 +3 m_1^2-2 m_2^2),
\nonumber\\
(i \sqrt{6}\,\w_3)^{-3} y_{\bw_3} &= 
2u^2
-864 u\, m_1 (4 \til m^2 - 8 \til m m_1 +3 m_1^2 - m_2^2)
\\
&\qquad\quad \mbox{}
-373248 m_1^3 
(2 \til m-m_1) 
(2 \til m-m_1-m_2) 
(2 \til m-m_1+m_2);
\nonumber\\[2mm]
\bw_4(\bm) &:= \tr_5 \, \tm + r_5\, (m_1+m_2),
\nonumber\\
(i \sqrt{6}\,\w_4)^{-2} x_{\bw_4} &=
-4u\, (2 \tm+3 S_1 )
+1728 S_2^2,
\nonumber\\
(i \sqrt{6}\,\w_4)^{-3} y_{\bw_4} &=
2u^2 - 864 u\, (2 \tm+S_1)S_2;
\nonumber\\[2mm]
r_5\,\bw_5(\bm) &:= r_5\,(m_1+m_2),
\nonumber\\
(i\sqrt{6})^2 x_{\bw_5} &=
u^2 +576 u\, \tm (S_1^2 - 3 S_2)  
+124416 (6 \tm^2 - S_1^2) S_2^2
\nonumber\\
(i\sqrt{6})^3 y_{\bw_5} &=
u^3 + 864 u^2\, \tm (S_1^2 - 3 S_2) 
- 186624 u\,  (4 \tm^2 [S_1^2-3S_2]
                      - S_1^2[S_1^2 - S_2]) S_2
\nonumber\\
&\qquad\ \mbox{}
- 161243136 \tm (4 \tm^2 - S_1^2) S_2^3
\nonumber
\end{align}
where $S_1:= m_1+m_2$ and $S_2:= m_1 m_2$.  Summing over the Weyl orbits of these poles in the $x$-plane in the MN ansatz and imposing the differential condition gives a 5-parameter family of 1-forms,
\begin{align}\label{}
a &= \frac{i}{3}\sqrt{\frac23} \, (3 r_3 + 3 r_4 + 4 r_5), &
c &= -\frac{4i}{3}\sqrt{\frac23} \, r_5, 
\nonumber\\
W &= 144 i \sqrt6\, \tm ( [\tr_2+2r_3-\tr_3] M_2 
-4\tr_2 \tm^2), &
\\
0 &= 2\tr_1 - 4 \tr_2 + 2 r_3 + 2\tr_3 
+ 2 r_4 - \tr_4+ 4 r_5, &
1 &= r_3 + r_4 + 2 r_5,
\nonumber
\end{align}
where the last relation comes from choosing the normalization $2a-c = 2i\sqrt{2/3}$.

\subsubsection{$\{ I_1^2,I_0^*\}$ with $\SU(3)$ flavor symmetry}

The $\{I_1^2,I_0^*\}$ deformation of the $IV^*$ singularity has curve
\begin{align}
y^2 = x^3 + u^2 x M_2 + 2 u^4 + u^3 M_3.
\end{align}
The spectrum of its mass dimensions implies that the curve is invariant under Weyl$(A_2)\simeq S_3$.  Choose a basis of linear masses, $\bm= \sum_{i=1}^3 m_i \be^i$ where $\sum m_i=0$, and define the basis of Weyl invariants
\begin{align}
N_2 &:= m_1m_2+m_2m_3+m_3m_1,&
N_3 &:= m_1m_2m_3.
\end{align}
Then we find that
\begin{align}\label{}
M_2 &= N_2, &
M_3 &= -N_3.
\end{align}
We find two pole solutions:
\begin{align}
r_1\, \bw_1(\bm) &= r_1\, (m_1 + m_2),
\nonumber\\
(\sqrt2/\w_1)^2\, x_{\bw_1} &= - 2 u\, (m_1 + m_2),
\nonumber\\
(\sqrt2/\w_1)^3\, y_{\bw_1} &= 4 u^2,
\\[2mm]
r_2\, \bw_2(\bm) &= r_2\, (m_1 + m_2),
\nonumber\\
(2\sqrt2/\w_2)^2\, x_{\bw_2} &=
16 u\,  ( m_1 + m_2 ) + 
( 2 m_1 + m_2)^2 (m_1 + 2 m_2)^2,
\nonumber\\
(2\sqrt2/\w_2)^3\, y_{\bw_2} &= 
32 u^2 
+ 24 u\, (2m_1^3 + 7 m_1^2 m_2 + 7 m_1 m_2^2 + 2 m_2^3)
+ (2 m_1^2 + 5 m_1 m_2 + 2 m_2^2)^3 .
\nonumber
\end{align}
Summing over their Weyl orbits in the MN ansatz and solving the differential equation, we find a 1-parameter family of one forms,
\begin{align}
a &= -\sqrt{2}\,(r_1-2r_2),&
c &=0, &
W &= -\frac{9}{2\sqrt2} N_3\, r_2, &
1 &= r_1 - 2 r_2,
\end{align}
where the last constraint comes from choosing the normalization $2a-c=-2\sqrt{2}$.

\subsubsection{$\{ I_1,I_1^*\}$ with $\U(1)$ flavor symmetry}

The $\{I_1,I_1^*\}$ deformation of the $IV^*$ singularity has the curve
\begin{align}
y^2 = x^3 - 3 u^2 x M_1^2+ 2 u^3 (u + M_1^3).
\end{align}
The fact that there is only a single mass of dimension 1 implies  that there is a $\U(1)$ flavor symmetry wiht trivial Weyl group.  Without loss of generality, we take $M_1:=m$ to be the linear mass.

The poles are given by
\begin{align}\label{IVsU1pole}
r_1\, \bw_1(\bm) &:= r_1\, m, 
\nonumber\\
(\sqrt2/m)^2\, x_{\bw_1}  &= 2u\,  m ,
\nonumber\\
(\sqrt2/m)^3\, y_{\bw_1} &= 4u^2 ;
\nonumber\\[2mm]
r_2\,\bw_2(\bm) &:= r_2\,  m, 
\nonumber\\
(\sqrt2/m)^2\, x_{\bw_2} &= -4 u\,  m,
\nonumber\\
(\sqrt2/m)^3\, y_{\bw_2} &= 4 u^2;
\\[2mm]
r_3\,  \bw_3(\bm) &:= r_3\, m, 
\nonumber\\
(2\sqrt2/m)^2\, x_{\bw_3} &=
32 m u + 81 m^4
\nonumber\\
(2\sqrt2/m)^3\, y_{\bw_3} &=
32 u^2+ 432 m^3 u + 729 m^6;
\nonumber\\[2mm]
r_4\, \bw_4(\bm) &:= r_4\, m, 
\nonumber\\
(3\sqrt2)^2\, x_{\bw_4} &=
16 u^2 + 18 m^3 u,
\nonumber\\
(3\sqrt2)^3\, y_{\bw_4} &= 
32 u^3 + 108 m^3 u^2.
\nonumber
\end{align}
Summing over these poles in the MN ansatz for the SW 1-form and imposing the differential condition we find a three-parameter family of solutions for the one-form:
\begin{align}
a &= \frac{\sqrt2}{36} (3r_1 - 6 r_2 + 12 r_3 - 5 r_4),& 
c &= \frac{2\sqrt2}{9} r_4,
\\
W &= \frac{9}{4\sqrt2} m^3 r_3,&
1 &= r_1 - 2 r_2 + 4 r_3 - 3 r_4,
\nonumber
\end{align}
where the last relation comes from choosing the normalization $2a-c=1/(3\sqrt2)$.

\subsection{Deformations of the $I_0^*$ singularity}\label{appA.4}

There are three deformations of the $I_0^*$ singularity: $I_0^*\to\{{I_1}^6\}$, $\{{I_2}^3\}$, and $\{{I_1}^2,I_4\}$.  We present here the curves and one forms for the second and third (submaximal) deformations.   Even though a curve and 1-form for the second deformation was constructed in \cite{Seiberg:1994aj}, we re-compute it using the MN ansatz for comparison.

Both the quadratic and linear ansatz are consistent for the $x$ pole position for the $I_0^*$ curve, so we will consider pole positions of the form \eqref{LinAnsatz} and \eqref{QuadAnsatz}.  Since the $I_0^*$ singularity depends on a dimensionless parameter, $\t$, the coefficients in the MN ansatz may also depend on it.  Since $\Delta(u)=2$, we have $\Delta(R)=2$, $\Delta(S)=4$, $\Delta(J)=2$, $\Delta(K)=4$, $\Delta(L)=6$, and $\Delta(R')=0$, $\Delta(S')=2$, $\Delta(K')=1$ and $\Delta(L')=3$.  Note that in this case there is no $u^2$ term in $y$ in the linear ansatz.

\subsubsection{$\{ {I_1}^2,I_4\}$ with $\Sp(2)$ flavor symmetry}

The $\{{I_1}^2,I_4\}$ deformation of the $I_0^*$ singularity has curve
\begin{align}\label{I0*C1sigM}
y^2 &= x^3
-\frac13 x [ u^2 (1+3\a^2) + 8 u m^2 \a^2 + 4 m^4 \a^4]
\\
&\qquad\text{}
-\frac2{27} \left[u^3 (9\a^2-1)
+3 u^2 m^2 \a^2 (5+3\a^2)
+24 u m^4 \a^4
+8 m^6 \a^6 \right].
\nonumber
\end{align}
Here $\a$ is the marginal coupling.
There is just a single dimension-2 mass, implying that the curve has a discrete Weyl$(C_1)\simeq \Z_2$ group of symmetries acting on the linear masses.  Thus without loss of generality we have taken this mass polynomial to be $m^2$ in terms of a linear mass, $m$.

We find four pole positions for which the curve factorizes,
\begin{align}\label{I0*C1pole}
r_1\, \bw_1(\bm) &:= r_1\, m,
\nonumber\\
m^{-2}\, x_{\bw_1} 
&= -\frac23 (u + m^2(3-2\a^2)),
\nonumber\\
m^{-3}\, y_{\bw_1} 
&= i \sqrt 2 (\a^2-1)\, m\, (u + 2 m^2) ;
\nonumber\\[2mm]
r_\pm\,\bw_\pm(\bm) &:= r_\pm\, m,
\nonumber\\
m^{-2}\, x_{\bw_\pm} 
&= \frac13 \left((1 \pm 3\a)\, u- 2 m^2 \a^2\right) ,
\nonumber\\
m^{-3}\, y_{\bw_\pm} 
&= i \sqrt2 \, \a (\a \pm 1)\, m\, u  
\nonumber\\[2mm]
r_2\, \bw_2(\bm) &:= r_2 \, m,
\nonumber\\
x_{\bw_2} 
&=-\frac{1}{6} (3 u^2 + 4 m^2 u + 4 m^4 \a^2),
\nonumber\\
y_{\bw_2} 
&=\frac{i}{2\sqrt2} u^2 (u+2m^2).
\nonumber
\end{align}

Combining these with their transforms under the $C_1$ Weyl group, we can solve for the one-form constraints.  Choosing, for convenience, a normalization $2a-c=-i/\sqrt2$ we get:
\begin{align}
a &= \frac{i}{4\sqrt2} (2 r_1 - r_+ - r_- - r_2),
& c &=\frac{i}{2\sqrt2}  r_2,
\nonumber\\
r_2 &= 1 + r_1 - \frac12 r_+ - \frac12 r_-,& W &= \frac{i}{2\sqrt2} m^2 (2 r_1 + \a r_+ - \a r_- ).
\end{align}
a 3-parameter family of solutions.  Rational $(r_1,r_\pm)$ then give solutions consistent with the interpretation as the low energy theory on the CB of an $\cN=2$ SCFT.

\subsubsection{$\{ {I_2}^3\}$ with $\Sp(2)$ flavor symmetry}

The $\{{I_2}^3\}$ deformation of the $I_0^*$ singularity was found in \cite{Seiberg:1994aj} to have a curve
\begin{align}\label{I0*C1sigSW}
Y^2 &= \prod_{j=1}^3 (\Xi-e_j U-e_j^2 M^2) .
\end{align}
The $e_j(\t)$ are modular forms of the marginal coupling, and $M=m/2$, $U=\til u$, $Y=y$, and $\Xi=x$ in the notation of \cite{Seiberg:1994aj}.  The curve is invariant under permutations of the $e_j$, and the $e_j$ satisfy $0 = \sum_j e_j $.  In Weierstrass form the curve becomes
\begin{align}\label{I0*C1sigSW2}
Y^2 = \prod_j 
\left[ X - U e_j - M^2 \left(e_j^2 - \frac13\sum_k e_k^2 
\right)\right]
\end{align}
in terms of a shifted coordinate $X = \Xi - (1/3) M^2 \sum_k e_k^2$.

We now solve for the one-form using the MN ansatz \eqref{MNform}, trying both a quadratic and a linear ansatz for the pole position.  We find five pole solutions for which the curve factorizes,
\begin{align}\label{I0*C1poleSW}
r_j\, \bw_j(\bm) &:= r_j\, M, \qquad\qquad 
\text{for} \ \ j = 1,2,3
\nonumber\\
M^{-2} X_{\bw_j} 
&= U e_j-M^2 \left(2 e_j^2-\frac16 \sum_k e_k^2 \right),
\nonumber\\
M^{-3} Y_{\bw_j} 
&= i \left(3 e_j^2 - \frac12\sum_k e_k^2 \right) M 
(U - M^2 e_j) ;
\nonumber\\[2mm]
r_4\, \bw_4(\bm) &:= r_4 \, M, 
\nonumber\\
M^{-2} X_{\bw_4} 
&=-\frac{1}{M^2} \left(\frac14 U^2 + \frac13 M^4 \sum_k e_k^2 \right),
\\
M^{-3} Y_{\bw_4} 
&=-\frac{i}{M^3} \prod_j \left(\frac12 U+M^2e_j\right).
\nonumber\\[2mm]
r_5\,\bw_5(\bm) &:= r_5 \, M,
\nonumber\\
i^{-2} X_{\bw_5} 
&=U^2 - \frac16 M^4 \sum_k e_k^2 ,
\nonumber\\
i^{-3} Y_{\bw_5} 
&= \prod_j (U-M^2e_j).
\nonumber
\end{align}
Sum over the Weyl$(C_1)$ orbits of these poles in the MN ansatz and solve the one-form constraint to find a 4-parameter family of solutions,  
\begin{align}
a &= -\frac{i}{8} \left(2 \sum_{j=1}^3 r_j + 3 r_4 \right) ,
& c &= \frac{i}{4} (r_4 + 2 r_5) ,
\nonumber\\
r_5 &= 1 - 2 r_4 - \sum_{j=1}^3 r_j ,& W &= \frac{i}{2} M^2\, \sum_{j=1}^3 e_j r_j  ,
\end{align}
where we have chosen, for convenience, the normalization $2a-c=-i/2$.  Rational $(r_j,r_4)$ give solutions consistent with the interpretation as the low energy theory on the CB of an $\cN=2$ SCFT.  The 1-form found in \cite{Seiberg:1994aj} corresponds to $r_j=\frac13$ and $r_4=0$.

\subsection{Deformation of the $IV$ singularity}

In this case only the quadratic ansatz \eqref{QuadAnsatz} for the pole positions can work, where the coefficients are polynomials in the linear masses and chiral deformation parameter, $M_{1/2}$.  $\Delta(u)=3/2$ implies $\Delta(R)=3/2$, $\Delta(S)=3$, $\Delta(J)=3/2$, $\Delta(K)=3$ and $\Delta(L)=9/2$. 

\subsubsection{$\{ I_1^4\}$ with $\SU(3)$ flavor symmetry}\label{c1}

The $\{ I_1^4\}$ deformation of the $IV$ singularity has curve
\begin{align}
y^2 = x^3 + x(u M_{1/2} + M_2 ) + (u^2 + M_3 ).
\end{align}
Choose a basis, $\bm= \sum_{i=1}^3 m_i \be^i$ where $\sum m_i=0$, of the linear masses so that Weyl$(A_2)$ acts by permutations of the three $m_i$.  A basis of Weyl invariants of the linear masses is
\begin{align}\label{A2Ndef}
N_2 &:= m_1 m_2 + m_2 m_3 + m_1 m_3 , & 
N_3 &:= m_1 m_2 m_3 .
\end{align}
Then either by solving the factorization condition for the MN ansatz for the SW one-form, or by demanding increased zero multiplicities of the curve's discriminant when $\ba(\bm)=0$ for $\ba$'s fixed by the Weyl group, we find that the parameters in the curve are related to the linear masses by
\begin{align}
M_3 &= - \frac{1}{1728} M^6_{1/2} - \frac{1}{12} M^6_{1/2} N_2 + N_3, &  
M_2 &=\frac{1}{48}M^4_{1/2} + N_2 .
\end{align}

The solution for the pole position in this case is:
\begin{align}
r_1\, \bw(\bm) &:= r_1\, (m_1-m_2),
\nonumber\\
(12)^2 x_\bw &= 
576 u^2 + u \bigl(-288 (m_1+ m_2) M_{1/2} 
+ 48 M^3_{1/2}\bigr)+144 (m_1-m_2)^2 (m_1 + m_2) \nonumber\\
&\qquad \mbox{}+12 (m_1^2 + 10 m_1 m_2 + 
m_2^2) M^2_{1/2} - 12(m_1 + m_2)M^4_{1/2}
+ M^6_{1/2} ,
\nonumber\\
(12)^3 y_\bw &= 
13824 u^3 
- u^2 \bigl(10368  (m_1+ m_2) - 1728   M^2_{1/2}\bigr) M_{1/2} 
\\
& \qquad \mbox{}
+ u \bigl( 
+ 5184 (m_1- m_2)^2  (m_1+ m_2) 
+ 1728 (m_1^2 + 4m_1m_2 + m_2^2)M^2_{1/2} 
\nonumber\\
& \qquad \mbox{} 
- 864 (m_1 + m_2) M^4_{1/2} 
+ 72 M^6_{1/2} \bigr)
- 864 (m_1 - m_2)^2 (m_1^2 + 4m_1m_2 + m_2^2)M_{1/2}
\nonumber\\
& \qquad \mbox{} 
+ 216 (m_1+ m_2) (m_1^2 - 6m_1m_2 +m_2^2)M^3_{1/2} 
+ 72(m_1^2 + 4m_1m_2 + m_2^2)M^5_{1/2} 
\nonumber\\
& \qquad \mbox{} 
- 18  (m_1+ m_2) M^7_{1/2}
+ M^9_{1/2}.
\nonumber
\end{align}
Summing over the Weyl orbit of this solution and imposing the arbitrary normalization of $2a-c=6$, we find the unique solution for the one-form:
\begin{align}
a = 1,\qquad c= -4.\qquad  W = \frac{1}{24}M_{1/2}^3, \qquad b= \frac{1}{3}, \qquad r_1 = -\frac{1}{2}.
\end{align}

\subsection{Deformation of the $III$ singularity}

Also in the case of deformations of the $III$ singularity we can only apply the quadratic ansatz \eqref{QuadAnsatz} for the pole positions.  $\D(u)=4/3$ implies $\D(R)=4/3$, $\D(S)=8/3$, $\D(J)=4/3$, $\D(K)=8/3$ and $\D(L)=4$. 

\subsubsection{$\{ I_1^3\}$ with $\SU(2)$ flavor symmetry}\label{c2}

The $\{I_1^3\}$ deformation of the $III$ singularity has curve
\begin{align}\label{2IIIcurves}
y^2 &= x^3 + u x + u M_{2/3} - M_2.
\end{align}
We find that
\begin{align}\label{}
M_2 = m^2-M^3_{2/3}
\end{align}
in terms of a linear mass $m$ on which Weyl$(A_1)$ acts by a sign change.  We find the pole solution
\begin{align}
r_1\, \bw_1(\bm) &:= r_1\, m_1,
\nonumber\\
(2i)^2 x_\bw & = 
u^2 
- 6 u M_{2/3} 
+ 8 m_1^2 M_{2/3} - 9M^4_{2/3},
\\
(2i)^3 y_\bw & = 
u^3 
+ 9 u^2 M^2_{2/3} 
- 3 u M_{2/3} (4 m_1^2 - 9 M^3_{2/3} )
+ 8 m_1^4 
- 36 m_1^2 M^3_{2/3} 
+ 27 M^6_{2/3}.
\nonumber
\end{align}
Summing over the Weyl orbit of this pole, we obtain the unique solution for the one-form (having normalized $2a-c=2i$):
\begin{align}
a =i\frac{5}{8},\qquad  c = -i\frac{3}{4},\qquad W=i \frac{9}{8}M^2_{2/3} , \qquad b = -i\frac{7}{8}, \qquad r_1=1.
\end{align}

\subsection{Deformation of the $II$ singularity}

For the $II$ singularity we do not need to introduce an ansatz for the pole positions since there are no poles in the only possible deformation of this singularity.

\subsubsection{$\{ I_1^2\}$ with no flavor symmetry}\label{c3}

The curve for the $II\to\{I_1^2\}$ deformation is
\begin{align}
y^2 = x^3 + x M_{4/5} + u.
\end{align}
There is no flavor symmetry, so no poles or residues in the one-form.  It is then straightforward to solve for the differential constraint for the one-form to find
\begin{align}
a=1, \qquad c=0, \qquad W=0, \qquad b=0.
\end{align} 

\bibliographystyle{JHEP}
\providecommand{\href}[2]{#2}\begingroup\raggedright\endgroup

\end{document}